\documentclass[runningheads,openany]{svmult}

\usepackage{palatino}
\usepackage{euler}
\usepackage{helvet}

\usepackage{graphicx}  % standard LaTeX graphics tool
\usepackage{physprbb}  
\usepackage{proc}  
\usepackage{epsfig}

\usepackage{amssymb}
\usepackage{axodraw}
\usepackage{rotating}

\makeindex             % used for the subject index
\usepackage{amsmath}   % useful for coding complex math
\usepackage{slashed}
\let\tocauthor\authorrunning

\begin{document}

\frontmatter

\thispagestyle{empty}
\parindent=0pt

{\Large\sc Blejske delavnice iz fizike \hfill Letnik~3, \v{s}t. 4}

\smallskip

{\large\sc Bled Workshops in Physics \hfill Vol.~3, No.~4}

\smallskip

\hrule

\hrule

\hrule

\vspace{0.5mm}

\hrule

\medskip
{\sc ISSN 1580--4992}

\vfill

\bigskip\bigskip
\begin{center}

{\bfseries 
{\Huge  Proceedings to the workshops\\
What comes beyond the Standard model 2000, 2001, 2002}

\vspace{5mm}
\centerline{\Large  Volume 2} 
\vspace{5mm}
\centerline{\Huge Proceedings --- PART I} }

\vfill

{\bfseries\large
Edited by

\vspace{5mm}
Norma Manko\v c Bor\v stnik\rlap{$^{1,2}$}

\smallskip

Holger Bech Nielsen\rlap{$^{3}$}

\smallskip

Colin D. Froggatt\rlap{$^{4}$}

\smallskip

Dragan Lukman\rlap{$^2$}

\bigskip

{\em\normalsize $^1$University of Ljubljana, $^2$PINT, %
$^3$ Niels Bohr Institute, $^4$ Glasgow University}

\vspace{12pt}

\vspace{3mm}

\vrule height 1pt depth 0pt width 54 mm}

\vspace*{3cm}

{\large {\sc  DMFA -- zalo\v{z}ni\v{s}tvo} \\[6pt]
{\sc Ljubljana, december 2002}}
\end{center}
\newpage
\thispagestyle{empty}
\parindent=0pt
\begin{flushright}
{\parskip 6pt
{\bfseries\large
                  The 5th Workshop \textit{What Comes Beyond  
                  the Standard Model}}

\bigskip\bigskip

{\bf was organized by}

{\parindent8pt
\textit{Department of Physics, Faculty of Mathematics and Physics,
University of Ljubljana}

\textit{Primorska Institute of Natural Sciences and Technology, Koper}}

\bigskip

{\bf and sponsored by}

{\parindent8pt
\textit{Ministry of Education, Science and Sport of Slovenia}

\textit{Department of Physics, Faculty of Mathematics and Physics,
University of Ljubljana}

\textit{Primorska Institute of Natural Sciences and Technology, Koper}

\textit{Society of Mathematicians, Physicists and Astronomers
of Slovenia}}}
\bigskip
\medskip

{\bf Organizing Committee}

\medskip

{\parindent9pt
\textit{Norma Manko\v c Bor\v stnik}

\textit{Colin D. Froggatt}

\textit{Holger Bech Nielsen}}

\end{flushright}

\setcounter{tocdepth}{0}
\tableofcontents
\cleardoublepage
\chapter*{Preface}
\addcontentsline{toc}{chapter}{Preface}

The series of workshops on "What Comes Beyond the Standard Model?" started 
in 1998 with the idea of organizing a real workshop, in which participants 
would spend most of the time in discussions, confronting different 
approaches and ideas. The picturesque town of Bled by the lake of the 
same name, surrounded by beautiful mountains and offering pleasant walks, 
was chosen to stimulate the discussions.

The idea was successful and has developed into an annual workshop. 
This year was a kind of small jubilee - the fifth workshop took place. 
Very open-minded and fruitful discussions have become the trade-mark of 
our workshop, producing several published works. It takes place in 
the house of Plemelj, which belongs to the Society of Mathematicians, 
Physicists and Astronomers of Slovenia.

These workshops have also inspired a series of EUROCONFERENCES, with 
the same name as the workshop (``WHAT COMES BEYOND THE STANDARD MO\-DEL''). 
The first meeting in the series entitled
``EUROCONFERENCE ON SYMMETRIES BEYOND THE STANDARD MODEL'' will take place
from 12 of July to 17 of July 2003 at hotel Histrion in Portoro\v z, 
Slovenia.  This series of conferences is also meant to confront the ideas, 
knowledge and experiences derived from different approaches to describing 
Nature beyond the Standard models of particle physics and cosmology.

In the fifth workshop, which took place from 13 to 24 of July 2002 at Bled, 
Slovenia, we have tried to answer some of the open questions which the 
Standard models leave unanswered, like:
 
\begin{itemize}
\item  Why has Nature made a choice of four (noticeable) dimensions while all 
the others, if existing, are hidden? And what are the properties of 
space-time in the hidden dimensions? 
\item  How could Nature make the decision about the breaking of symmetries 
down to the noticeable ones, coming from some higher dimension d?
\item  Why is the metric of space-time Minkowskian and how is the choice 
of metric connected with the evolution of our universe(s)? 
\item  Where does the observed asymmetry between matter and antimatter 
originate from?
\item  Why do massless fields exist at all? Where does the weak scale 
come from?
\item  Why do only left-handed fermions carry the weak charge? Why does 
the weak charge break parity?
\item  What is the origin of Higgs fields? Where does the Higgs mass come from?
\item  Where does the small hierarchy come from? (Or why are some Yukawa 
couplings so small and where do they come from?) 
\item  Do Majorana-like particles exist?
\item  Where do the generations come from? 
\item  Can all known elementary particles be understood as different states of 
only one particle, with a unique internal space of spins and charges?
\item  How can all gauge fields (including gravity) be unified and quantized?
\item  Why do we have more matter than antimatter in our universe?
\item  What is our universe made out of (besides the baryonic matter)?
\item  What is the role of symmetries in Nature?
\item  What is the origin of the field which caused inflation?
\end{itemize}

We have discussed these questions for ten days. Some results of this effort 
appear in these Proceedings, not only of the last workshop, but also of 
two earlier workshops. 
The discussion will continue next year; at the EURESCONFERENCE and in the 
Workshop, which will take place after the conference, from 17 of July to 
28 the of July, again at Bled, again in the house of Josip Plemelj.

The organizers are grateful to all the participants for the lively discussions 
and the good working atmosphere.\\[2 cm]

\parbox[b]{40mm}{%
                 \textit{Norma Manko\v c Bor\v stnik}\\
                 \textit{Holger Bech Nielsen}\\
		 \textit{Colin Froggatt}\\
		 \textit{Dragan Lukman} }
\qquad\qquad\qquad\qquad\qquad\qquad\quad
\textit{Ljubljana, December 2002}

\newpage
\thispagestyle{empty}

{\bf Workshops organized at Bled}

\begin{description}
\item[$\triangleright$]
What Comes beyond the Standard Model (June 29--July 9, 1998)
        \item[$\triangleright$]
Hadrons as Solitons (July 6-17, 1999)
        \item[$\triangleright$]
What Comes beyond the Standard Model (July 22--31, 1999)
        \item[$\triangleright$]
Few-Quark Problems (July 8-15, 2000)
        \item[$\triangleright$]
What Comes Beyond the Standard Model (July 17--31, 2000)
        \item[$\triangleright$]
Statistical Mechanics of Complex Systems (August 27--September 2, 2000)
        \item[$\triangleright$]
What Comes beyond the Standard Model (July 17--27, 2001)
        \item[$\triangleright$]
Studies of Elementary Steps of Radical Reactions in Atmospheric Chemistry
(August 25--28, 2001)
        \item[$\triangleright$]
What Comes Beyond the Standard Model (July 13--24, 2002)

\end{description}

\cleardoublepage
%%%%%%%%%%%%%%%%%%%%%%%%%%%%%%%%%%%%%%%%%%%%%%%%%%%%%%%%%%%%%

\mainmatter
\parindent=20pt
\setcounter{page}{5}
\setcounter{page}{1}
\title*{Derivation of Lorentz Invariance and Three Space Dimensions %
 in Generic Field Theory}
\author{C~D.~Froggatt${}^{\dag}$\thanks{E-mail: c.froggatt@physics.gla.ac.uk}
and H.~B.~Nielsen${}^{\ddag}$\thanks{E-mail: hbech@mail.desy.de; hbech@nbi.dk}}
\institute{%
${}^{\dag}$ Department of Physics and Astronomy,
Glasgow University,
Glasgow G12 8QQ, Scotland\\
${}^{\ddag}$ Deutsches Elektronen-Synchrotron DESY,
Notkestra{\ss}e 85,
D-22603 Hamburg, Germany
and
The Niels Bohr Institute,
Blegdamsvej 17, Copenhagen {\O}, Denmark}

\titlerunning{Derivation of Lorentz Invariance and Three Space Dimensions}
%% in Generic Field Theory}
\authorrunning{C~D.~Froggatt and H.~B.~Nielsen}
\maketitle

\begin{abstract}
A very general quantum field theory, which is not even
assumed to be Lorentz invariant, is studied in the limit of very
low energy excitations. Fermion and Boson field theories are
considered in parallel. Remarkably, in both cases it is argued
that, in the free and lowest energy approximation, a relativistic
theory with just three space and one time dimension emerges for
each particle type separately. In the case of Fermion fields it is
in the form of the Weyl equation, while in the case of the Bosons
it is essentially in the form of the Maxwell equations.
\end{abstract}
%%%%%%%%%%%%%%%%%%%%%%%%%%%%%%%%%%%%%%%%%%%%%%%%%%%%%%%

\section{Introduction}

Since many years ago~\cite{RDold}, we have worked on the project of
``deriving'' all the known laws of nature, especially the symmetry
laws~\cite{book}, from the assumption of the existence of
exceedingly complicated fundamental laws of nature. However the
derivations are such {\em that it practically does not
matter what these exceedingly complicated laws are in detail, just
provided we only study them in some limits } such as the low energy
limit. This is the project which we have baptized ``Random dynamics'',
in order to make explicit the idea that we are thinking of
the fundamental laws of nature as being given by a particular model
pulled out at random from a very large class of
models. In this way, one can overcome the immediate reproach to the
project that it is easy to invent model-proposals which, indeed, do not
deliver the laws of nature as we know them today. We only make the
claim that sufficiently
complicated and generic models should work, not very special ones that
could potentially be constructed so as  not to work.
Also it should be stressed
that there is a lot of interpretation involved,
as to which elements in the ``random'' model
are to be identified phenomenologically with what.
As a consequence, the project tends to
be somewhat phenomenological itself, honestly speaking.
However, in principle,
we should only use the phenomenology to find out which quantities in the
``random dynamics'' model are to be identified with which physically defined
quantities (concepts).

One of the most promising steps, in developing this random dynamics
pro\-ject, was~\cite{RDold,book} to start without assuming
Lorentz invariance but to assume that we already have several
known laws such as quantum mechanics, quantum
field theory and momentum conservation. Lorentz invariance was then
``derived'', at least for a single species of Weyl particles which
emerged at low energy. However this ``derivation'' of Lorentz invariance
might not actually be the most interesting result from this step in
random dynamics; it is after all not such an overwhelming success,
since it only works for one particle species on its own and does not,
immediately at least, lead to Lorentz invariance if several particle
species are involved. It may rather be the prediction of the number
of space dimensions which is more significant.
Actually the fundamental model is assumed to have an arbitrary number of
dimensions and has momentum degrees of freedom in all these dimensions,
but the velocity components in all but three
dimensions turn out to be zero. In this way the extra dimensions
are supposedly not accessible. So the prediction is effectively
that there are just three spatial dimensions (plus one time)!

In these early studies only a fermionic field theory (without Lorentz
symmetry) was considered, while Bosons were left out of consideration;
we then sometimes speculated that the Bosons could at least be
partly composed from Fermions and thus inherit their Lorentz symmetry.
Indeed, even in more recent work, it is the Fermions that play the main
role~\cite{NormaHolger,RughHolger}. For a summary of other recent 
theoretical models and experimental tests of Lorentz invariance 
breaking see, for example, reference \cite{Kostelecy}.  

It is the purpose of the present paper to review the work with Fermions
stressing a new feature aimed at solving a certain technical
problem---the use of the ``Homolumo-gap-effect'' to be explained
below---and to extend the work to the case of bosonic fields, which is
a highly non-trivial extension.

In the following section we shall put forward our very general
field theory model and then, in section 3, we shall write down in
parallel the equations of motion for Bosons and Fermions
respectively. It turns out that we obtain a common equation of
motion for the ``fields'' in ``momentum''
representation---momentum here being really thought of as a rather
general parameterisation of the degrees of freedom, on which the
Hamiltonian and commutation rules depend smoothly. This equation
of motion involves an antisymmetric matrix which depends on the
``momenta''. The behaviour of the eigenvalue spectrum of such an
antisymmetric real matrix is studied in section 4, with the help
of some arguments based on the Homolumo-gap-effect which are
postponed till section 5. The conclusions are put into section 6.

\section{A random dynamics model}

Since it is our main purpose to derive Lorentz symmetry 
together with $3 + 1$
dimensions, we must start from a model that does not
assume Lorentz invariance nor the precise number of space dimensions
from the outset.
We would, of course, eventually hope to avoid having to assume
momentum conservation or even the existence of the concept of
momentum. However this assumption is less crucial than the others,
since the derivation of Lorentz invariance is highly non-trivial
even if momentum conservation is assumed. Therefore, ``for
pedagogical reasons'', we shall essentially assume translational
symmetry and momentum conservation in our model---in practice
though we shall actually allow a small departure from
translational symmetry. That is to say we consider the model
described in terms of a Fock space, corresponding to having
bosonic or fermionic particles that can be put into single
particle states which are momentum eigenstates. This gives rise to
bosonic and fermionic
fields $\phi(\vec{p})$ and $\psi(\vec{p})$ annihilating these
particles. We shall formulate the model in terms of fields that
are essentially real or obey some Hermiticity conditions, which
mean that we can treat the fields $\phi(\vec{p})$ and
$\psi(\vec{p})$ as Hermitian fields. In any case, one can always
split up a non-Hermitian field into its Hermitian and
anti-Hermitian parts. This is done since, in the spirit of the
random dynamics project, we do not want to assume any charge
conservation law from the outset.

\subsection{Technicalities in a general momentum description}

In the very general type of model we want to set up, without any
assumed charge conservation, it is natural to
use a formalism which is suitable for neutral particles like,
say, $\pi^0$ mesons.  However,
when one constructs a second quantized formalism
from a single particle Fock-space description,
in which there can be different numbers of particles in the different single
particle states\footnote{In the fermionic case there can be 0 or 1
particle in a particular single particle state, while in the bosonic
case there can also be many.}, one at first gets
{\em ``complex''} i.e.~{\em non-Hermitian}
second quantized fields. In order to describe say the $\pi^0$-field,
one must put
restrictions on the allowed Fock-space states, so that one cannot
just completely freely choose how many particles there should be in each
single particle state. Basically one ``identifies'' particles and
antiparticles (= holes), so that they are supposed to be in analogous
states (in the Fermion case, it is the Majorana condition that must be
arranged). Field creation of a particle with momentum $\vec{p}$ is
brought into correspondence with annihilation of a particle
with momentum $-\vec{p}$.

In our general description of bosonic or fermionic second
quantized particles, we want to use a formalism of this $\pi^0$ or
Majorana type. We can always return to
a charged particle description by introducing a doubling of the
number of components for such a field; we can simply make a
non-Hermitian (i.e.~essentially charged) field component from two
Hermitian ones, namely the Hermitian (``real'') and
anti-Hermitian (``purely imaginary'') parts, each of which are
then Majorana or $\pi^0$-like. Let us recall here that the $\pi^0$
field is Hermitian when written as a field depending on the position
variable $\vec{x}$, while it is not Hermitian in momentum space.
In fact, after Fourier transformation, the property of
Hermiticity or reality in position space becomes the property,
in momentum representation, that the fields at $\vec{p}$
and $-\vec{p}$ are related by
Hermitian/complex conjugation:
\begin{equation}
\label{reflection}
\phi(\vec{p}) = \phi^{\dagger}(-\vec{p})
\end{equation}

For generality, we should also like to have
Hermitian momentum dependent fields, which corresponds to having
a similar reflection symmetry in position space, saying that
the values of the fields at $\vec{x}$ and $-\vec{x}$
are related by Hermitian/complex conjugation.
To make the ``most general'' formalism for our study, we should
therefore impose Hermiticity both in momentum {\em and} in
position representation. We then
have to accept that we also have a reflection symmetry in both
position and momentum space. In this paper, we shall
in reality only
consider this most general formalism for bosonic fields.
For this purpose, let us denote the $\pi^0$ field and its
momentum conjugate field by $\phi_0(\vec{p})$ and $\pi_0(\vec{p})$
respectively. Then, in standard relativistic quantum field theory,
the non-vanishing equal time commutation
relations between their real and imaginary parts are as
follows:
\begin{eqnarray}
\label{pi0commutators}
\left[ \mathrm{Re}\,\phi_0(\vec{p}),
\mathrm{Re}\,\pi_0(\vec{p'}) \right]
=\frac{i}{2} \left(\delta (\vec{p} - \vec{p'})
+\delta(\vec{p} + \vec{p'}) \right) \\
\left[ \mathrm{Im}\,\phi_0(\vec{p}),
\mathrm{Im}\,\pi_0(\vec{p'}) \right]
=\frac{i}{2} \left(\delta (\vec{p} - \vec{p'})
-\delta(\vec{p} + \vec{p'}) \right)
\end{eqnarray}
We note that the appearance of the $\delta(\vec{p} + \vec{p'})$
function as well as the $\delta(\vec{p} - \vec{p'})$ function is a
consequence of the reflection symmetry (\ref{reflection}).

Now the reader should also notice that we are taking the point of
view that many of the observed laws of nature are only laws of
nature in the limit of {\em ``the poor physicist''}, who is
restricted to work with the lowest energies and only with a small
range of momenta compared to the fundamental (Planck) scale. In
the very generic and not rotational invariant type of model which
we want to consider, it will now typically happen that the small
range of momenta to which the physicist has access is not centred
around zero momentum---in the presumably rather arbitrary choice
of the origin for momentum---but rather around some momentum,
$\vec{p}_0$ say. This momentum $\vec{p_0}$ will generically be
large compared to the momentum range accessible to the poor
physicist; so the reflection symmetry in momentum space and the
associated $\delta(\vec{p} + \vec{p'})$ terms in commutators will
not be relevant to the poor physicist and can be ignored. However,
in our general field theory model, there can be a remnant
reflection symmetry in position space. Indeed we shall see below
that what may be considered to be a mild case of momentum
non-conservation does occur for the Maxwell equations derived in
our model: there is the occurrence of a reflection centre
somewhere, around which the Maxwell fields should show a parity
symmetry in the state of the fields. If we know, say, the electric
field in some place, then we should be able to conclude from this
symmetry what the electric field is at the mirror point. If, as is
most likely, this reflection point is far out in space, it would
be an astronomical challenge to see any effect of this lack of
translational symmetry. In this sense the breaking of
translational symmetry is very ``mild''.

\subsection{General Field Theory Model}
At the present stage in the development of our work, it is assumed
that we {\em only work to the free field approximation} and thus
the Hamiltonian is taken to be bilinear in the Hermitian fields
$\psi(\vec{p})$ and $\phi(\vec{p})$. Also, because of the assumed
rudiment of momentum conservation in our model, we only consider
products of fields taken for the same momentum $\vec{p}$.
In other words our Hamiltonian takes the following form:
\begin{equation}
H_F =\frac{1}{2} \int d\vec{p} \:
\sum_{i,j}\psi_i(\vec{p})\psi_j(\vec{p})
H_{ij}^{(F)}(\vec{p})
\end{equation}
and
\begin{equation}
H_B =\frac{1}{2} \int d\vec{p} \:
\sum_{i,j}\phi_i(\vec{p})
\phi_j(\vec{p})H_{ij}^{(B)}(\vec{p}).
\end{equation}
for Fermions and Bosons respectively.
Here the coefficient functions $H_{ij}^{(F)}(\vec{p})$ and
$H_{ij}^{(F)}(\vec{p})$ are non-dynamical in the free field
approximation and just reflect the general features of ``random''
laws of nature expected in the random dynamics project. That is to
say we do not impose Lorentz invariance conditions on these
coefficient functions, since that is what is hoped to {\em come
out} of the model. We should also not assume that the $\vec{p}$
vectors have any sort of Lorentz transformation properties {\it a
priori} and they should not even be assumed to have, for instance,
3 spatial dimensions. Rather we start out with $D>3$
spatial dimensions;
then one of our main achievements will be to show that the
velocity components in all but a three dimensional subspace are
zero. It is obvious that, in these expressions, the coefficient
functions  $H_{ij}^{(F)}(\vec{p})$ and $H_{ij}^{(F)}(\vec{p})$ can
be taken to have the symmetry properties:
\begin{equation}
 H_{ij}^{(F)}(\vec{p})=-  H_{ji}^{(F)}(\vec{p})\quad \quad
\hbox{ and}\quad\quad
H_{ij}^{(B)}(\vec{p})=
H_{ji}^{(B)}(\vec{p}).
\end{equation}

However, it should be borne in mind that {\it a priori} the fields are
arbitrarily normalised and that we may use the Hamiltonians
to define the normalisation of the fields, if we so choose.
In fact an important ingredient in the formulation of the present
work is to assume that a linear transformation has been made on the
various field components $\phi_i(\vec{p})$, i.e.~a transformation
on the component index $i$, such that the symmetric coefficient
functions $H_{ij}^{(B)}(\vec{p})$ become equal to the unit matrix:
\begin{equation}
H_{ij}^{(B)}(\vec{p}) = \delta_{ij} \qquad
\hbox{(by normalisation for all $\vec{p}$)}
\end{equation}
Thereby, of course, the commutation relations
among these components $\phi_i(\vec{p})$ are modified and
we cannot simultaneously arrange for them to be trivial. So for the Bosons
we choose a notation in which the non-trivial behaviour of
the equations of motion, as a function of  the momentum
$\vec{p}$, is put into
the commutator expression\footnote{Note that we are here
ignoring possible terms of the form
$iB_{ij}(\vec{p}) \delta(\vec{p} + \vec{p'})$ as irrelevant
to the poor physicist, according to the discussion after equation
(\ref{pi0commutators}).}
\begin{equation}
[\phi_i(\vec{p}), \phi_j(\vec{p'})]=iA_{ij}(\vec{p}) \delta(\vec{p} - \vec{p'})
\label{commutator}
\end{equation}
It follows that the information which we would, at first,
imagine should be contained in the Hamiltonian is, in fact,
now contained in the antisymmetric matrices $A_{ij}(\vec{p})$.

For the Fermions, on the other hand, we shall keep to the more
usual formulation. So we normalize the anti-commutator to be the
unit matrix and let the more
nontrivial dependence on $\vec{p}$ sit in the Hamiltonian
coefficient functions $H_{ij}^{(F)}(\vec{p})$. That is to say that
we have the usual equal time anti-commutation relations:
\begin{equation}
\{ \psi_i(\vec{p}), \psi_j(\vec{p'})\} =
\delta_{ij} \delta(\vec{p} - \vec{p'}).
\end{equation}

The component indices $i$, $j$ enumerate the very general discrete
degrees of freedom in the model. These degrees of freedom might, at the
end, be identified with Hermitian and anti-Hermitian components, spin
components, variables versus conjugate momenta or even totally
different types of particle species, such as flavours etc.
It is important to realize that this model
is so general that it has, in that sense, almost no assumptions built
into it---except for our free approximation, the above-mentioned rudimentary
momentum conservation and some general features of second quantized models.
It follows from the rudimentary momentum conservation in our model that the
(anti-)commutation relations have a $\delta(\vec{p} - \vec{p'})$ delta
function factor in them.

Obviously the Hermiticity of the Hamiltonians for the second quantized
systems means that the matrices $H_{ij}^{(F)}(\vec{p})$ and
$H_{ij}^{(B)}(\vec{p})$ are Hermitian
and thus have purely imaginary and real matrix elements respectively.
Similarly, after the extraction of the $i$ as a conventional factor in
equation (\ref{commutator}), the matrix $A_{ij}(\vec{p})$
has real matrix elements and is antisymmetric.

\section{Equations of motion for the general fields}
\vspace{.3cm}

\noindent
We can easily write down the equations of motion for the field components
in our general quantum field theory, both in the fermionic case:
\begin{equation}
\dot{\psi}_i(\vec{p}) = i[H_F, \psi_i(\vec{p})] =
 i \sum_k\psi_k(\vec{p})H^{(F)}_{ki}(\vec{p})
\end{equation}
and in the bosonic case:
\begin{equation}
\label{blhbeq8} \dot{\phi}_i(\vec{p}) = i[H_B, \phi_i(\vec{p})] = -
\sum_k\phi_k(\vec{p})A_{ki}(\vec{p}).
\end{equation}

Since $H^{(F)}_{ij}(\vec{p})$ has purely imaginary matrix elements,
we see that both the bosonic and the fermionic equations of motion
are of the form
\begin{equation}
\dot{\xi_i}(\vec{p})= \sum_k A_{ik}\xi_k(\vec{p})
\end{equation}
In the fermionic case we have extracted a factor of $i$,
by making the definition
\begin{equation}
H^{(F)}_{ij}(\vec{p}) = i A_{ij}(\vec{p}).
\end{equation}
Also the Boson field $\phi$ and the Fermion field $\psi$ have both
been given the neutral name $\xi$ here.

\section{Spectrum of an antisymmetric matrix depending on $\vec{p}$}

An antisymmetric matrix $A_{ij}(\vec{p})$ with real matrix elements is
anti-Hermitian and thus has purely imaginary eigenvalues. However, if
we look for a time dependence ansatz of the form
\begin{equation}
\xi_i(\vec{p},t)= a_i(\vec{p}) \exp(-i\omega t),
\end{equation}
the eigenvalue equation for the frequency $\omega$ becomes
\begin{equation}
\omega a_i= \sum_j iA_{ij}(\vec{p})a_j.
\end{equation}
Now the matrix $iA_{ij}(\vec{p})$ is Hermitian and the
eigenvalues $\omega$ are therefore real.

It is easy to see, that if $\omega$ is an eigenvalue, then so also is
$-\omega$. In fact we could imagine calculating the eigenvalues by
solving the equation
\begin{equation}
\det{(iA - \omega)} =0
\end{equation}
We then remark that transposition of the matrix $(iA-\omega)$ under the
determinant sign will not change the value of the determinant, but
corresponds to changing the sign of $\omega$ because of the antisymmetry
of the matrix $iA$.
So non-vanishing eigenvalues occur in pairs.

In order to compare with the more usual formalism, we should
really keep in mind that the creation operator for a particle with
a certain $\omega$-eigenvalue is, in fact, the annihilation
operator for a particle in the eigenstate with the opposite value
of the eigenvalue, i.e.~$-\omega$. Thus, when thinking in usual
terms, we can ignore the negative $\omega$ orbits as being already
taken care of via their positive $\omega$ partners. The unpaired
eigenstate, which is formally a possibility for $\omega=0$, cannot
really be realized without some little ``swindle''. In the bosonic
case it would correspond to a degree of freedom having, say, a
generalized coordinate but missing the conjugate momentum. In the
fermionic case, it would be analogous to the construction of a set
of $\gamma$-matrices in an odd dimension, which is strictly
speaking only possible because one allows a relation between them
(the product of all the odd number of them being, say, unity) or
because one allows superfluous degrees of freedom.
It is obviously difficult to construct such a set of
$\gamma$-matrices in complete analogy with the case of
an even number of fields, since then the
number of components in the representation of the $n$
gamma-matrices would be $2^{n/2}$, which can hardly make sense for
$n$ odd. Nevertheless, we shall consider the possibility of an
unpaired $\omega=0$ eigenstate in the bosonic case below.

Now the main point of interest for our study is how the second quantized
model looks close to its
ground state.  The neighbourhood of this ground state is supposed to
be the only regime which we humans can study in our ``low energy''
experiments, with small momenta compared to the fundamental (say
Planck) mass scale.
With respect to the ground state of such a second quantized world machinery,
it is well-known that there is a difference between the fermionic and the
bosonic case. In the fermionic case, you can at most have one Fermion in
each state and must fill the states up to some specific value of the
single particle energy---which is really $\omega$. However, in the bosonic
case, one can put a large number of Bosons into the same orbit/single
particle state, if that should pay energetically.

\subsection{ The vacuum}
If we allow for the existence of a chemical potential, which
essentially corresponds to the conservation of the number of
Fermions, we shall typically get the ground state to have Fermions
filling the single particle states up to some special value of the
energy called the Fermi-energy $\omega_{FS}$ ($FS$ standing for
``Fermi-surface''). For Bosons, on the other hand, we will always
have zero particles in all the orbits, except perhaps in the zero
energy ground state; it will namely never pay energetically to put
any bosons into positive energy orbits.

\subsection{The lowest excitations}

So for the investigation of the lowest excitations, i.e.~those that a
``poor physicist'' could afford to work with, we should look for
the excitations very near to the Fermi-surface in the fermionic
case. In other
words, we should put Fermions into the orbits with energies very little
above the Fermi-energy,
or make holes in the Fermi-sea at values of the orbit-energies very
little below the Fermi-energy. Thus, for excitations accessible to
the ``poor physicist'', it is only necessary to study the behaviour
of the spectrum for the Bosons having a value of $\omega$ near to zero,
and for the Fermions having a value of $\omega$ near the Fermi-energy
$\omega_{FS}$.

\subsubsection{Boson case: levels approaching a group of
$\omega =0$ levels}
In section 5 we shall argue that, if the model
has adjustable degrees of freedom (``garbage variables''), they
would tend to make the $\omega = 0$ eigenvalue multiply
degenerate. However, for simplicity, we shall first consider here
the case where there {\em is} just a single zero-eigenvalue
$\omega$-level. We should mention that the true generic situation
for an even number of fields is that there are normally no
zero-eigenvalues at all. So what we shall study here, as the
representative case, really corresponds to the case with an odd
number of fields. In this case there will normally be just one
(i.e.~non-degenerate) $\omega=0$ eigenvalue. However it can happen
that, for special values of the ``momentum parameters'', a pair of
eigenvalues---consisting of eigenvalues of opposite sign of
course---approach zero. It is this situation which we believe to
be the one of relevance for the low energy excitations.

We shall concentrate our interest on a small region in the
momentum parameter space, around a point $\vec{p}_0$
where the two levels with
the numerically smallest non-zero eigenvalues merge together with
a level having zero eigenvalue. Using the well-known fact that, in
quantum mechanics, perturbation corrections from faraway levels
have very little influence on the perturbation of a certain level,
we can ignore all the levels except the zero eigenvalue and this
lowest non-zero pair. So if, for simplicity, we think of this case
of just one zero eigenvalue except where it merges with the other
pair, we need only consider three states and that means, for the
main behaviour, we can calculate as if there were only the three
corresponding fields. This, in turn, means that we can treat the
bosonic model in the region of interest, by studying the spectrum
of a (generic) antisymmetric $3\times 3$ matrix with real
elements, or rather such a matrix multiplied by $i$. Let us
immediately notice that such a matrix is parameterised by
\underline{three} parameters. The matrix and thus the spectrum, to
the accuracy we are after, can only depend on three of the
momentum parameters. In other words the dispersion relation will
depend trivially on all but 3 parameters in the linear
approximation. By this linear approximation, we here mean the
approximation in which the ``poor physicist'' can only work with
a small region in momentum parameter space also---not only in
energy. In this region we can trust the lowest order Taylor
expansion in the differences of the momentum parameters from their
starting values (where the nearest levels merge). Then the
$\omega$-eigenvalues---i.e.~the dispersion relation---will not
vary in the direction of a certain subspace of co-dimension three.
Corresponding to these directions the velocity components of the
described Boson particle will therefore be zero! The Boson, as
seen by the ``poor physicist'', can only move inside a three
dimensional space; in other directions its velocity must remain
zero. It is in this sense we say that the three-dimensionality of
space is explained!

\subsection{Maxwell equations}
\label{Maxwelleq}

The form of the equations of motion for the fields, in this low excitation
regime where one can use the lowest order Taylor expansion in the momentum
parameters, is also quite remarkable: after a linear transformation in
the space of ``momentum parameters'', they can be transformed into the
{\em Maxwell equations} with the fields being complex (linear)
combinations of the magnetic and electric fields.

We can now easily identify the linear combinations of the momentum
parameters minus their values at the selected merging point, which
should be interpreted as true physical momentum components. They
are, in fact, just those linear combinations which occur as matrix
elements in the $3\times 3$ matrix $A$ describing the development
of the three fields $\phi_j$ relevant to the ``poor physicist''.
That is to say we can choose the definition of the ``true momentum
components'' $\vec{k}$ as such linear functions of the deviations,
$\vec{p} -\vec{p}_0$, of the momentum parameters from the merging
point that the antisymmetric matrix $A$ reduces to
\begin{equation}
\label{blhbeq14}
A = \left ( \begin{array}{c|c|c}0 & k_3 & -k_2\\
-k_3 & 0 & k_1\\ k_2 & -k_1 & 0\\ \end{array} \right)
\end{equation}
with eigenvalues $-i\omega = 0,\pm i\sqrt{k_1^2+k_2^2+k_3^2}$.

In the here chosen basis for the momenta, we can make a Fourier
transform of the three fields $\phi_j(\vec{k})$ into the
$\vec{x}$-representation.
These new position space fields $\phi_j(\vec{x})$ are no
longer Hermitian. However,
it follows from the assumed Hermiticity
of the $\phi_j(\vec{k})$ that, in the $\vec{x}$-representation,
the real parts of the fields $\phi_j(\vec{x})$ are even, while the
imaginary parts are odd functions of $\vec{x}$.
We now want to identify these
real and imaginary parts as magnetic and electric fields
$B_j(\vec{x})$ and $E_j(\vec{x})$ respectively: $\phi_j(\vec{x}) =
iE_j(\vec{x}) + B_j(\vec{x})$. However the symmetry of these
Maxwell fields means that they must be in a configuration/state
which goes into itself under a parity reflection in the origin.
This is a somewhat strange feature which seems necessary for the
identification of our general fields with the Maxwell fields; a
feature that deserves further investigation. For the moment let
us, however, see that we do indeed get the Maxwell equations in
the free approximation with the proposed identification.

By making the inverse Fourier transformation back to momentum
space, we obtain the following identification of the fields
$\phi_j(\vec{k})$ in our general quantum field theory with the
electric field $E_j(\vec{k})$ and magnetic field $B_j(\vec{k})$
Fourier transformed into momentum space:
\begin{equation}
\label{identification}
\left ( \begin{array}{c}\phi_1(\vec{k})\\
\phi_2(\vec{k})\\\phi_3(\vec{k})\\ \end{array} \right)=
\left ( \begin{array}{c}iE_1(\vec{k}) +  B_1(\vec{k})\\
iE_2(\vec{k}) + B_2(\vec{k})\\i E_3(\vec{k}) +B_3(\vec{k})\\
\end{array} \right).
\end{equation}
We note that the Fourier transformed electric field $E_j(\vec{k})$
in the above ansatz (\ref{identification}) has to be purely
imaginary, while the magnetic field $B_j(\vec{k})$ must be purely
real.

By using the above identifications, eqs.~(\ref{blhbeq14}) and
(\ref{identification}), the equations of motion (\ref{blhbeq8}) take
the following form
\begin{equation}
\left ( \begin{array}{c}i\dot{E}_1(\vec{k}) +  \dot{B}_1(\vec{k})\\
i\dot{E}_2(\vec{k}) + \dot{B}_2(\vec{k})\\i \dot{E}_3(\vec{k})
+ \dot{B}_3(\vec{k})\\
\end{array}
\right)  =\left ( \begin{array}{c|c|c}0 & k_3 & -k_2\\
-k_3 & 0 & k_1\\ k_2 & -k_1 & 0\\ \end{array} \right)
\left ( \begin{array}{c}iE_1(\vec{k}) +  B_1(\vec{k})\\
iE_2(\vec{k}) + B_2(\vec{k})\\i E_3(\vec{k}) +B_3(\vec{k})\\
\end{array} \right).
\end{equation}
We can now use the usual Fourier transformation identification in
quantum mechanics to transform these equations
to the $\vec{x}$-representation,
simply from the definition of $\vec{x}$ as the Fourier transformed
variable set associated with $\vec{k}$,
\begin{equation}
k_j  = i^{-1} \partial_j
\end{equation}
Thus in $\vec{x}$-representation the equations of motion become
\begin{equation}
\left ( \begin{array}{c}i\dot{E}_1(\vec{x}) +  \dot{B}_1(\vec{x})\\
i\dot{E}_2(\vec{x}) + \dot{B}_2(\vec{x})\\i \dot{E}_3(\vec{x})
+\dot{B}_3(\vec{x})\\
\end{array}
\right) =\left ( \begin{array}{c|c|c}0 & -i\partial_3 & i\partial_2\\
i\partial_3 & 0 & -i\partial_1\\ -i\partial_2 & i\partial_1 & 0\\
\end{array}
\right)\left ( \begin{array}{c}iE_1(\vec{x}) +  B_1(\vec{x})\\
iE_2(\vec{x}) + B_2(\vec{x})\\i E_3(\vec{x}) +B_3(\vec{x})\\
\end{array} \right).
\end{equation}
The imaginary terms in the above equations give rise to the
equation:
\begin{equation}
\label{iprop} \dot{\vec{E}}(\vec{x}) = \mathrm{curl}\, \vec{B}
\end{equation}
while the real parts give the equation:
\begin{equation}
\label{realpart} \dot{\vec{B}}(\vec{x}) =- \mathrm{curl}\, \vec{E}
\end{equation}
These two equations are just the Maxwell equations in the absence
of charges and currents, except that strictly speaking we miss two
of the Maxwell equations, namely
\begin{equation}
\label{missing} \mathrm{div}\, \vec{E}(\vec{x}) =0 \quad
\hbox{ {and} } \quad \mathrm{div}\, \vec{B}(\vec{x}) =0.
\end{equation}
However, these two missing equations are derivable from the other
Maxwell equations in time differentiated form. That is to say, by
using the result that the divergence of a curl is zero, one can
derive from the other equations that
\begin{equation}
\label{timederived}
\mathrm{div}\, \dot{\vec{E}}(\vec{x}) =0 \quad \hbox{
{and}} \quad \mathrm{div}\, \dot{\vec{B}}(\vec{x}) =0
\end{equation}
which is though not quite sufficient. Integration of the resulting
equations (\ref{timederived}) effectively replaces the $0$'s on
the right hand sides of equations (\ref{missing}) by terms
constant in time, which we might interpret as some constant
electric and magnetic charge distributions respectively. In our
free field theory approximation, we have potentially ignored such
terms. So we may claim that, in the approximation to which we have
worked so far, we have derived the Maxwell equations sufficiently
well.

\section{Homolumo-gap and analogue for bosons}

The Homolumo-gap effect refers to a very general feature of
systems of Fermions, which possess some degrees of freedom that
can adjust themselves so as to lower the energy as much as
possible. The effect is so general that it should be useful for
almost all systems of Fermions, because even if they did not have
any extra degrees of freedom to adjust there would, in the Hartree
approximation, be the possibility that the Fermions could
effectively adjust themselves. The name Homolumo gap was
introduced in chemistry and stands for the gap between `` the
highest occupied'' HO ``molecular orbit'' MO and the ``lowest
unoccupied'' LU ``molecular orbit'' MO.  The point is simply that
if the filled (occupied) orbits (single particle states) are
lowered the whole energy is lowered, while it does not help to
lower the empty orbits. It therefore pays energetically to make
the occupied orbits go down in energy and separate from the unfilled
ones; thus a gap may appear or rather {\em there will be a general
tendency to get a low level density near the Fermi-surface}. This
effect can easily be so strong that it causes a symmetry to break
\cite{Teller}; symmetry breaking occurs if some levels, which are
degenerate due to the symmetry, are only partially filled so that
the Fermi-surface just cuts a set of degenerate states/orbits. It
is also the Homolumo-gap effect which causes the deformation of
transitional nuclei, which are far from closed shell
configurations. We want to appeal to this Homolumo gap
effect, in subsection \ref{Weylderivation}, as a justification for
the assumption that the Fermi-surface gets close to those places
on the energy axis where the level density is minimal.

However we first want to discuss a similar effect, where the degrees
of freedom of a system of Bosons adjust themselves to lower the
total energy. As for the Fermion systems just discussed, this
lowering of the total energy is due to the adjustment of a sum
over single particle energies---the minimisation of the zero-point
energy of the bosonic system. We consider the effect of this
minimisation to be the analogue for Bosons of the Homolumo-gap
effect.

\subsection{The analogue for bosons}
In the ``derivation'' of the Maxwell equations given in subsection
\ref{Maxwelleq}, we started by introducing the assumption of the
existence of a zero frequency, $\omega=0$, eigenvalue by taking
the number of Hermitian fields and thereby the order of the
antisymmetric matrix $A_{ij}$ to be {\em odd.} We now turn to our
more general assumption of the existence of multiply degenerate
$\omega=0$ eigenvalues. Honestly we can only offer a rather
speculative argument in favour of our assumption that there should
be several eigenvalues which are zero, even in the case when the
total number of fields is not odd. For quite generic matrices, as
would be the cleanest philosophy, it is simply not true that there
would be zero eigenvalues for most momenta in the case of an even
number of fields. However, let us imagine that there are many
degrees of freedom of the whole world machinery that could adjust
themselves to minimize the energy of the system and could also
influence the matrix $A_{ij}(\vec{p})$. Then one could, for
instance, ask how it would be energetically profitable to adjust
the eigenvalues, in order to minimize the zero-point energy of the
whole (second quantized) system. This zero-point energy is
formally given by the integral over all (the more than three
dimensional) momentum space; let us just denote this integration
measure by $d\vec{p}$, so that:
\begin{equation}
\label{zpe}
E_{{\it zero-point}} = \int  d\vec{p}
\sum_{{\it eigenvalue\,\, pair   \,\, k}}
|\omega_k(\vec{p})|/2 
\end{equation}
Provided some adjustment took place in minimizing this quantity,
there would {\em a priori} be an argument in favour of having
several zero eigenvalues, since they would contribute the least to
this zero-point energy $E_{{\it zero-point}}$. At first sight,
this argument is not very strong, since it just favours making the
eigenvalues small and not necessarily making any one of them
exactly zero. However, we underlined an important point in favour
of the occurrence of exactly zero eigenvalues, by putting the
numerical sign explicitly into the integrand
$|\omega_k(\vec{p})|/2$ in the expression (\ref{zpe}) for the
zero-point energy. The important point is that the numerical value
function is not an ordinary analytic function, but rather has a
kink at $\omega_k(\vec{p})=0$. This means that, if other
contributions to the energy of the whole system are
smooth/analytic, it could happen that the energy is lowered when
$\omega_k(\vec{p})$ is lowered numerically for both signs of
$\omega_k(\vec{p})$; here we consider $\omega_k(\vec{p})$ to be a
smooth function of the adjusting parameters of the whole world
machinery (we could call them ``garbage parameters''). For a
normal analytic energy function this phenomenon could of course
never occur, except if the derivative just happened (is fine-tuned
one could say) to be equal to zero at $\omega_k(\vec{p}) =0$. But
with a contribution that has the numerical value singularity
behaviour it is possible to occur with a finite probability
(i.e.~without fine-tuning), because it is sufficient that the
derivative of the contribution to the total energy from other
terms
%than the special zero point contribution being discussed
is numerically lower than the derivative of the zero-point term
discussed. Then, namely, the latter will dominate the sign of the
slope and the minimum will occur {\em exactly} for zero
$\omega_k(\vec{p})$.

In this way, we claim to justify our assumption that the matrix
$A_{ij}(\vec{p})$ will have several exactly zero eigenvalues and
thus a far from maximal rank; the rank being at least piecewise
constant over momentum space. We shall therefore now study
antisymmetric matrices with this property in general and look for
their lowest energy excitations.

\subsection{Using several zero eigenvalues to derive Maxwell equations}

As in subsection \ref{Maxwelleq}, we assume that when a single
pair of opposite sign eigenvalues approach zero as a function of the
momentum, we can ignore the faraway eigenvalues. Then, using the
approximation of only considering the fields corresponding to the
two eigenvalues approaching zero and the several exact zero
eigenvalues, we end up with an effective $(n+2)$ x $(n+2)$ matrix
$A_{ij}(\vec{p})$ obeying the constraint of being of rank two (at
most). Now we imagine that we linearize the momentum dependence of
$A_{ij}(\vec{p})$ on $\vec{p}$ around a point in momentum space,
say $\vec{p}_0$, where the pair of eigenvalues approaching zero
actually reach zero, so that the matrix is totally zero,
$A_{ij}(\vec{p}_0)=0$, at the starting point for the Taylor
expansion. That is to say that, corresponding to different basis
vectors in momentum space, we get contributions to the matrix
$A_{ij}(\vec{p})$ linear in the momentum difference
$\vec{p}-\vec{p}_0$. Now any non-zero antisymmetric matrix is
necessarily of rank at least 2. So the contribution from the first
chosen basis vector in momentum space will already give a matrix
$A_{ij}$ of rank 2 and contributions from other momentum
components should not increase the rank beyond this. A single
basis vector for a set of linearly parameterised antisymmetric
real matrices can be transformed to just having elements (1,2) 
and (2,1) nonzero and the rest zero. In order to avoid a further
increase in the rank of the matrix by adding other linear
contributions, these further contributions must clearly not
contribute anything to matrix elements having both column and row
index different from $1$ and $2$. However this is not sufficient
to guarantee that the rank remains equal to 2. This is easily
seen, because we can construct 4 x 4 antisymmetric matrices, which
are of the form of having 0's on all places (i,j) with both i and
j different from 1 and 2 and have nonzero determinant.

So let us consider 4 by 4 sub-determinants of the matrix $A_{ij}$
already argued to be of the form
\begin{equation}
\label{mform}
\left ( \begin{array}{c|c|c|c|c}0 & A_{12} & A_{13} &\cdots & A_{1n}\\
-A_{12} & 0 & A_{23}& \cdots & A_{2n}\\ -A_{13} & -A_{23} & 0& \cdots &0\\
: & : & 0 & \cdots &0\\
-A_{1n} & -A_{2n} & 0 & \cdots & 0 \end{array} \right).
\end{equation}
Especially let us consider a four by four sub-determinant along the
diagonal involving columns and rows 1 and 2. The determinant is
for instance
\begin{equation}
\label{sub4}
det
\left ( \begin{array}{c|c|c|c}0 & A_{12} & A_{13} & A_{15}\\
-A_{12} & 0 & A_{23}& A_{25}\\ -A_{13} & -A_{23} & 0&0\\
-A_{15} & -A_{25} & 0 & 0 \end{array} \right ) =-\left( det \left (
\begin{array}{c|c} A_{13} & A_{15}\\
 A_{23}& A_{25} \end{array} \right ) \right )^2.
\end{equation}
In order that the matrix $A_{ij}$ be of rank 2, this determinant
must vanish and so we require that the 2 by 2 sub-matrix
\begin{equation}
\left (
\begin{array}{c|c} A_{13} & A_{15}\\
 A_{23}& A_{25}\\ \end{array} \right)
\end{equation}
must be degenerate, i.e.~of rank 1 only. This means that the
two columns in it are proportional, one to the other. By
considering successively several such selected four by four
sub-matrices, we can easily deduce that all the two columns
\begin{equation}
\left ( \begin{array}{c} A_{13}\\ A_{23}  \end{array} \right),
\left ( \begin{array}{c} A_{14}\\ A_{24}  \end{array} \right), \cdots
\left ( \begin{array}{c} A_{1n}\\ A_{2n}  \end{array} \right)
\end{equation}
are proportional. This in turn means that we can transform them
all to zero, except for say
\begin{equation}
\left ( \begin{array}{c} A_{13}\\ A_{23} \end{array} \right),
\end{equation}
by going into a new basis for the fields
$\phi_k(\vec{p}-\vec{p_0})$. So, finally, we have
transformed the formulation of the fields in such a way that only
the upper left three by three corner of the $A$ matrix is
non-zero. But this is exactly the form for which we argued in
subsection \ref{Maxwelleq} and which was shown to be interpretable
as the Maxwell equations, and moreover the Maxwell equations for
just three spatial dimensions!

\subsection{The Weyl equation derivation \label{Weylderivation}}

Let us now turn to the application of the Homolumo-gap effect to a
system of Fermions in our general field theory model. We shall
assume that the Homolumo-gap effect turns out to be strong enough
to ensure that the Fermi-surface just gets put to a place where
the density of levels is very low. Actually it is very realistic
that a gap should develop in a field theory with continuum
variables $\vec{p}$ labeling the single particle states. That is
namely what one actually sees in an insulator; there is an
appreciable gap between the last filled {\em band} and the first
empty band. However, if the model were totally of this insulating 
type, the poor physicist would not ``see'' anything, because he is 
supposed to be unable to afford to raise a particle from the filled
band to the empty one. So he can only see something if there are
at least {\em some} Fermion single particle states with energy
close to the Fermi-surface.

We shall now divide up our discussion of what happens near the
Fermi-surface according to the number of components of the Fermion
field that are relevant in this neighborhood.
Let us denote by $n$ the number
of Fermion field components, which contribute significantly to
the eigenstates near the Fermi-surface in the small region
of momentum space we choose to consider.

The eigenvalues $\pm\omega$ of $iA_{ij}$ -- which come in
pairs -- correspond to  eigenstates with complex components. Thus
it is really easiest in the fermionic case to ``go back'' to
a complex field notation, by constructing complex fields out of
twice as big a number of real ones. So now we consider the
level-density near the Fermi-surface for $n$ complex Fermion field
components.

\subsection{The case of $n=0$ relevant levels near Fermi-surface}

The $n=0$ case must, of course, mean that there are no levels at
all near the Fermi-surface in the small momentum range considered.
This corresponds to the already mentioned insulator case. The 
poor physicist sees nothing from such regions in momentum space 
and he will not care for such regions at all. Nonetheless
this is the generic situation close to the Fermi surface and
will apply for most of the momentum space.

\subsection{The case of $n$ = 1 single relevant level near the Fermi-surface}

In this case the generic situation will be that, as a certain
component of the momentum is varied, the level will vary
continuously in energy. This is the kind of behaviour observed
in a metal. So there will be a rather smooth density
of levels and such a situation is not favoured by the Homolumo gap
effect, if there is any way to avoid it.

\subsection{The case of $n$ = 2 relevant levels near the Fermi-surface}

In this situation a small but almost trivial calculation is
needed. We must estimate how a Hamiltonian,
described effectively as a 2 by 2 Hermitian matrix $H$ with
matrix elements depending on the momentum $\vec{p}$,
comes to look in the generic case---\i.e. when nothing is
fine-tuned---and especially how the level density behaves. That
is, however, quite easily seen, when one remembers that the three
Pauli matrices and the unit 2 by 2 matrix together form a basis
for the four dimensional space of two by two matrices. All
possible Hermitian 2 by 2 matrices can be expressed as linear
combinations of the three Pauli matrices $\sigma^j$ and the
unit 2 by 2 matrix $\sigma^0$ with real coefficients.
We now consider a linearized Taylor
expansion\footnote{A related discussion of the redefinition 
of spinors has been given in the context of the low 
energy limit of a Lorentz violating QED model \cite{Colladay}.} 
of the momentum dependence of such matrices, by taking
the four coefficients to these four matrices to be arbitrary
linear functions of the momentum minus the ``starting momentum''
$\vec{p_0}$, where the two levels become degenerate with 
energy $\omega(\vec{p_0})$.
That is to say we must take the Hermitian 2 by 2
matrix to be
\begin{equation}
H = \sigma^a V^i_a (p_i - p_{0i}) + 
\sigma^0 \omega(\vec{p_0}). \label{Weyl}
\end{equation}
This can actually be interpreted as the Hamiltonian for a
particle obeying the Weyl equation, by defining
\begin{equation}
P_1 = V^i_1(p_i - p_{0i}) \qquad  P_2 = V^i_2(p_i - p_{0i}) 
\qquad P_3 = V^i_3(p_i - p_{0i}) 
\end{equation}
\begin{equation}
H_{new} = H - \sigma^0 V^i_0(p_i - p_{0i}) -
\sigma^0 \omega(\vec{p_0}) 
= \vec{\sigma}\cdot\vec{P} 
\end{equation}
\begin{equation}
\omega_{new} = \omega - V^i_0(p_i - p_{0i}) - 
\omega(\vec{p_0}) 
\label{Hnew}
\end{equation}
and supposing that the $V^i_0$ are not too large\footnote{If the 
$V^i_0$ are very large, there is a risk that different sides 
of the upper light-cone fall above and below the value 
$\omega(\vec{p_0})$ of the energy at the tip of the cone.} 
compared to the other $V^i_a$'s. 
The renormalisation of the energy, eq.~(\ref{Hnew}),
is the result of transforming away a constant velocity $V_0^i$ in
D dimensions carried by all the Fermions, using the change of
co-ordinates $x^{\prime i} = x^i - tV_0^i$, and measuring the 
energy relative to $\omega(\vec{p_0})$.
Note that the ``starting momentum" $\vec{p_0}$ will generically be of the 
order of a fundamental (Planck scale) momentum, which 
cannot be significantly changed by a ``poor physicist". 
So the large momentum $\vec{p_0}$ effectively plays the role 
of a conserved charge at low energy, justifying the use 
of complex fermion fields and the existence of a Fermi 
surface.

A trivial calculation
for the Weyl equation, $H_{new} \psi = \omega_{new} \psi$, leads
to a level density with a thin neck, behaving like
\begin{equation}
 \rho \propto \omega_{new}^2
\end{equation}
According to our strong assumption about homolumo-gap effects, we
should therefore imagine that the Fermi-surface in this case would
adjust itself to be near the $\omega_{new} =0$ level. Thereby there
would then be the fewest levels near the Fermi-surface.

\subsection{The cases $n\ge 3$}

For $n$ larger than $2$ one can easily find out\footnote{HBN would 
like to thank S.~Chadha for a discussion of this $n\ge 3$ case 
many years ago.} that, in the
neighbourhood of a point where the $n$ by $n$ general Hamiltonian
matrix deviates by zero from the unit matrix, there are
generically branches of the dispersion relation for the particle
states that behave in the metallic way locally, as in the case
$n=1$. This means that the level density in such a neighborhood
has contributions like that in the $n=1$ case, varying rather
smoothly and flatly as a function of $\omega$. So these cases are
not so favourable from the Homolumo-gap point of view.

\subsection{Conclusion of the various $n$ cases for the Fermion model}

The conclusion of the just given discussion of the various
$n$-cases is that, while of course the $n=0$ case is the ``best''
case from the point of view of the homolumo-gap, it would not be
noticed by the ``poor physicist'' and thus would not be of any
relevance for him. The next ``best'' from the homolumo-gap point
of view is the case $n=2$ of just two complex components
(corresponding to 4 real components) being relevant near the
Fermi-surface. Then there is a neck in the distribution of the
levels, which is not present in the cases $n=1$ and $n>2$.

So the ``poor physicist'' should in practice observe the case
$n=2$, provided the homolumo-gap effect is sufficiently 
strong (a perhaps suspicious assumption).

Now, as we saw, this case of $n=2$ means that the Fermion field
satisfies a Weyl equation, formally looking like the Weyl equation
in just 3+1 dimensions! It should however be noticed that there
are indeed more spatial dimensions, by assumption, in our model.
In these extra spatial dimensions, the Fermions have the same
constant velocity which we were able to renormalise to zero, because
the Hamiltonian only depends on the three momentum
components $\vec{P}$ in the Taylor expandable region accessible to
the ``poor physicist''. The latter comes about because there are
only the three non-trivial Pauli matrices that make the single
particle energy vary in a linear way around the point of expansion.
In this sense the number of spatial dimensions comes out as equal to
the number of Pauli matrices.

\section{Conclusion, r\'{e}sum\'{e}, discussion}

We have found the remarkable result that, in the free
approximation, our very general quantum
field theory, which does {\em not have Lorentz invariance put in},
leads to {\em Lorentz invariance in three plus one dimensions}
for {\em both Bosons and Fermions}. In the
derivation of this result, we made use of what we called the
homolumo-gap effect and its ``analogue for Bosons'' and that
experimentalists only have access to energies low compared to the
fundamental scale. The derivation of three spatial dimensions
should be understood in the sense that our model, which has at
first a space of D dimensions, leads to a dispersion relation
(\i.e.~a relation between energy and momentum) for which the
derivative of the energy ${\omega}$ w.r.t.~the momentum in $D-3$
of the dimensions is independent of the momentum.
Then, in the remaining $3$ dimensions, we get
the well-known Lorentz invariant dispersion relations both in the
Bosonic and the Fermionic cases. In fact we obtained the Weyl
equation and the Maxwell equations, in the fermionic and the bosonic
cases respectively, as ``generic'' equations of motion -- after the
use of the homolumo gap and its analogue. These
Maxwell and spin one half equations of motion are in remarkable
accord with the presently observed (\i.e. ignoring the Higgs particle)
fundamental particles!

\subsection{Some bad points and hopes}
In spite of this remarkable success of our model in the free
approximation, we have to admit there are a number of flaws:

1) The three space dimensions selected by each type of particle
{\em are a priori overwhelming likely {\em not} to be the same
three}. That is to say we would have to hope for some
speculative mechanism that could align the three dimensions used
by the different species of particles, so as to be the same three
dimensions.

2) Although we have hopes of introducing interactions, 
it is not at all clear how these 
interactions would come to look and whether e.g.~they would 
also be Lorentz invariant---according to point 1) one would
{\it a priori}
say that they do not have much chance to be Lorentz invariant.

3) There are extra dimensions in the model, although they do not
participate in the derived Lorentz invariance which is only a 3+1
Lorentz invariance. Rather the velocity components in the extra 
dimension directions are  constant, independent of the momentum 
of the particles. We can really by convention renormalise them 
to zero and claim that we do not see the extra dimensions, 
because we cannot move in these directions. 
But from point 1) there is the worry
that these directions (in which we have no movement)
are different for the different types of particles.

The best hope for rescuing the model from these problems might be to
get rid of momentum conservation in the extra directions. We might
hope to get some attachment of the particles to a fixed position
in the extra directions much like attachments to branes, but then
one would ask how this could happen in a natural way.
Of course the point of view most in the spirit of the random 
dynamics project would be that {\it a priori} we did not even 
have momentum conservation, but that
it also just arose as the result of some Taylor expansion. This
becomes very speculative but it could easily happen that it is
much easier to get a translational invariance symmetry develop,
along the lines suggested in section 6.2.3 of our book \cite {book}, 
for the momentum directions in which we have rapid motion than 
in the directions in which we have zero velocity. If we crudely
approximated the particles by non-relativistic ones, the rapid
motion would mean low mass while the zero motion
would mean a very huge mass. The uncertainty principle would, 
therefore, much more easily allow these
particles to fall into the roughness valleys\footnote{By
roughness valleys we refer to the (local) minima or valleys
in a potential representing a non-translational invariant
potential set up so as break momentum conservation.} in the translational
invariance violating potential in the extra directions, where the
non-relativistic mass is much larger than in the 3 space directions.
A particle would be very much spread out by uncertainty in the
3 directions and, thus, only feel a very smoothed out roughness
potential, if translational invariance is broken in these
directions. In this way translational invariance could develop
in just 3 dimensions.

A breakdown of the translational invariance---or, as just
suggested, a lack of its development---in the extra
dimensions would be very helpful in solving the  above-mentioned
problems. This is because there would then effectively only be 3
space dimensions and all the different types of particles would,
thus, use the same set of 3 dimensions. It must though be admitted that
they would still have different metric tensors, or metrics we
should just say. We had some old ideas \cite{book} for solving
this problem, but they do not quite work in realistic models.

\subsection{Where did the number three for the space dimension come from?}

One might well ask why we got the prediction of just three
for the number of spatial dimensions. In fact we
have derived it differently, although in many ways analogously,
for Bosons and for Fermions:

\noindent {\bf Bosons:} \  For Bosons we obtained this result by
considering the simultaneous approach of a pair of equal and
opposite eigenvalues of the real antisymmetric matrix $A_{ij}$ to
the supposedly existing zero frequency, $\omega = 0$, level(s).
Thus the rank of the matrix $A$ relevant to this low energy range
is just {\em two}, except at the point around which we expand
where it has rank zero. Then we argued that we could transform
such a matrix in such a way that it effectively becomes a 3 by 3
matrix---still antisymmetric and real. So the matrix $A$ has
effectively three independent matrix elements and each can vary
with the components of the momentum. However, in the low energy
regime, this dependence can be linearized and $A$ only depends on
three linearly independent components of the momentum. It is these
three dimensions in the directions of which we have non-zero
velocity (or better non-constant velocity) for the Boson---the
photon, say, in as far as it obeys the Maxwell equations. We thus
got the number three as the number of independent matrix elements
in the antisymmetric matrix $A$, obtained after transforming away
most of this rank two matrix.

\noindent {\bf Fermions:} \  In this case we went to a complex
notation, although we still started from the same type of
antisymmetric real matrix as in the bosonic case. The homolumo-gap
argumentation suggested that just $n=2$ complex components in the
field should be ``relevant'' near the Fermi surface, after ruling
out the trivial $n=0$ case as unobservable by anybody. This number
of relevant components then meant that just $n^2 -1=4-1=3$
non-trivial linearly independent $n$ by $n$ matrices could be
formed. These three matrices could, of course, then be used as the
coefficients for three momentum components in the linearized
(Taylor expanded) momentum dependence of the Hamiltonian. In this
way the number three arose again.

So there is an analogy at least in as far as, for both Bosons
and Fermions, it is the number of linearly independent matrices of
the type finally used, which remarkably predicts the observable
number of spatial dimensions to be 3. However in the bosonic case
it is real three by three matrices which we ended up with, while
in the fermionic case it is Hermitian 2 by 2 matrices with the
unit matrix omitted. The unit matrix is not counted because it
does not split the levels and basically could be transformed away
by the shift of a vierbein, \i.e.~by adjusting the meaning of the
momentum and energy components.

A strange prediction of the Boson model is that, at first at
least, we get a parity symmetric state of the world for the
Maxwell fields. That is to say for every state of the
electromagnetic---or generalized Yang Mills---field there is
somewhere, reflected in the origin of position space, a
corresponding reflected state. In principle we could test such an
idea, by looking to see whether we could classify galaxies found on
the sky into pairs that could correspond to mirror images---in the
``origin''---of each other. Really we hope that this illness of
our model might easily repair itself.

\section{Acknowledgements}
C.D.F.~thanks Jon Chkareuli and H.B.N.~thanks Bo Sture Skagerstam 
for helpful discussions. C.D.F.~thanks PPARC for a travel grant to 
attend the 2001 Bled workshop and H.B.N.~thanks the Alexander von 
Humboldt-Stiftung for the For\-schungs\-preis.
 
%\vspace{1cm}

%%
% noncompactness
\title*{%
Unitary Representations, Noncompact Groups $SO(q, d-q)$ and More Than One Time}
\author{ N. Manko\v c Bor\v stnik$^{1,2}$, H. B. Nielsen$^3$ and D. Lukman$^2$}
\institute{%
${}^1$ Department of Physics, University of
Ljubljana, Jadranska 19, 1111 Ljubljana, Slovenia\\
${}^2$ Primorska Institute for Natural Sciences and Technology, 
C. Mare\v zganskega upora 2, Koper 6000, Slovenia\\
${}^3$ Department of Physics, Niels Bohr Institute, Blegdamsvej 17,
Copenhagen, DK-2100}

\titlerunning{Unitary Representations, Noncompact Groups $SO(q, d-q)$ \ldots}
\authorrunning{N. Manko\v c Bor\v stnik, H. B. Nielsen and D. Lukman}
\maketitle

\begin{abstract}
Since the Lorentz group and accordingly the Poincar\' e group are
noncompact groups, the question arises if and under which
conditions can one define an inner product of two state vectors of
a chosen irreducible representation space of the Lorentz group or
accordingly of the Poincar\' e group for let us say spinors which is
invariant under the Lorentz transformations and unitary for any
dimension $d$ and any signature $q + (d-q)$ (with $q$ time and $d-q$
space dimensions) and what can one define as a probability density.
This contribution is analyzing some aspects of a possible answer to this
question.
\end{abstract}
%%\pacs{
%% 04.50.+h, 11.10.Kk,11.30.-j,12.10.-g
%%}
 
\section{Introduction}\label{introduction}

There are many experiences and accordingly many papers in the
literature and books about representations of the Lorentz and the
Poincar\' e group in one time and three space dimensions.  The answer
to the question when can one define for, let us say, spinors a
current, which transforms under the Lorentz transformations as a
$d$-vector with a zero component, which can be interpreted as a
probability density for three space and one time dimensions and yet
spinors obey equations of motion, are known since Dirac and
Wigner\cite{wigner,weinberg,luciano}.

For the Minkowski metric we may write an inner product for two generic
states $|\phi\rangle$ and $|\psi\rangle$ of a chosen irreducible
representation of the Lorentz group and accordingly of the
Poincar\' e group as
\begin{eqnarray}
\langle\phi||\psi\rangle = \langle\phi|{\cal O}\psi\rangle.
\label{1+3inner}
\end{eqnarray}
In this contribution we shall pay attention to mainly  ``spinors``. (We call
``spinors`` the representations, for which the relation $[S^{ab}, S^{ac}]_- 
= 1/2 \eta^{aa}\eta^{bc}$ is fulfilled. For these cases
${\cal O} = \gamma^0 $, which is the time-like member of $d$
$\gamma^a$ matrices, fulfilling the Clifford algebra
\begin{eqnarray}
\{\gamma^a,\gamma^b\}_{+} = 2\eta^{ab}.
\label{Clifford}
\end{eqnarray}
Although the matrices $\gamma^a$ are not among the infinitesimal
generators of either the Lorentz or the Poincar\'e group they still
are needed to define the inner product for spinors so that it is
unitary and Lorentz invariant.  However, as it is well known, the
infinitesimal generators of the Lorentz group $S^{ab}$ - the internal
part of - are for spinors expressible in terms of the $\gamma^a$
matrices
\begin{eqnarray}
S^{ab} = \frac{i}{4}[\gamma^a,\gamma^b]_- 
\label{sab}
\end{eqnarray}
and one of invariants\cite{norma93} of the Poincar\' e group - the operator of handedness $\Gamma$ - can 
be written as 
\begin{eqnarray}
\Gamma :&=& \;
(i)^{d/2}\;\prod_a \quad (\sqrt{\eta^{aa}} \gamma^a), \quad {\rm if } \quad d = 2n, 
\label{hand}
\end{eqnarray}
with $\Gamma^{\dagger} = \Gamma$ and $\Gamma^2 = I.$
Can one define the inner product, which is
unitary and Lorentz invariant for any signature and any spin? The answer to 
this question could help to understand, why Nature has made a choice of one time and three
space coordinates at the low energy regime.
(For  discussions on this topic the reader  can look at 
references\cite{holgerbefore,norma01,holgernorma2002,holger1,holgerbefore}, arguing
that it is the internal space, which influences the choice of
signature and also the choice of dimensions.)
%There are additional open problems, like
%the one of the Lorentz invariance and yet the unitarity of the inner product 
%(with respect to the noncompact 
%Lorentz group) just mentioned. 
%We would like to have the inner product to be Lorentz invariant and unitary
%for fields of any spin living in any dimension $d$ of any signature $d +(d-q)$.

We also would like to know all equations of motion which fields should
obey for a signature $q, (d-q)$. %We would further like to understand
%the appearance of a ray for all possible signatures.  
We would like to
know a possible quantization procedure, starting from Lagrange
formalism, Hamilton equations of motion, Poisson brackets, first
quantization procedure and the corresponding uncertainty principle,
Fourier transformations, second quantization procedure, an interaction
scheme among fields and many other properties for more than one time
signature (if all these procedures are at all applicable to more than
one time signature).

In this paper, we shall mainly discuss the question whether or not
one can define the inner product, which is Lorentz invariant and yet
unitary, for any signature. Since the use of the
Bargmann-Wigner\cite{bargmannwigner,norma01} prescription
enables to construct states for any spin out of states of spinors, we
shall look for the inner product only for spinors.

\section{Proposal for inner product for spinors in any dimension and for any signature}
\label{generalization}

The answer to the question, whether or not one can define an inner
product, which is Lorentz invariant and yet unitary for a spinor in
$d$-dimensional spaces for a generic signature $q,(d-q)$, is rather
simple.

We first have to notice that due to the Hermiticity property of the $\gamma^a$ matrices,
\begin{eqnarray}
\gamma^{a\dagger} = \gamma^a \eta^{aa},
\label{gammaher}
\end{eqnarray}
the inner part of the generators of the Lorentz transformations has
the Hermiticity properties as follows
\begin{eqnarray}
S^{ab\dagger} = \eta^{aa}\eta^{bb} S^{ab}.
\label{sabher}
\end{eqnarray}
The generators of the Lorentz transformations are the sum of the spin
(the inner part) and the generators of the Lorentz transformations in
ordinary part of the space-time, that is $L^{ab} = x^a p^b - x^b p^a$,
where $p^a $ is a conjugate momentum operator to $x^a$. $L^{ab}$ are
Hermitian operators, $L^{ab\dagger} = L^{ab}$, while $S^{ab}$, as we
see in Eq.{\ref{sabher}}, are in general not.  Since the Hermiticity or
anti-Hermiticity property of $S^{ab}$ depends on the signature, a
general Lorentz transformation operator
\begin{eqnarray}
U(\omega) = e^{-\frac{i}{2} \omega_{ab}S^{ab}}
\label{genLT}
\end{eqnarray}
has not the property of an unitary operator in the sense that
$U^{\dagger} $ would be equal to $U^{-1}$.

In order that two observers, belonging to two Lorentz transformed
coordinate systems (with respect to each other), will measure the same
probabilities for a physical event, it must be that the inner product
should not change with Lorentz transformations
\begin{eqnarray}
\langle U\phi||U\psi\rangle = \langle\phi||\psi\rangle = \langle\phi|{\cal O}\psi\rangle.
\label{1+3inneruni}
\end{eqnarray}
With ${\cal O} = \gamma^0$ and $d=4$, this requirement is certainly
fulfilled for the Minkowski metric when a spinor is involved, since
$S^{ab\dagger} \gamma^0 = \gamma^0 S^{ab}$ for each $a,b$ and
accordingly $U^{\dagger} \gamma^0 = \gamma^0 U^{-1}$.  We then have
(this is the well known fact for $d=1+3$, of course) that 
($\psi(x) =\langle x|\psi\rangle$)
\begin{eqnarray}
\psi(x)^{\dagger} \gamma^0 \underbrace{\gamma^a \gamma^b... \gamma^n}_{n \; times}\psi(x)
\label{tensorn}
\end{eqnarray}
transforms as a tensor of the rank $n$  under a Lorentz
transformation. We find accordingly that
\begin{eqnarray}
\psi(x)^{\dagger} \gamma^0 \gamma^a \psi(x)
\label{dvector}
\end{eqnarray}
is a $4$ vector and then $\psi(x)^{\dagger}  \psi(x)$ is as a 
positive definite quantity interpreted as a probability density.

When the signature $q + (d-q)$ is under consideration and $q\ge 1$ the 
inner product as defined in Eq.(\ref{1+3inner})
still guarantees its Lorentz invariance and the unitarity
\begin{eqnarray}
\langle U\phi||U \psi\rangle = \langle U\phi|{\cal O} U\psi\rangle = 
\langle\phi||\psi\rangle = \langle\phi|{\cal O}\psi\rangle.
\label{q+d-qinneruni}
\end{eqnarray}
provided now that ${\cal O}$ is for spinors generalized to
\begin{eqnarray}
{\cal O} &=& (i)^{(q-1)/2} \;\prod_{a \in \eta^{aa} =1} \; \gamma^{a}, \qquad {\rm for \; q\; odd},
\nonumber\\
{\cal O} &=& (i)^{q/2} \;\prod_{a \in \eta^{aa} =1} \; \gamma^{a}, \qquad {\rm for \; q\; even}. 
\label{q+d-qo}
\end{eqnarray}
We can check that ${\cal O}$ has the property 
${\cal O}^{\dagger} = {\cal O}$, which is needed in order that 
$\langle \psi|| \psi\rangle = (\langle \psi|| \psi\rangle)^{\dagger}$.
%(for spinors and $1+ (d-1)$, 
%$\langle \psi|| \psi\rangle = (\langle \psi||\psi\rangle)^{\dagger}=O$, 
%since odd ${\cal O}$ always transforms left to right handedness, 
%accordingly  a Clifford odd projector $\delta (\gamma^a p_a)$ is essential)
%
The product in Eq.(\ref{q+d-qo}) includes all $\gamma^a$, 
with the index $a$ (in the increasing order)  belonging to the time-like coordinates. 
Then $U^{\dagger} {\cal O} = {\cal O} U^{-1}$ and the inner 
product for two Lorentz transformed observers is the same.

It is still true that 
\begin{eqnarray}
\psi(x)^{\dagger} \gamma^0 \underbrace{\gamma^a \gamma^b... \gamma^n}_{n \; times}\psi(x)
\label{dvector1}
\end{eqnarray}
transforms as a tensor of the rank $n$ with respect to a Lorentz transformation. For a vector, for example, it
follows that $(U \psi(x))^{\dagger} 
{\cal O} \gamma^a (U\psi(x)) = \Lambda^a{ }_b \;\psi(x)^{\dagger} {\cal O} \gamma^b \psi(x)$.
(It is usually  $\psi(x)^{\dagger} {\cal O} \gamma^0 \psi(x)$, 
which in $1+(d-1)$ plays the role of a probability density and not 
$\psi(x)^{\dagger} {\cal O}  \psi(x)$ !)

When there are more than one time coordinate the question, what can 
one interpret as a probability density, arises, which in this contribution we are not yet
able to answer.
Only for one time coordinate (for any $d$) the time component of the 
vector $\psi(x)^{\dagger} {\cal O} \gamma^a \psi(x)$
has the form, which we use to interpret as a probability density, 
namely $\psi(x)^{\dagger} \psi(x)$.

\section{Inner product for spinors obeying equations of motion}
\label{innereigen3p1}

Eq.(\ref{q+d-qinneruni}) defines the inner product for all possible
states of a Hilbert space in $q + (d-q)$ dimensional space, without
specifying details. 

Let us first treat scalars, which obey no equation of motion.
We expect ${\cal O} = I$. 
Let a label $i$ define (see Weinberg\cite{weinberg}
Vol I, Eqs.(2.5.1,2.5.19)) all the internal
degrees of freedom (we pay attention first of all on spin degree of
freedom). 
%
%\begin{eqnarray}
%P^a \psi_{p,i} = p^a \psi_{p,i}.
%\label{peigenstates}
%\end{eqnarray}
%
States $\psi_{i}$  depend on momenta
$p^a=(p^0,p^1,..,p^d)$ with all possible components of $p^a$. If eigenstates
of the momentum operator, they  can be
orthonormalized as follows
\begin{eqnarray}
\label{planewaves}
\langle \psi_{p,i}| \psi_{p,k}\rangle =
\delta_{ik}\; \delta^d(p-p').  
\end{eqnarray}

Of a particular interest are eigenstates of the
 equations of motion operator $(p^ap_a =0)\Phi_i$, which depend  on components $p^a$
defined on  a surface on which $p^ap_a=0$. Let us treat for a
while massless particles only.  Let $q=1$ and $\vec{p}$ is a $d-1$
vector in the subspace with the Euclidean signature.  If one starts with a vector
$\psi_{p,k}$
which depends on a generic $p^a$,  one
has to project $\psi_{p,k}$ first on a $(d-1)$-dimensional
surface on which $(p^0)^2 = (\vec{p})^2$. We can achieve this by a projector-like operator
$\delta(p^a p_a)$ since
 $\delta(p^a p_a)\delta(p^a p_a) = {\cal
  N}\; \delta(p^a p_a)$, with ${\cal N}$ which is $\infty$ (but can be
taken into account in the normalization procedure). Accordingly,
$\delta(p^a p_a) \Psi_{p,i}$ is (up to a normalization factor) a projection of $\Psi_{p,i}$ on a state
defined on a surface on which $(p^0)^2 = (\vec{p})^2$.  A state $\Psi_i$, which
depend on a generic $p^a$ can then with the requirement that only the projection on the surface
of $p^ap_a=0$ makes sense
be orthonormalized as follows 
\begin{eqnarray}
\langle \hat{\Psi}_i| \hat{\Psi}_k \rangle &=& 
\; \int \; d^d p \; \langle \Psi_i|p\rangle \delta(P^aP_a)
\langle p |\Psi_k \rangle \mu \nonumber\\
  &=&  \delta_{ik}\; \int \; {\cal N}\; \frac{d^{d-1}p}{2\sqrt{\vec{p}^2}}  \; 
 (\hat{\Psi}^{\dagger}_i\ \hat{Psi_k} )\bigg|_{ (p^0)^2  = 
(\vec{p})^2},
\label{orthonor}
\end{eqnarray}
with the weight function $\mu$ which is equal to $\mu = {\cal N}
\theta(p^0)$ for $p^0 \ge 0$, or $\mu = {\cal N} (1- \theta(p^0))$
for $p^0 \le 0$, with ${\cal N}$ which is a normalization constant.
In Eq.(\ref{orthonor}) the projector was of course needed only once,
we put it on $\langle p|\Psi_i\rangle$. We may interpret
$(\hat{\Psi}^{\dagger}_i\hat{\Psi_k} )\bigg|_{ (p^0)^2 = (\vec{p})^2}$ as a
probability density in the Euclidean part of the Hilbert space, if we
chose ${\cal N} = 2p^0$. The  notation $\hat{\Psi}_i$ is to point out that
the projection of the state $\Psi_i$ on a state
defined on a surface on which $(p^0)^2 = (\vec{p})^2$ was performed.

Eqs.(\ref{1+3inner},\ref{orthonor}) are valid for any dimension $d$
with the Minkowski signature ($1 + (d-1)$).  We easily prove
Eq.(\ref{orthonor}) by integrating it over $p^0$.

Let us assume again Minkowski signature $1+(d-1)$ and treat spinors
instead of scalars. Again we are interested on spinors, which obey
equations of motion. Massless spinors obey equations of motion
$\gamma^a P_a =0$. For spinors $\delta(\gamma^a P_a) $ is a kind of ``projector``
as it is
 $\delta(P^a P_a)$ in the case of scalars.%: $\delta(\gamma^a
%P_a)\delta(\gamma^a P_a) = {\cal N} \delta(\gamma^a P_a) $, with
%${\cal N} = \infty$.
We find\footnote{%
We are using the formula 
$\delta(f(x))=\delta(x)/\bigl(\frac{\partial f}{\partial x}\bigr)_{x=a}$
which is valid if $f$ has a single simple zero $a$. This formula can
be proved in the following way: we substitute Taylor expansion for 
$f$ around simple zero $a$, 
$$f(x)\approx f(a)+ (x-a) \cdot 
\biggl(\frac{\partial f}{\partial x}\biggr)_{x=a} +\ldots,$$
in the argument of $\delta$ function 
$$\int h(x) \delta(f(x))\,dx = 
\int h(x) \delta\biggl((x-a)\cdot\frac{\partial f}{\partial x}\biggr)\,dx$$
which immediately yields the required formula. }
\begin{eqnarray}
\delta(\gamma^a P_a) = 2 (\gamma^a P_a) \; \delta (P^a\gamma_a P^b
\gamma_b) 
= 2 (\gamma^a P_a) \; \delta(P^aP_a).
\label{deltaspinor0}
\end{eqnarray}
We immediately see that for massless spinors the ``projector``
$\delta(\gamma^a P_a)$ not only projects states $\Psi_i$ on a subspace
of states which are eigenstates of the operator $P^a P_a =0$ but also
transforms a left handed spinor with $\Gamma= -1$ into a right handed
spinor with $\Gamma = 1$. We stay within a Weyl representation of a
chosen handedness, if we first multiply a state $\Psi_i$ with a scalar
operator $\gamma^a n_a$, where $n_a$ is a vector with a property $n^a
n_a =1$ and then make use of the ``projector`` $\delta(\gamma^a P_a)$. We
can write an inner product for two states, which are defined in the
entire momentum space, but for which we are only interested in the
parts, which are the projections of the two states on the subspace, on
which the equations of motion are fulfilled, as follows
 \begin{eqnarray}
 \langle \hat{\Psi}_i|\hat{\Psi}_k \rangle &=& \int\; d^d p \langle \gamma^c n_c \Psi_i|p\rangle\; {\cal O}\;
 2 \gamma^a p_a \; \delta(p^b p_b)\; \gamma^d n_d \langle p| \Psi_k \rangle\\
 &=& \int \; d^d p \langle \Psi_i|p\rangle \; {\cal O}\; 2 (p^0 \gamma^0 + \vec{\gamma} \vec{p})\; \delta(p^bp_b) 
 \langle p|\Psi_k\rangle\\
 &=&{\cal N} \;\int \; d^{d-1} p \;(\hat{\Psi}_{i}^{\dagger}\hat{\Psi}_{k})\bigg|_{ (p^0)^2  = 
(\vec{p})^2},  
 \label{deltaspinor}
 \end{eqnarray}
since $(n_a\gamma^a)^{\dagger} {\cal O} = {\cal O} n_a \gamma^a$, $n_a \gamma^a n_b \gamma^b =I$, $p^a \gamma_a
n^b \gamma_b = n^b\gamma_b (p^0 \gamma^0 + \vec{p} \vec{\gamma})$, while 
$(p^0 \gamma^0 + \vec{\gamma} \vec{p}) \delta(P^aP_a)\Psi_{p,k} = 2 p^0 \gamma^0 \delta (p_a p^a) \Psi_{p,k} $.
Again, $\hat{\Psi}_i$ is the notation, which point out that the projection of a state
$\Psi_i$ on a state defined on a surface on which $(p^0)^2 = (\vec{p})^2$ is what we are interested in.

We have obtained a meaningful definition of the probability density,
which is $\hat{\Psi}_{i}^{\dagger}\hat{\Psi}_{k}\bigg|_{ (p^0)^2 = (\vec{p})^2}$.

Have we learned enough from the case $ 1+(d-1)$ to proceed to the
general $q+(d-q)$?  To generalize the inner product for spinors 
from $q=1$ to any $q$ we would need to know all the equations of
motion, since equations of motion occur as projectors in the inner
product (Eq.(\ref{deltaspinor})).  Up to now we only have taken into account one equation of
motion, the Weyl one. Are there more than one?  Having more than one
time parameter, one would expect more than one equation of motion.  We
see, for example, that for two time parameters one immediately can
find two momenta, which are orthogonal to each other and yet both
fulfil the condition $p^ap_a=0$, namely, for example, $p^a_0
=(p^0,0,\cdots,0,p^0)$ and $p^a_1 =(0, p^1,0, \cdots,p^1,0)$. For
general $q$ one finds $min \{q, d-q\}$ vectors $p^a$, which all are
orthogonal to each other. We further see that
\begin{eqnarray}
(\sum_{\beta =1}^{min \{q, d-q \}} \; \alpha_{\beta}\; \gamma^a p_{\beta a})^2 =0
\label{moreweyl}
\end{eqnarray}
leads to a solution
\begin{eqnarray}
p^a_{\beta}p_{\beta a} =0,\;p^a_{\beta}p_{\gamma a} = 0, \quad {\rm for \; all} \; \beta,\gamma \in{(1,min\{q,d-q\})},
\label{moreweylorth}
\end{eqnarray}
with $q$ vectors $p^a_{\beta}$, which are orthogonal to each other. This should motivate us to look for more
than one equations of motion, namely for $min \{q,d-q\}$ equations of motion.

\section{Conclusion}
\label{conclusion}

In this contribution we have tried to confront the problems which occur when the signature of time-space  
is $q + (d-q)$, with $q > 1$. We have studied in particular a definition of an inner product
of two generic spinors in $d$-dimensional space, belonging to a chosen irreducible representation
of the Lorentz group or accordingly of the Poincar\ 'e group. We expect that the inner product is
unitary and Lorentz invariant,
while the Lorentz group $SO(q,d-q)$ is noncompact. We also expect that the probability density
has some meaning.  One of the motivations for studying these problems is to find out, why Nature has 
made a choice of one time and three space dimensions, with some problems already discussed
in ref.\cite{holgernormawhy2002}. 

For $q=1$ the answers to these questions are known since Dirac and Wigner. The problem of the unitarity 
of the inner product was solved by defining $\bar{\Psi}= \Psi^{\dagger}\gamma^0$. Since usually not the 
whole Hilbert
space, but only the projection of this space on the surface on which the equations of
motion are fulfilled, is of the interest, the inner product has to manifest this requirement.

In this contribution we generalized the  definition of $\bar{\psi}$ for spinors to any $q$ of $d$ dimensional space, 
by introducing the operator ${\cal O} = \prod_{a \;{\rm is\; a\; time-like\; index}} \;\; \gamma^a $.
We further study the case with $q=1$, by introducing the projector $\delta(\gamma^a p_a) $ which projects
the vectors $\gamma^d n_d \langle p|\Psi_k \rangle$, with $\Psi_k$, which is a vector in 
the d-dimensional space, on the surface on which the Weyl equation is fulfilled and transforms it from
one handedness to another, so that the projected vector has the same handedness as the starting one, namely
$\Psi_k$. We did that to be 
able to generalize the inner product for any $q$. The problem, which we haven't yet 
solved, is however to find out all the equations of motion which a  spinor should obey in 
$d$-dimensional space of the signature $q+ (d-q)$. All the equations of motion should to our 
understanding namely enter into the definition of the inner product in a way the Weyl 
equation does in Eq.(\ref{deltaspinor}).

\section{Acknowledgement } 
This work was supported by Ministry of Education, Science and Sport of
Slovenia and Ministry of Science of Denmark.  The authors would like
to thank to the communicator and his referees for stimulating
suggestions to further clarify the paper.

%% small charges, 25.11.2002
\newcommand{\SO}{\mathrm{SO}}
\newcommand{\SU}{\mathrm{SU}}
\newcommand{\unit}{\mathrm{U}}

\title*{%
Weyl Spinor of $\SO(1,13)$, Families of Spinors of the 
Standard Model and Their Masses}
\author{A. Bor\v stnik Bra\v ci\v c$^{1,2}$ and 
N. Manko\v c Bor\v stnik$^{2,3}$} 
\institute{%
${}^1$ Educational Faculty, University of Ljubljana,
 Kardeljeva plo\v s\v cad 17, 1000 Ljubljana\\
${}^2$ Primorska Institute for Natural Sciences and Technology, 
C. Mare\v zganskega upora 2, Koper 6000, Slovenia\\
${}^3$ Department of Physics, University of
Ljubljana, Jadranska 19, 1111 Ljubljana}

\titlerunning{Weyl Spinor of $\SO(1,13)$, Families of Spinors \ldots}
\authorrunning{A. Bor\v stnik Bra\v ci\v c and N. Manko\v c Bor\v stnik}
\maketitle

\begin{abstract} 
We propose some possible answers to the open questions of the Standard electroweak model, using the
approach of one of us\cite{norma92,norma93,norma97,norma01} unifying spins and charges. 
We demonstrate that
one (!) Weyl left handed multiplet of the group $\SO(1,13)$ contains, if represented in a way 
to demonstrate the $\SU(3), \SU(2)$ and $\unit(1)$'s substructure, the spinors (quarks and leptons) 
and the ``antispinors``(antiquarks and antileptons) of the Standard
model (with the right handed weak chargeless neutrino and the left handed weak chargeless 
antineutrino in addition), that the weak charge breaks parity while the colour charge does not, 
comment on a possible break of the group $\SO(1,13) $ which leads to  
spins, charges and flavours of  leptons and quarks and antileptons and antiquarks, comment on the
appearance of spin connections and vielbeins as gauge fields 
connected with charges and as  Yukawa couplings and accordingly as masses of families. 
We demonstrate the appearance of families, suggesting symmetries of mass matrices and argue for
the appearance of the fourth family, with all the properties (besides the masses) 
of the three known families (all in agreement with ref.\cite{okun-sc3}). 
We also comment on small charges of observed spinors (and ``antispinors``) 
and on anomaly cancellation.

\end{abstract}

\section{Introduction}
\label{introduction-sc3}

The Standard electroweak model assumes the left handed weak charged doublets which are either colour triplets 
(quarks) or colour singlets (leptons) and the right handed weak chargeless singlets which are again 
either colour triplets
or colour singlets. And the corresponding ``antispinors`` (antiquarks and antileptons). 
How can it be  at all that charges are connected with the 
handedness of the group $\SO(1,3)$ which concerns spins?
Why does the weak charge break parity while the
colour charge does not? Why there are families of quarks and leptons 
and where do families of spinors (fermions) come from? 
Why are quarks and leptons massless - until (in the concept of the Standard model) gaining a (small) mass at low 
energies through the 
vacuum expectation value(s) of Higgs fields and Yukawa couplings 
(which agrees with the experimental data so well) or where the Yukawa couplings come from? 
How particles and antiparticles appear?
Why the observed representations of known 
charges are so small\footnote{This paper started as an answer to the Holger Bech Nielsen 
presentation of an attemt to argue for small charges at the workshop BLED 2002, ``What comes
beyond the Standard model? ``. My  (S.N.M.B.) answer to the attempt was that the approach of mine
unifying spins and charges offers a natural explanation for small charges. 
To show this we present the work  (done mainly five years ago with A.B.B.), which is an attempt
to answer also some of other open questions of the Standard model.}: 
singlets, doublets and at most triplets? 
Why does the  anomaly cancellation  occur?
Why are spinors (almost) massless at low (observed) energies?  
Can it not be that indeed all the internal degrees of freedom - spins and charges - are only one unified degree 
of freedom manifesting at low energies as spins and charges? 

We are proposing a possible answer to (some of) these questions 
by using the approach of one of us\cite{norma92,norma93,norma95,norma97,norma01,holgernorma2002,pikanormaproceedings} 
and the technique\cite{normasuper94,holgernorma00,holgernorma2002}, developed within this 
approach. We demonstrate in section \ref{spinor} that a left handed $\SO(1,13)$ Weyl spinor multiplet includes, 
if the representation is interpreted
in terms of the subgroups $\SO(1,3)$, $\SU(2)$, $\SU(3)$ and the sum of the two $\unit(1)$'s,  spinors and ``antispinors`` of
the Standard model - that is left handed $\SU(2)$ doublets and right handed  $\SU(2)$ singlets of $\SU(3)$ charged 
quarks and $\SU(3)$ chargeless leptons, while ``antispinors`` are oppositely charged and have opposite handedness.
(Right handed neutrinos and left handed antineutrinos - both weak chargeless - are also included, so that 
the multiplet
has 64 members, half with spin up and half with spin down.)
The representation for spinors alone is anomaly free (and so is for ``antispinors``) (section \ref{conclusions-sc3}). 
And it is also mass 
protected. Besides, the approach offers a possible 
explanation 
for families of spinors and their masses (sections \ref{mechanism}, \ref{Yukawa}).

Our gauge group is $\SO(1,13)$ - the smallest complex Lorentz group with a left handed Weyl spinor
containing the needed representations of the Standard model.  Accordingly the gauge fields are spin 
connections and vielbeins\cite{norma93sc3,norma01},which 
in four dimensional subspace manifest as the gauge fields of the known charges and the Yukawa couplings
(section \ref{Yukawa}).
 
We shall mainly comment on representations
of the group $\SO(1,13)$ and of subgroups of this group (section \ref{spinor}) and on a possible way of 
breaking the $\SO(1,13)$ symmetry (section \ref{break})
leading  to the Standard model degrees of freedom.
But we also shall demonstrate that the action for a Weyl (massless) spinor in a d-dimensional space 
with a gravitational gauge field present
can manifest in a four dimensional subspace as an action for a massive (Dirac) spinor, with the Yukawa 
couplings following from the gauge field (\ref{Yukawa}), and with the appearance of families (\ref{familiessub},
\ref{mechanism}) of quarks and leptons.

We define the handedness of the group $\SO(1,d-1)$ and  the subgroups  $\SO(d')$ of this group (\ref{spinor}),
with $d=2n$ and $d'=2k$ in terms of appropriate products of the generators of the Lorentz transformations in 
internal 
space\cite{norma93sc3,normasuper94}. We demonstrate that handedness of the group $\SO(1,d-1)$ and of subgroups play an
essential role for spinors (\ref{conclusions-sc3}). 

We use the technique\cite{holgernorma2002sc3}, which enables to follow explicitly the appearance of 
charged and chargeless states and antistates of an irreducible
(left handed Weyl) representation of the group $\SO(1,13)$ (\ref{technique}). It helps to understand the ano\-maly 
freedom of the representation, the
smallness of the representation, the appearance of the complex representation - needed to distinguish
between spinors and ``antispinors``.

We also introduce (\ref{mechanism}, \ref{Yukawa}) the operators, transforming one family into another and present
the mass matrices for the four families of quarks and leptons, suggested by our approach (\ref{sec:mass}). We also comment on
more than four families, assuming the symmetry of mass matrices, proposed by the approach.

In this paper we 
pay attention on spinor representations only, that is on the smallest representation besides the trivial one  
(which is a scalar with respect to all the internal degrees of freedom, leading to trivial noninteracting fields 
without spins and 
in our approach accordingly also without charges).

\section{Spinor representations in d-dimensional space expressed in terms of Clifford algebra elements }
\label{spinor}

In this section we demonstrate, using the 
technique\cite{normasuper94,norma01,holgernorma00,holgernorma2002,pikanormaproceedings}
which represents spinors in terms of products of the Clifford
algebra elements $\gamma^a$, that one left handed $\SO(1,13)$ multiplet contains states with
the properties needed
to  describe quarks and leptons of one family of the Standard model. 
We also demonstrate the appearance of families.

For this purpose we first make in subsection \ref{lorentz} a choice of basic states of the Lorentz group. 
We choose them to be
eigenstates of a chosen Cartan subalgebra elements ($d/2$ for $d$ even) of the Lorentz algebra  with  
$d(d-1)/2$ elements. We are interested in all possible invariants of the group $\SO(1,13)$ and of subgroups. 
We look only on maximal regular subgroups of the group $\SO(1,13)$, that is on subgroups  the sum of ranks of
which is the rank of the group $\SO(1,13)$, which  is $d/2=7$ paying attention on even dimensional
subspaces, since only in even dimensional subspaces the mass protection mechanism 
works\cite{norma01sc3,holgernorma2002}. We accordingly pay attention 
(\ref{subgroups}) on the
$\SO(d')$ subgroups, with $d=2k$ and their subgroups $\SU(2)$ and $\SU(3)$ and $\unit(1)$'s.
We define the handedness, which is a Casimir of the group $\SO(1,13)$ or any of subgroups $\SO(1,d'-1)$ or
$\SO(d')$.

In subsection \ref{technique} we present a simple and transparent technique to describe spinor basic states  
in terms of products of $\gamma^a$'s.
We use this technique in subsection \ref{left} to represent one left handed $\SO(1,13)$ Weyl 
spinor and to demonstrate its properties.

\subsection{Lorentz group $\SO(1,13)$ and subgroups $\SO(1,3), \SU(2), \SU(3), \unit(1) $}
\label{lorentz}
%handedness

Let  operators $\gamma^a$ close the Clifford algebra
\begin{equation}
\{\gamma^a, \gamma^b \}_+ = 2\eta^{ab}, \quad {\rm for} \quad a,b \quad \in \{0,1,2,3,5,\cdots,d \},
\label{clif}
\end{equation}
for any $d$, even or odd, and let the Hermiticity property of $\gamma^a$'s be
\begin{eqnarray}
\gamma^{a+} = \eta^{aa} \gamma^a,
\label{cliffher}
\end{eqnarray}
in order that 
$\gamma^a$ be unitary as usual, i.e. ${\gamma^a}^{\dagger}\gamma^a=1$.

 The operators 
\begin{equation}
S^{ab} = \frac{i}{4} [\gamma^a, \gamma^b ] := \frac{i}{4} (\gamma^a \gamma^b - \gamma^b \gamma^a)
\label{sabsc3}
\end{equation}
close the algebra of the Lorentz group 
\begin{equation}
\{S^{ab},S^{cd}\}_- = i (\eta^{ad} S^{bc} + \eta^{bc} S^{ad} - \eta^{ac} S^{bd} - \eta^{bd} S^{ac})
\label{loralg}
\end{equation}
and also fulfill the  spinor algebra 
$\{S^{ab},S^{ac}\}_+ = \frac{1}{2} \eta^{aa} \eta^{bc}.$
Recognizing from Eq.(\ref{loralg}) that two operators $S^{ab}, S^{cd}$ with all indices different 
commute, we readily select the Cartan subalgebra of the algebra of the Lorentz group with $m =  d/2$  
for $ d $ even commuting operators.

We define one of the Casimirs of the Lorentz group which determines the handedness of an
irreducible representation of the Lorentz group\footnote{To see the definition of the 
operator $\Gamma$ for any spin in even-dimensional spaces see references\cite{norma93sc3,%
normasuper94,bojannorma2001,holgernorma00}.} 
\begin{eqnarray}
\Gamma^{(d)} :&=& 2^{d/2} \; \prod_a \sqrt{\eta^{aa}} \quad S^{03} S^{12} S^{56} \cdots S^{d-1\; d} \nonumber\\
 &=& \;
(i)^{d/2}\;\prod_a \quad (\sqrt{\eta^{aa}} \gamma^a), \quad {\rm if } \quad d = 2n, 
\label{handsc3}
\end{eqnarray}
for any integer $n$. We understand the product of $\gamma^a$'s in the ascending order with respect to 
the index $a$: $\gamma^0 \gamma^1\cdots \gamma^d$. 
It follows for any choice of the signature $\eta^{aa}$  that 
$\Gamma^{(d)}$ is Hermitean and its square is equal to the unity operator
\begin{eqnarray}
\Gamma^{(d)\dagger}= \Gamma^{(d)},\quad
\Gamma^{(d)2} = I.
\label{prophand1}
\end{eqnarray}
One also finds that in even-dimensional spaces $\Gamma^{(d)}$ anticommutes (while in odd-dimensional spaces
$\Gamma^{(d)}$ commutes) with $\gamma^a$'s
$
(\{\Gamma^{(d)},\gamma^a\}_{+} = 0 \;{\rm for} \; d \; {\rm even}$). 
%\{\Gamma^{(d)},\gamma^a\}_{-} = 0, \; {\rm for} \; d \; {\rm odd}$
%
Accordingly, $\Gamma^{(d)}$ always commutes with the generators of the Lorentz algebra
\begin{eqnarray}
\{\Gamma^{(d)}, S^{ab}\}_- = 0.
\label{prophand2}
\end{eqnarray}

For spinors it is easy to see that eigenstates of the Cartan subalgebra  are   
eigenstates 
of the operator of handedness as well.

We shall select operators belonging to the Cartan subalgebra of $7$ elements of $\SO(1,13)$ as follows
\begin{eqnarray}
S^{03}, S^{12}, S^{56}, \cdots, S^{13\; 14}.
\label{cartan}
\end{eqnarray}
We present the operators of handedness for the Lorentz group $\SO(1,13)$ and the subgroups $\SO(1,3), \SO(1,7),
\SO(1,9), \SO(6)$ and $\SO(4)$
\begin{eqnarray}
\Gamma^{(1,13)} &=& \; 2^{7}i \; S^{03} S^{12} S^{56} \cdots S^{13 \; 14},
\nonumber\\
\Gamma^{(1,3)} \; &=& \; - 4i \; S^{03} S^{12},
\nonumber\\
\Gamma^{(1,7)} \;&=& \; - 2^{4} i \; S^{03} S^{12} S^{56} S^{78},
\nonumber\\
\Gamma^{(1,9)}\; &=& \; 2^{5} i \; S^{03} S^{12} S^{9\;10} S^{11\;12} S^{13 \; 14},
\nonumber\\
\Gamma^{(6)}\;\;\; &=& \; 8 \; S^{9 \;10} S^{11\;12} S^{13 \; 14},
\nonumber\\
\Gamma^{(4)}\;\;\; &=&\; 4 \; S^{56} S^{78}.
\label{gammas}
\end{eqnarray}

\subsection{Subgroups of $\SO(1,13)$}
\label{subgroups}

From the point of view of small charge representations one would look for 
the $(\SU(2) \times \SU(2))^k$ (isomorphic
to $(\SO(4))^k$) 
content of the group $\SO(1,13)$. Namely, the $\SU(2)\times \SU(2)$ content of  $\SO(4)$ (or $\SO(1,3)$) 
demonstrates  either doublets with respect to the first $\SU(2)$ and singlets with respect to the second or 
opposite.
But $14$ is not a multiple of $4$. Accordingly, one can look for the regular
subgroups $\SO(1,7)$ (with the $(\SU(2) \times \SU(2))^2$ content of it) and $\SO(6)$ of the group $\SO(1,13)$, 
with the sum of the  ranks of subgroups ($4$ and $3$, respectively)
equal to the rank of $\SO(1,13)$. 

The group $\SO(1,13)$ and the subgroup $\SO(6)$ (isomorphic to $\SU(4)$) 
have complex representations as all  groups of the type $\SO(2(2k+1)),$ for any $k$, 
have\cite{georgi}. Complex representations are needed to distinguish 
between spinors and ``antispinors``\footnote{The concept of spinors and ``antispinors`` can be understood, if
looking at the properties of one Weyl representation in terms of subgroups of $\SU(3), \SU(2)$ and two
$\unit(1)$'s in Table I. We started with one Weyl spinor only, which then in terms of the
representations of the subgroups demonstrates as spinors with respect to the subgroup $\SO(1,3)$ with
the $\SU(3)$ and $\unit(1)$'s charges (quarks and leptons) and as spinors with respect to the subgroup $\SO(1,3)$
with the $\SU(3)$ and $\unit(1)$'s anticharges (antiquarks and antileptons) and we call these spinors with 
anticharges ``antispinors``. }. We shall use indices $0,1,2,3$ to describe the spin part of 
the internal degrees of freedom, indices $5,6,7,8$ to describe 
the weak and one $\unit(1)$ charge and indices $9,10,11,12$ to describe the colour and another $\unit(1)$ 
charge. We shall use the sum of the two $\unit(1)$'s generators to describe the 
hypercharge $Y$ of the Standard model and
the difference as an additional hypercharge $Y'$.

The generators of the subgroups $\SO(1,3)$, 
 $\SU(2), \SU(3)$ and $\unit(1)$'s, needed to determine
the spin, the weak charge, the colour charge and the hyper 
charges content of $\SO(1,13)$  can be written in terms of the generators $S^{ab}$ 
\begin{eqnarray}
\tau^{Ai} = \sum_{a,b} \;c^{ai}{ }_{ab} \; S^{ab},
\nonumber\\
\{\tau^{Ai}, \tau^{Bj}\}_- = i \delta^{AB} f^{Aijk} \tau^{Ak},
\label{tau}
\end{eqnarray}
with $A=1,2,3,4,5,6$ representing the corresponding subgroups and $f^{Aijk}$ the corresponding structure constants.
Coefficients $c^{Ai}{ }_{ab}$ have to be determined so that the commutation relations of Eq.(\ref{tau}) 
hold\cite{norma97}.

One expresses  the left  handed ($\Gamma^{(1,3)}=-1$) and the right handed ($\Gamma^{(1,3)}=1$) content of
$\SO(1,3)$ representation with the help of the operators
\begin{eqnarray}
\tau^{11}: = \frac{1}{2} ( {\mathcal S}^{23} - i {\mathcal S}^{01} ),\quad
\tau^{12}: = \frac{1}{2} ( {-\mathcal S}^{13} - i {\mathcal S}^{02} ),\quad
\tau^{33}: = \frac{1}{2} ( {\mathcal S}^{12} - i {\mathcal S}^{03} )
\nonumber\\
\tau^{21}: = \frac{1}{2} ( {\mathcal S}^{23} + i {\mathcal S}^{01} ),\quad
\tau^{22}: = \frac{1}{2} ( {-\mathcal S}^{13} + i {\mathcal S}^{02} ),\quad
\tau^{23}: = \frac{1}{2} ( {\mathcal S}^{12} + i {\mathcal S}^{03} ).
\label{su12s}
\end{eqnarray}
Equivalently the $\SU(2)\times \SU(2)$ content of the compact group $\SO(4)$ follows
\begin{eqnarray}
\tau^{31}: = \frac{1}{2} ( {\mathcal S}^{58} - {\mathcal S}^{67} ),\quad
\tau^{32}: = \frac{1}{2} ( {\mathcal S}^{57} + {\mathcal S}^{68} ),\quad
\tau^{33}: = \frac{1}{2} ( {\mathcal S}^{56} - {\mathcal S}^{78} )
\nonumber\\
\tau^{41}: = \frac{1}{2} ( {\mathcal S}^{58} + {\mathcal S}^{67} ), \quad
\tau^{42}: = \frac{1}{2} ( {\mathcal S}^{57} - {\mathcal S}^{68} ), \quad
\tau^{43}: = \frac{1}{2} ( {\mathcal S}^{56} + {\mathcal S}^{78} ).
\label{su12w}
\end{eqnarray}
We shall choose $\tau^{3i}, i=1,2,3$ to describe the weak charge and $\tau^{43}$ 
to describe the  $\unit(1)$ content of $\SO(4)$.
We find
$\;\{\tau^{Ai}, \tau^{Bj}\} = i \delta^{AB}\; \epsilon^{ijk} \tau^{3k}$,
with $A=1,2,3,4$ and $i=1,2,3$, except for $A=4$, since $\tau^{43}$ is used to determine the $\unit(1)$
content of representations.

We express 
generators of  subgroups $\SU(3)$ and $\unit(1)$ of the group
$\SO(6)$ in terms of the generators ${\mathcal S}^{ab}$ as follows
\begin{eqnarray}
\tau^{51}: &=& \frac{1}{2} ( {\mathcal S}^{9\;12} - {\mathcal S}^{10\;11} ),\quad
\tau^{52}: = \frac{1}{2} ( {\mathcal S}^{9\;11} + {\mathcal S}^{10\;12} ),\quad
\tau^{53}: = \frac{1}{2} ( {\mathcal S}^{9\;10} - {\mathcal S}^{11\;12} ),
\nonumber
\\
\tau^{54}: &=& \frac{1}{2} ( {\mathcal S}^{9\;14} - {\mathcal S}^{10\;13} ),\quad
\tau^{55}: = \frac{1}{2} ( {\mathcal S}^{9\;13} + {\mathcal S}^{10\;14} ),\quad
\tau^{56}: = \frac{1}{2} ( {\mathcal S}^{11\;14} - {\mathcal S}^{12\;13}
),
\nonumber\\
\tau^{57}: &=& \frac{1}{2} ( {\mathcal S}^{11\;13} + {\mathcal S}^{12\;14} ),
\quad
\tau^{58}: = \frac{1}{2\sqrt{3}} ( {\mathcal S}^{9\;10} + {\mathcal
S}^{11\;12} - 2{\mathcal S}^{13\;14})
\nonumber\\
\tau^{61}: &=& -\frac{1}{3}( {\mathcal S}^{9\;10} + {\mathcal S}^{11\;12}
+ {\mathcal S}^{13\;14} ).
\label{su3u1so6}
\end{eqnarray}
We find
$\; \{\tau^{5i}, \tau^{5j}\} = i f^{ijk} \tau^{5k}, \quad
\{\tau^{61}, \tau^{5i}\} = 0, {\mathrm \;\; for\;\; each \;\;i},\;$
with coefficients $f^{ijk}$ which are the structure constants of the group
$\SU(3)$. 

We define two superpositions of the two $\unit(1)$'s generators as follows
\begin{eqnarray}
Y = \tau^{61} + \tau^{43}, \quad  Y' = \tau^{61} - \tau^{43}. 
\label{yyprime}
\end{eqnarray}

The choice of subgroups of the group $\SO(1,13)$ makes possible to comment one weyl spinor of the 
group $\SO(1,13)$ in terms of the groups of the Standard model.

\subsection{Technique for generating spinor representations in terms of $\gamma^a$'s}
\label{technique}

In this subsection we very briefly present the technique from refs.\cite{normasuper94,pikanormaproceedings,%
holgernorma2002}, asking the
reader to see for proofs and details the reference\cite{holgernorma2002sc3}.
This simple technique makes  spinor representations and accordingly all their properties very transparent.
We define, namely, spinors  as polynomials  of the Clifford algebra objects $\gamma^a$'s, which are 
eigenstates of the chosen Cartan subalgebra and assume that $\gamma^a$'s and accordingly also $S^{ab}$'s
are applied from the left hand side.

Let $S^{ab}$ be any of the elements of the Lorentz algebra (Eq.(\ref{sabsc3})) of the Lorentz group.
Then the states obtained by operating by the operators
\begin{eqnarray}
\stackrel{ab}{(\pm \sqrt{-\eta^{aa}\eta^{bb}})}: &=& 
\frac{1}{\sqrt{2}}(\gamma^a \pm \sqrt{-\eta^{aa} \eta^{bb}}\; \gamma^b),
\nonumber\\  
\stackrel{ab}{[\pm \sqrt{-\eta^{aa}}]}: &=& \frac{1}{\sqrt{2}}(1 \pm \sqrt{-\eta^{aa}\eta^{bb}} \;\gamma^a
\gamma^b)
\label{eigensab}
\end{eqnarray}
on $|\psi>$ are eigenstates of the operator $S^{ab}$
with the eigenvalues 
$$\pm (i/2) \eta^{bb} \; \sqrt{-\eta^{aa} \eta^{bb}}$$ 
and 
$$\pm (i/2) (-\eta^{aa}\eta^{bb}) \; \sqrt{-\eta^{aa} \eta^{bb}},$$ 
respectively. $|\psi> $ is
any state which the above operators will not transform into zero. The  
reader can find the proof in ref.\cite{holgernorma2002sc3}.

One also notices that the operators $\gamma^a$  transforms the 
first two operators of Eq.(\ref{eigensab}) into the second two operators of Eq.(\ref{eigensab}),
or opposite, if being multiplied
from the left
\begin{eqnarray}
\gamma^a \stackrel{ab}{(\pm \sqrt{-\eta^{aa}\eta^{bb}})}: &=& 
\frac{1}{\sqrt{2}}(\eta^{aa} \pm \sqrt{-\eta^{aa} \eta^{bb}}\;
\gamma^a \gamma^b),\nonumber\\  
\gamma^{a}\stackrel{ab}{[\pm \sqrt{-\eta^{aa}}]}: &=& \frac{1}{\sqrt{2}}(\gamma^a \pm \eta^{aa}
\sqrt{-\eta^{aa}\eta^{bb}} \;\gamma^b)
\label{eigensabgamma}
\end{eqnarray}
and similar transformation follows also for the operator $\gamma^b$. We leave the reader to work out
that  the state obtained by the application of $\gamma^a$ or $\gamma^b$ has the opposite eigenvalue
of the operator $S^{ab}$ than the starting state.

Let $S^{ab}$ and $S^{cd}$ be the two elements of the Cartan subalgebra (Eq.(\ref{cartan})). Then the
two vectors 
\begin{eqnarray}
(\gamma^a + \sqrt{-\eta^{aa} \eta^{bb}}\;\; \gamma^b)\; (\gamma^c + \sqrt{-\eta^{cc}\eta^{dd}} \;\;
\gamma^d)|\psi\rangle,
\nonumber\\
(1 + \frac{1}{\eta^{aa}} \sqrt{-\eta^{aa} \eta^{bb}}\;\; \gamma^a \gamma^b) \;  (1 + \frac{1}{\eta^{cc}}
\sqrt{-\eta^{cc}\eta^{dd}} \;\;\gamma^c \gamma^d)|\psi\rangle
\label{eigenirred}
\end{eqnarray}
belong to the same irreducible representation, where $|\psi\rangle$ is any state which will 
 not be transformed into zero. This can easily be checked by applying $S^{ac}, S^{bd}, S^{ad}$ or $S^{bc}$
on one of the two vectors to obtain the second one by taking into account Eq.(\ref{eigensabgamma}). 
The reader can  find the proof in ref.\cite{holgernorma2002sc3}. We also point out that the second state has
the opposite eigenvalues for both $S^{ab}$ and $S^{cd}$ than the first one.

We would like to pay attention  that $\gamma^a$ (or $\gamma^b$) being applied from the 
left on the operator $\stackrel{ab}{(\pm)}$
transforms it to the operator $\stackrel{ab}{[\mp]}$, to which an opposite value of $S^{ab}$ belongs.

Let us make a choice of a starting state  of a Weyl representation of the group $\SO(1,13)$, which is the
eigenstate of all the members of the Cartan subalgebra (Eq.(\ref{cartan})) and is left handed
($\Gamma^{(1,13)} =-1$), 
as follows
\begin{eqnarray}
\stackrel{03}{(+i)}\stackrel{12}{(+)}|\stackrel{56}{(+)}\stackrel{78}{(+)}
||\stackrel{9 \;10}{(+)}\stackrel{11\;12}{(-)}\stackrel{13\;14}{(-)} |\psi \rangle  = \nonumber\\
(\gamma^0 -\gamma^3)(\gamma^1 +i \gamma^2)| (\gamma^5 + i\gamma^6)\nonumber\\
 (\gamma^7 +i \gamma^8)||
(\gamma^9 +i\gamma^{10})(\gamma^{11} -i \gamma^{12})(\gamma^{13}-i\gamma^{14})|\psi \rangle. 
\label{start}
\end{eqnarray}

The signs "$|$" and "$||$" are to point out the  $\SO(1,3)$ (up to $|$), $\SO(1,7)$ (up to $||$)
and $\SO(6)$ (between $|$ and $||$) substructure of the starting state of the left handed multiplet of
$\SO(1,13)$ which has $2^{14/2-1}= 64 $ vectors. Again $|\psi\rangle$ is any state, which is not transformed 
to zero. From now on we shall not write down $|\psi \rangle$ anylonger.
One easily finds that the eigenvalues of the chosen
Cartan subalgebra elements of Eq.(\ref{cartan}) are $+i/2, 1/2, 1/2,1/2,1/2,-1/2,-1/2$, 
respectively. This state is a right handed spinor with respect to $\SO(1,3)$ ($\Gamma^{(1,3)} =1$, 
Eq.(\ref{gammas})), with spin up 
($S^{12} =1/2$), it is $\SU(2)$  
singlet ($\tau^{33} = 0$, Eq.(\ref{su12w})), and it is the member 
of  the $\SU(3)$ triplet (Eq.(\ref{su3u1so6})) with ($\tau^{53} =1/2, \tau^{58} = 1/(2 \sqrt{3})$),
it has $\tau^{43} = 1/2$ and $\tau^{6,1}= 1/2$. We further find
according to Eq.(\ref{gammas}) that $\Gamma^{(4)} =1, \Gamma^{(1,7)}= 1, \Gamma^{(6)} = -1$ and 
$\Gamma^{(1,9)} = -1$.

To obtain all the states of one Weyl spinor one only has to apply on the starting state of Eq.(\ref{start})
the generators $S^{ab}$. 

The generators $S^{01}, S^{02}, S^{31}, S^{32}$ transform spin up state (the 
$\stackrel{03}{(+i)}\stackrel{12}{(+)} $ part of the starting state (Eq.(\ref{start}) with
$S^{12}=1/2$ and $S^{03}=i/2$) into spin 
down state ($\stackrel{03}{[-i]}\stackrel{12}{[-]}$, which has
$S^{12}=-1/2$ and $S^{03} = -i/2$), leaving all the other parts of the state and accordingly also all the other properties of 
this state unchanged.  None of the generators $S^{mn},$ with $m,n= 0,1,2,3$, can change a right handed
$\SO(1,3)$ Weyl spinor ($\Gamma^{(1,3)} =1$) into a left handed
$\SO(1,3)$ Weyl spinor ($\Gamma^{(1,3)} =-1$).

The generators $S^{57}, S^{58}, S^{67}, S^{68}$ transform one $\SU(2)$ singlet into another singlet
($\stackrel{56}{(+)}\stackrel{78}{(+)}$ into $\stackrel{56}{[-]}\stackrel{78}{[-]}$), changing
at the same time the value of 
$\tau^{43}$ from $1/2$ to $-1/2$. ($\tau^{3i}$ can not do that, of course.)

The generators $S^{mh}, m=0,1,2,3$ ; $\;h =5,6,7,8$, transform a right handed 
$\SU(2)$ singlet  ($\Gamma^{(1,3)}=1$) with spin up into a member of the left handed ($\Gamma^{(1,3)}=-1$)
$\SU(2)$ doublet, 
with spin up ($S^{05}$, for example, changes 
$\stackrel{03}{(+i)}\stackrel{12}{(+)} \stackrel{56}{(+)}\stackrel{78}{(+)}$ into $\stackrel{03}{[-i]}
\stackrel{12}{(+)}\stackrel{56}{[-]}\stackrel{78}{(+)}$) or spin down ($ S^{15}$, for example,
changes $\stackrel{03}{(+i)}\stackrel{12}{(+)}\stackrel{56}{(+)}\stackrel{78}{(+)}$ into the
$\stackrel{03}{(+i)}\stackrel{12}{[-]}\stackrel{56}{[-]}\stackrel{78}{(+)}$).
$S^{57}, S^{58}, S^{67}, S^{68}$
transform one state of a doublet into another state of the same doublet
($\stackrel{56}{(+)}\stackrel{78}{(+)}$ into $\stackrel{56}{[-]}\stackrel{78}{[-]}$), as also $\tau^{3i}$ do.
The $\SU(3)$ quantum numbers $\tau^{53}$
and $\tau^{58}$ as well as $\tau^{61}$ stay unchanged.

The generators $S^{kl}$, with $k,l = 9,10,11,12,13,14$, transform one member of the triplet 
into the other
two members ($S^{9\;11}$, $S^{9\;12}$, $S^{10\;11}$,  $S^{10\;12}$,  
$S^{9\;13}$, $S^{9\;14}$, $S^{10\;13}$, $S^{10\;14}$ 
transform $\stackrel{9 \;10}{(+)}\stackrel{11\;12}{(-)}\stackrel{13\;14}{(-)}$ either into 
$\stackrel{9 \;10}{[-]} \; \stackrel{11\;12}{[+]} \; \stackrel{13\;14}{(-)}$ with $\tau^{53} = -1/2$, 
$\tau^{58}= 1/(2\sqrt{3})$ and $\tau^{61} = 1/6$ or into 
$\stackrel{9 \;10}{[-]}\; \stackrel{11\;12}{(-)}\; \stackrel{13\;14}{[+]}$ with $\tau^{53} = 0$,
$\tau^{58}= -1/\sqrt{3}$ and $\tau^{61}= 1/6$) or 
they transform a triplet into a singlet ($S^{11\;13}$, $ S^{11\;14}$, 
$ S^{12\;13}$, $ S^{12\;14}$
transform $\stackrel{9 \;10}{(+)}\; \stackrel{11\;12}{(-)} \; \stackrel{13\;14}{(-)}$ 
into $\stackrel{9 \;10}{(+)} \; \stackrel{11\;12}{[+]} \; \stackrel{13\;14}{[+]}$ with  $\tau^{53}=0=\tau^{58}$ and
$\tau^{61}=-1/2$).

The generators $S^{h,k}$, with $h=0,1,2,3,5,6,7,8$ and $k=9,10,11,12,13,14$, transform a triplet of 
Eq.(\ref{start}) into an antitriplet. $S^{7\;13}$ for example, applied on the starting state, transforms it into
the right handed member of the $\SU(2)$ (anti) doublet with $\tau^{33}= 1/2, \tau^{43}=0, \tau^{53}=1/2,
\tau^{58}= -1/(2\sqrt{3})$ and $\tau^{61}= -1/6$. Both $\Gamma^{(4)}$ and $\Gamma^{(6)}$ change sign.

We present the discussed left handed Weyl (spinor) representation of 
the group $\SO(1,13)$ with 64 states
in the next subsection in Table I.

\subsection{Left handed weak charged  and right handed weak chargeless quarks and leptons and corresponding
antiquarks and antileptons}
\label{left}

We built a left handed ($\Gamma^{1,13}=-1$) spinor (Weyl) representation  of $2^{14/2-1}= 64$ states 
on the state of Eq.(\ref{start}),
which is, as we discussed in the previous subsection \ref{technique}, a triplet state with respect to 
the group $\SU(3)$ and it is
a right handed ($\Gamma^{(1,3)} =1$) $ \SU(2)$ singlet by simply applying the operators $S^{ab}$ as 
discussed in the previous subsection, which means that we pairwisely change  $(+)(+)$ (or $(+i)(+)$)
into $[-][-]$ (or into $[-i][-]$). The $64$-plet is presented in Table I.  The reader 
can find in Table I also the Standard model names of the representation as well as the two 
hypercharges $Y$ and $Y'$ of Eq.(\ref{yyprime}). 

We first generate the octet of $\SO(1,7)$ by
applying on a starting state the generators $S^{ab},$ with $a,b = 0,1,2,3,5,6,7,8.$ This octet 
(with $\Gamma^{(1,7)}=1$) will have
the $\SU(3)$ charge equal to ($\tau^{53}=1/2$, $\tau^{58}=1/(2\sqrt{3}))$  and the $\unit(1)$ charge equal to 
$\tau^{61}= 1/6$, while  $\Gamma^{(6)}= -1$. It contains  {\em two right handed ($\Gamma^{(1,3)}=1$)
$\SU(2)$ singlets} ($ \Gamma^{(4)}=1$), each with spin up and spin down, and {\em one left handed 
($\Gamma^{(1,3)}=-1$) $\SU(2)$ doublet ($\Gamma^{(4)}=-1$)}, 
each state of the doublet appears with spin up and spin down. This part of the $64$ multiplet is presented
in Octet I of Table I. 

By applying the generators $\tau^{5i}$ (or $ S^{hk}$, $h = 9,10$ and $k = 11,12,13,14$) the handedness
$\Gamma^{(6)}= -1$ can  of course not change, but   
the $\SU(3)$ charge within the triplet changes. The same octet for each of the triplet $\SU(3)$ charges repeats
(Octets II and III on Table I). The triplet representation of $3\times 8$ states represents {\em right handed
weak chargeless quarks and left handed weak charged quarks} of one family.

By applying the generators $S^{hk}$, with $h,k = 11,12,13,14$, the $\SU(3)$ triplet changes into a $\SU(3)$
singlet with $\tau^{53}=0=\tau^{58}$. $\tau^{61}$ changes to $-1/2$, while $\Gamma^{(6)} =-1$
can not change. The octet, the same as in the above triplet case, describes {\em weak chargeless right handed
leptons and weak charged left handed leptons} (Octet IV of Table I).

If we now apply on the starting state (Eq,(\ref{start})) generators $S^{lh},$ with $l=0,1,2,3,5,6,7,8$ and
$h=11,12,13,14$, the triplet transforms into antitriplet. Let us take, for example, $S^{0 \;11}$. This
generator, applied on the starting state, generates an antitriplet with $\tau^{53}=-1/2, 
\tau^{58}= -1/(2\sqrt{3})$, $\tau^{61}=-1/6$.  $\Gamma^{(6)}$ changes sign since the octet
changes properties and goes to an antioctet.
One finds (Antioctet V of Table I) {\em two left handed ($\Gamma^{(1,3)}=-1$)
$\SU(2)$ singlets} ($ \Gamma^{(4)}=1$), each with spin up and spin down, and {\em one right handed 
($\Gamma^{(1,3)}=1$) $\SU(2)$ doublet ($\Gamma^{(4)}=-1$)}, 
each state of the doublet appears with spin up and spin down. As in the case of the $\SU(3)$ triplets
the application of $\tau^{5i}$ (or $S^{hk},$ with $h=11,12$ and $k = 9,10,13,14$, so that
only one out of three factors determining the colour charge is of the ``-`` type, that is 
either $(-)$ or $[-]$)  generates 
the other two antitriplets (Antioctets 
VI and VII of Table I). 

When $S^{hk}; h,k = 9,10,11,12 $ changes two``+`` into ``-``, the antisinglet states
are generated, with the same octet content as the antitriplet states have and the same $\Gamma^{(6)}$, but with
$\tau^{53}=0=\tau^{58}$, while $\tau^{61} = 1/2$. One finds  {\em weak charged right handed
leptons and weak chargeless left handed leptons} (Antioctet VIII of Table I).

\begin{sidewaystable}
\centering
\begin{tabular}{|r|c||c||c|c||c|c|c||c|c|c||r|r|}
\hline
i&$$&$|^a\psi_i>$&$\Gamma^{(1,3)}$&$ S^{12}$&$\Gamma^{(4)}$&
$\tau^{33}$&$\tau^{43}$&$\tau^{53}$&$\tau^{58}$&$\tau^{61}$&$Y$&$Y'$\\
\hline\hline
&& ${\rm Octet\; I},\;\Gamma^{(1,7)} =1,\;\Gamma^{(6)} = -1,$&&&&&&&&&& \\
&& ${\rm of \; quarks}$&&&&&&&&&&\\
\hline\hline
1&$u_{R}^{c1}$&$\stackrel{03}{(+i)}\stackrel{12}{(+)}|\stackrel{56}{(+)}\stackrel{78}{(+)}
||\stackrel{9 \;10}{(+)}\stackrel{11\;12}{(-)}\stackrel{13\;14}{(-)}$
&1&1/2&1&0&1/2&1/2&$1/(2\sqrt{3})$&1/6&2/3&-1/3\\
\hline 
2&$u_{R}^{c1}$&$\stackrel{03}{[-i]}\stackrel{12}{[-]}|\stackrel{56}{(+)}\stackrel{78}{(+)}
||\stackrel{9 \;10}{(+)}\stackrel{11\;12}{(-)}\stackrel{13\;14}{(-)}$
&1&-1/2&1&0&1/2&1/2&$1/(2\sqrt{3})$&1/6&2/3&-1/3\\
\hline
3&$d_{R}^{c1}$&$\stackrel{03}{(+i)}\stackrel{12}{(+)}|\stackrel{56}{[-]}\stackrel{78}{[-]}
||\stackrel{9 \;10}{(+)}\stackrel{11\;12}{(-)}\stackrel{13\;14}{(-)}$
&1&1/2&1&0&-1/2&1/2&$1/(2\sqrt{3})$&1/6&-1/3&2/3\\
\hline 
4&$d_{R}^{c1}$&$\stackrel{03}{[-i]}\stackrel{12}{[-]}|\stackrel{56}{[-]}\stackrel{78}{[-]}
||\stackrel{9 \;10}{(+)}\stackrel{11\;12}{(-)}\stackrel{13\;14}{(-)}$
&1&-1/2&1&0&-1/2&1/2&$1/(2\sqrt{3})$&1/6&-1/3&2/3\\
\hline
5&$d_{L}^{c1}$&$\stackrel{03}{[-i]}\stackrel{12}{(+)}|\stackrel{56}{[-]}\stackrel{78}{(+)}
||\stackrel{9 \;10}{(+)}\stackrel{11\;12}{(-)}\stackrel{13\;14}{(-)}$
&-1&1/2&-1&-1/2&0&1/2&$1/(2\sqrt{3})$&1/6&1/6&1/6\\
\hline
6&$d_{L}^{c1}$&$\stackrel{03}{(+i)}\stackrel{12}{[-]}|\stackrel{56}{[-]}\stackrel{78}{(+)}
||\stackrel{9 \;10}{(+)}\stackrel{11\;12}{(-)}\stackrel{13\;14}{(-)}$
&-1&-1/2&-1&-1/2&0&1/2&$1/(2\sqrt{3})$&1/6&1/6&1/6\\
\hline
7&$u_{L}^{c1}$&$\stackrel{03}{[-i]}\stackrel{12}{(+)}|\stackrel{56}{(+)}\stackrel{78}{[-]}
||\stackrel{9 \;10}{(+)}\stackrel{11\;12}{(-)}\stackrel{13\;14}{(-)}$
&-1&1/2&-1&1/2&0&1/2&$1/(2\sqrt{3})$&1/6&1/6&1/6\\
\hline
8&$u_{L}^{c1}$&$\stackrel{03}{(+i)}\stackrel{12}{[-]}|\stackrel{56}{(+)}\stackrel{78}{[-]}
||\stackrel{9 \;10}{(+)}\stackrel{11\;12}{(-)}\stackrel{13\;14}{(-)}$
&-1&-1/2&-1&1/2&0&1/2&$1/(2\sqrt{3})$&1/6&1/6&1/6\\
\hline\hline 
\end{tabular}
\\[1mm]
\begin{tabular}{|r|c||c||c|c||c|c|c||c|c|c||r|r|}
\hline
i&$$&$|^a\psi_i>$&$\Gamma^{(1,3)}$&$ S^{12}$&$\Gamma^{(4)}$&
$\tau^{33}$&$\tau^{43}$&$\tau^{53}$&$\tau^{58}$&$\tau^{61}$&$Y$&$Y'$\\
\hline\hline
&& ${\rm Octet\; II},\;\Gamma^{(1,7)} =1,\;\Gamma^{(6)} = -1,$&&&&&&&&&& \\
&& ${\rm of \; quarks}$&&&&&&&&&&\\
\hline\hline
9&$u_{R}^{c2}$&$\stackrel{03}{(+i)}\stackrel{12}{(+)}|\stackrel{56}{(+)}\stackrel{78}{(+)}
||\stackrel{9 \;10}{[-]}\stackrel{11\;12}{[+]}\stackrel{13\;14}{(-)}$
&1&1/2&1&0&1/2&-1/2&$1/(2\sqrt{3})$&1/6&2/3&-1/3\\
\hline 
10&$u_{R}^{c2}$&$\stackrel{03}{[-i]}\stackrel{12}{[-]}|\stackrel{56}{(+)}\stackrel{78}{(+)}
||\stackrel{9 \;10}{[-]}\stackrel{11\;12}{[+]}\stackrel{13\;14}{(-)}$
&1&-1/2&1&0&1/2&-1/2&$1/(2\sqrt{3})$&1/6&2/3&-1/3\\
\hline
11&$d_{R}^{c2}$&$\stackrel{03}{(+i)}\stackrel{12}{(+)}|\stackrel{56}{[-]}\stackrel{78}{[-]}
||\stackrel{9 \;10}{[-]}\stackrel{11\;12}{[+]}\stackrel{13\;14}{(-)}$
&1&1/2&1&0&-1/2&-1/2&$1/(2\sqrt{3})$&1/6&-1/3&2/3\\
\hline 
12&$d_{R}^{c2}$&$\stackrel{03}{[-i]}\stackrel{12}{[-]}|\stackrel{56}{[-]}\stackrel{78}{[-]}
||\stackrel{9 \;10}{[-]}\stackrel{11\;12}{[+]}\stackrel{13\;14}{(-)}$
&1&-1/2&1&0&-1/2&-1/2&$1/(2\sqrt{3})$&1/6&-1/3&2/3\\
\hline
13&$d_{L}^{c2}$&$\stackrel{03}{[-i]}\stackrel{12}{(+)}|\stackrel{56}{[-]}\stackrel{78}{(+)}
||\stackrel{9 \;10}{[-]}\stackrel{11\;12}{[+]}\stackrel{13\;14}{(-)}$
&-1&1/2&-1&-1/2&0&-1/2&$1/(2\sqrt{3})$&1/6&1/6&1/6\\
\hline
14&$d_{L}^{c2}$&$\stackrel{03}{(+i)}\stackrel{12}{[-]}|\stackrel{56}{[-]}\stackrel{78}{(+)}
||\stackrel{9 \;10}{[-]}\stackrel{11\;12}{[+]}\stackrel{13\;14}{(-)}$
&-1&-1/2&-1&-1/2&0&-1/2&$1/(2\sqrt{3})$&1/6&1/6&1/6\\
\hline
15&$u_{L}^{c2}$&$\stackrel{03}{[-i]}\stackrel{12}{(+)}|\stackrel{56}{(+)}\stackrel{78}{[-]}
||\stackrel{9 \;10}{[-]}\stackrel{11\;12}{[+]}\stackrel{13\;14}{(-)}$
&-1&1/2&-1&1/2&0&-1/2&$1/(2\sqrt{3})$&1/6&1/6&1/6\\
\hline
16&$u_{L}^{c2}$&$\stackrel{03}{(+i)}\stackrel{12}{[-]}|\stackrel{56}{(+)}\stackrel{78}{[-]}
||\stackrel{9 \;10}{[-]}\stackrel{11\;12}{[+]}\stackrel{13\;14}{(-)}$
&-1&-1/2&-1&1/2&0&-1/2&$1/(2\sqrt{3})$&1/6&1/6&1/6\\
\hline\hline
\end{tabular}
\caption{Parts I and II of Table I}
\end{sidewaystable}
\begin{sidewaystable}
\centering
\begin{tabular}{|r|c||c||c|c||c|c|c||c|c|c||r|r|}
\hline
i&$$&$|^a\psi_i>$&$\Gamma^{(1,3)}$&$ S^{12}$&$\Gamma^{(4)}$&
$\tau^{33}$&$\tau^{43}$&$\tau^{53}$&$\tau^{58}$&$\tau^{61}$&$Y$&$Y'$\\
\hline\hline
&& ${\rm Octet\; III},\;\Gamma^{(1,7)} =1,\;\Gamma^{(6)} = -1,$&&&&&&&&&& \\
&& ${\rm of \; quarks}$&&&&&&&&&&\\
\hline\hline
17&$u_{R}^{c3}$&$\stackrel{03}{(+i)}\stackrel{12}{(+)}|\stackrel{56}{(+)}\stackrel{78}{(+)}
||\stackrel{9 \;10}{[-]}\stackrel{11\;12}{(-)}\stackrel{13\;14}{[+]}$
&1&1/2&1&0&1/2&0&$-1/\sqrt{3}$&1/6&2/3&-1/3\\
\hline 
18&$u_{R}^{c3}$&$\stackrel{03}{[-i]}\stackrel{12}{[-]}|\stackrel{56}{(+)}\stackrel{78}{(+)}
||\stackrel{9 \;10}{[-]}\stackrel{11\;12}{(-)}\stackrel{13\;14}{[+]}$
&1&-1/2&1&0&1/2&0&$-1/\sqrt{3}$&1/6&2/3&-1/3\\
\hline
19&$d_{R}^{c3}$&$\stackrel{03}{(+i)}\stackrel{12}{(+)}|\stackrel{56}{[-]}\stackrel{78}{[-]}
||\stackrel{9 \;10}{[-]}\stackrel{11\;12}{(-)}\stackrel{13\;14}{[+]}$
&1&1/2&1&0&-1/2&0&$-1/\sqrt{3}$&1/6&-1/3&2/3\\
\hline 
20&$d_{R}^{c3}$&$\stackrel{03}{[-i]}\stackrel{12}{[-]}|\stackrel{56}{[-]}\stackrel{78}{[-]}
||\stackrel{9 \;10}{[-]}\stackrel{11\;12}{(-)}\stackrel{13\;14}{[+]}$
&1&-1/2&1&0&-1/2&0&$-1/\sqrt{3}$&1/6&-1/3&2/3\\
\hline
21&$d_{L}^{c3}$&$\stackrel{03}{[-i]}\stackrel{12}{(+)}|\stackrel{56}{[-]}\stackrel{78}{(+)}
||\stackrel{9 \;10}{[-]}\stackrel{11\;12}{(-)}\stackrel{13\;14}{[+]}$
&-1&1/2&-1&-1/2&0&0&$-1/\sqrt{3}$&1/6&1/6&1/6\\
\hline
22&$d_{L}^{c3}$&$\stackrel{03}{(+i)}\stackrel{12}{[-]}|\stackrel{56}{[-]}\stackrel{78}{(+)}
||\stackrel{9 \;10}{[-]}\stackrel{11\;12}{(-)}\stackrel{13\;14}{[+]}$
&-1&-1/2&-1&-1/2&0&0&$-1/\sqrt{3}$&1/6&1/6&1/6\\
\hline
23&$u_{L}^{c3}$&$\stackrel{03}{[-i]}\stackrel{12}{(+)}|\stackrel{56}{(+)}\stackrel{78}{[-]}
||\stackrel{9 \;10}{[-]}\stackrel{11\;12}{(-)}\stackrel{13\;14}{[+]}$
&-1&1/2&-1&1/2&0&0&$-1/\sqrt{3}$&1/6&1/6&1/6\\
\hline
24&$u_{L}^{c3}$&$\stackrel{03}{(+i)}\stackrel{12}{[-]}|\stackrel{56}{(+)}\stackrel{78}{[-]}
||\stackrel{9 \;10}{[-]}\stackrel{11\;12}{(-)}\stackrel{13\;14}{[+]}$
&-1&-1/2&-1&1/2&0&0&$-1/\sqrt{3}$&1/6&1/6&1/6\\
\hline\hline
\end{tabular}
\\[1mm]
\begin{tabular}{|r|c||c||c|c||c|c|c||c|c|c||r|r|}
\hline
i&$$&$|^a\psi_i>$&$\Gamma^{(1,3)}$&$ S^{12}$&$\Gamma^{(4)}$&
$\tau^{33}$&$\tau^{43}$&$\tau^{53}$&$\tau^{58}$&$\tau^{61}$&$Y$&$Y'$\\
\hline\hline
&& ${\rm Octet\; IV},\;\Gamma^{(1,7)} =1,\;\Gamma^{(6)} = -1,$&&&&&&&&&& \\
&& ${\rm of \; leptons}$&&&&&&&&&&\\
\hline\hline
25&$\nu_{R}$&$\stackrel{03}{(+i)}\stackrel{12}{(+)}|\stackrel{56}{(+)}\stackrel{78}{(+)}
||\stackrel{9 \;10}{(+)}\stackrel{11\;12}{[+]}\stackrel{13\;14}{[+]}$
&1&1/2&1&0&1/2&0&$0$&-1/2&0&-1\\
\hline 
26&$\nu_{R}$&$\stackrel{03}{[-i]}\stackrel{12}{[-]}|\stackrel{56}{(+)}\stackrel{78}{(+)}
||\stackrel{9 \;10}{(+)}\stackrel{11\;12}{[+]}\stackrel{13\;14}{[+]}$
&1&-1/2&1&0&1/2&0&$0$&-1/2&0&-1\\
\hline
27&$e_{R}$&$\stackrel{03}{(+i)}\stackrel{12}{(+)}|\stackrel{56}{[-]}\stackrel{78}{[-]}
||\stackrel{9 \;10}{(+)}\stackrel{11\;12}{[+]}\stackrel{13\;14}{[+]}$
&1&1/2&1&0&-1/2&0&$0$&-1/2&-1&0\\
\hline 
28&$e_{R}$&$\stackrel{03}{[-i]}\stackrel{12}{[-]}|\stackrel{56}{[-]}\stackrel{78}{[-]}
||\stackrel{9 \;10}{(+)}\stackrel{11\;12}{[+]}\stackrel{13\;14}{[+]}$
&1&-1/2&1&0&-1/2&0&$0$&-1/2&-1&0\\
\hline
29&$e_{L}$&$\stackrel{03}{[-i]}\stackrel{12}{(+)}|\stackrel{56}{[-]}\stackrel{78}{(+)}
||\stackrel{9 \;10}{(+)}\stackrel{11\;12}{[+]}\stackrel{13\;14}{[+]}$
&-1&1/2&-1&-1/2&0&0&$0$&-1/2&-1/2&-1/2\\
\hline
30&$e_{L}$&$\stackrel{03}{(+i)}\stackrel{12}{[-]}|\stackrel{56}{[-]}\stackrel{78}{(+)}
||\stackrel{9 \;10}{(+)}\stackrel{11\;12}{[+]}\stackrel{13\;14}{[+]}$
&-1&-1/2&-1&-1/2&0&0&$0$&-1/2&-1/2&-1/2\\
\hline
31&$\nu_{L}$&$\stackrel{03}{[-i]}\stackrel{12}{(+)}|\stackrel{56}{(+)}\stackrel{78}{[-]}
||\stackrel{9 \;10}{(+)}\stackrel{11\;12}{[+]}\stackrel{13\;14}{[+]}$
&-1&1/2&-1&1/2&0&0&$0$&-1/2&-1/2&-1/2\\
\hline
32&$\nu_{L}$&$\stackrel{03}{(+i)}\stackrel{12}{[-]}|\stackrel{56}{(+)}\stackrel{78}{[-]}
||\stackrel{9 \;10}{(+)}\stackrel{11\;12}{[+]}\stackrel{13\;14}{[+]}$
&-1&-1/2&-1&1/2&0&0&$0$&-1/2&-1/2&-1/2\\
\hline\hline
\end{tabular}
\caption{Parts III and IV of Table I}
\end{sidewaystable}
\begin{sidewaystable}
\centering
\begin{tabular}{|r|c||c||c|c||c|c|c||c|c|c||r|r|}
\hline
i&$$&$|^a\psi_i>$&$\Gamma^{(1,3)}$&$ S^{12}$&$\Gamma^{(4)}$&
$\tau^{33}$&$\tau^{43}$&$\tau^{53}$&$\tau^{58}$&$\tau^{61}$&$Y$&$Y'$\\
\hline\hline
&& ${\rm Antioctet\; V},\;\Gamma^{(1,7)} =-1,\;\Gamma^{(6)} = 1,$&&&&&&&&&& \\
&& ${\rm of \; antiquarks}$&&&&&&&&&&\\
\hline\hline
33&$\bar{d}_{L}^{\bar{c1}}$&$\stackrel{03}{[-i]}\stackrel{12}{(+)}|\stackrel{56}{(+)}\stackrel{78}{(+)}
||\stackrel{9 \;10}{[-]}\stackrel{11\;12}{[+]}\stackrel{13\;14}{[+]}$
&-1&1/2&1&0&1/2&-1/2&$-1/(2\sqrt{3})$&-1/6&1/3&-2/3\\
\hline 
34&$\bar{d}_{L}^{\bar{c1}}$&$\stackrel{03}{(+i)}\stackrel{12}{[-]}|\stackrel{56}{(+)}\stackrel{78}{(+)}
||\stackrel{9 \;10}{[-]}\stackrel{11\;12}{[+]}\stackrel{13\;14}{[+]}$
&-1&-1/2&1&0&1/2&-1/2&$-1/(2\sqrt{3})$&-1/6&1/3&-2/3\\
\hline
35&$\bar{u}_{L}^{\bar{c1}}$&$\stackrel{03}{[-i]}\stackrel{12}{(+)}|\stackrel{56}{[-]}\stackrel{78}{[-]}
||\stackrel{9 \;10}{[-]}\stackrel{11\;12}{[+]}\stackrel{13\;14}{[+]}$
&-1&1/2&1&0&-1/2&-1/2&$-1/(2\sqrt{3})$&-1/6&-2/3&1/3\\
\hline 
36&$\bar{u}_{L}^{\bar{c1}}$&$\stackrel{03}{(+i)}\stackrel{12}{[-]}|\stackrel{56}{[-]}\stackrel{78}{[-]}
||\stackrel{9 \;10}{[-]}\stackrel{11\;12}{[+]}\stackrel{13\;14}{[+]}$
&-1&-1/2&1&0&-1/2&-1/2&$-1/(2\sqrt{3})$&-1/6&-2/3&1/3\\
\hline
37&$\bar{d}_{R}^{\bar{c1}}$&$\stackrel{03}{(+i)}\stackrel{12}{(+)}|\stackrel{56}{(+)}\stackrel{78}{[-]}
||\stackrel{9 \;10}{[-]}\stackrel{11\;12}{[+]}\stackrel{13\;14}{[+]}$
&1&1/2&-1&1/2&0&-1/2&$-1/(2\sqrt{3})$&-1/6&-1/6&-1/6\\
\hline
38&$\bar{d}_{R}^{\bar{c1}}$&$\stackrel{03}{[-i])}\stackrel{12}{[-]}|\stackrel{56}{(+)}\stackrel{78}{[-]}
||\stackrel{9 \;10}{[-]}\stackrel{11\;12}{[+]}\stackrel{13\;14}{[+]}$
&1&-1/2&-1&1/2&0&-1/2&$-1/(2\sqrt{3})$&-1/6&-1/6&-1/6\\
\hline
39&$\bar{u}_{R}^{\bar{c1}}$&$\stackrel{03}{(+i)}\stackrel{12}{(+)}|\stackrel{56}{[-]}\stackrel{78}{(+)}
||\stackrel{9 \;10}{[-]}\stackrel{11\;12}{[+]}\stackrel{13\;14}{[+]}$
&1&1/2&-1&-1/2&0&-1/2&$-1/(2\sqrt{3})$&-1/6&-1/6&-1/6\\
\hline
40&$\bar{u}_{R}^{\bar{c1}}$&$\stackrel{03}{[-i]}\stackrel{12}{[-]}|\stackrel{56}{[-]}\stackrel{78}{(+)}
||\stackrel{9 \;10}{[-]}\stackrel{11\;12}{[+]}\stackrel{13\;14}{[+]}$
&1&-1/2&-1&-1/2&0&-1/2&$-1/(2\sqrt{3})$&-1/6&-1/6&-1/6\\
\hline\hline 
\end{tabular}
\\[1mm]
\begin{tabular}{|r|c||c||c|c||c|c|c||c|c|c||r|r|}
\hline
i&$$&$|^a\psi_i>$&$\Gamma^{(1,3)}$&$ S^{12}$&$\Gamma^{(4)}$&
$\tau^{33}$&$\tau^{43}$&$\tau^{53}$&$\tau^{58}$&$\tau^{61}$&$Y$&$Y'$\\
\hline\hline
&& ${\rm Antioctet\; VI},\;\Gamma^{(1,7)} =-1,\;\Gamma^{(6)} = 1,$&&&&&&&&&& \\
&& ${\rm of \; antiquarks}$&&&&&&&&&&\\
\hline\hline
41&$\bar{d}_{L}^{\bar{c2}}$&$\stackrel{03}{[-i]}\stackrel{12}{(+)}|\stackrel{56}{(+)}\stackrel{78}{(+)}
||\stackrel{9 \;10}{(+)}\stackrel{11\;12}{(-)}\stackrel{13\;14}{[+]}$
&-1&1/2&1&0&1/2&1/2&$-1/(2\sqrt{3})$&-1/6&1/3&-2/3\\
\hline 
42&$\bar{d}_{L}^{\bar{c2}}$&$\stackrel{03}{(+i)}\stackrel{12}{[-]}|\stackrel{56}{(+)}\stackrel{78}{(+)}
||\stackrel{9 \;10}{(+)}\stackrel{11\;12}{(-)}\stackrel{13\;14}{[+]}$
&-1&-1/2&1&0&1/2&1/2&$-1/(2\sqrt{3})$&-1/6&1/3&-2/3\\
\hline
43&$\bar{u}_{L}^{\bar{c2}}$&$\stackrel{03}{[-i]}\stackrel{12}{(+)}|\stackrel{56}{[-]}\stackrel{78}{[-]}
||\stackrel{9 \;10}{(+)}\stackrel{11\;12}{(-)}\stackrel{13\;14}{[+]}$
&-1&1/2&1&0&-1/2&1/2&$-1/(2\sqrt{3})$&-1/6&-2/3&1/3\\
\hline 
44&$\bar{u}_{L}^{\bar{c2}}$&$\stackrel{03}{(+i)}\stackrel{12}{[-]}|\stackrel{56}{[-]}\stackrel{78}{[-]}
||\stackrel{9 \;10}{(+)}\stackrel{11\;12}{(-)}\stackrel{13\;14}{[+]}$
&-1&-1/2&1&0&-1/2&1/2&$-1/(2\sqrt{3})$&-1/6&-2/3&1/3\\
\hline
45&$\bar{d}_{R}^{\bar{c2}}$&$\stackrel{03}{(+i)}\stackrel{12}{(+)}|\stackrel{56}{(+)}\stackrel{78}{[-]}
||\stackrel{9 \;10}{(+)}\stackrel{11\;12}{(-)}\stackrel{13\;14}{[+]}$
&1&1/2&-1&1/2&0&1/2&$-1/(2\sqrt{3})$&-1/6&-1/6&-1/6\\
\hline
46&$\bar{d}_{R}^{\bar{c}}$&$\stackrel{03}{[-i])}\stackrel{12}{[-]}|\stackrel{56}{(+)}\stackrel{78}{[-]}
||\stackrel{9 \;10}{(+)}\stackrel{11\;12}{(-)}\stackrel{13\;14}{[+]}$
&1&-1/2&-1&1/2&0&1/2&$-1/(2\sqrt{3})$&-1/6&-1/6&-1/6\\
\hline
47&$\bar{u}_{R}^{\bar{c2}}$&$\stackrel{03}{(+i)}\stackrel{12}{(+)}|\stackrel{56}{[-]}\stackrel{78}{(+)}
||\stackrel{9 \;10}{(+)}\stackrel{11\;12}{(-)}\stackrel{13\;14}{[+]}$
&1&1/2&-1&-1/2&0&1/2&$-1/(2\sqrt{3})$&-1/6&-1/6&-1/6\\
\hline
48&$\bar{u}_{R}^{\bar{c2}}$&$\stackrel{03}{[-i]}\stackrel{12}{[-]}|\stackrel{56}{[-]}\stackrel{78}{(+)}
||\stackrel{9 \;10}{(+)}\stackrel{11\;12}{(-)}\stackrel{13\;14}{[+]}$
&1&-1/2&-1&-1/2&0&1/2&$-1/(2\sqrt{3})$&-1/6&-1/6&-1/6\\
\hline\hline 
\end{tabular}
\caption{Parts V and VI of Table I}
\end{sidewaystable}
\begin{sidewaystable}
\centering
\begin{tabular}{|r|c||c||c|c||c|c|c||c|c|c||r|r|}
\hline
i&$$&$|^a\psi_i>$&$\Gamma^{(1,3)}$&$ S^{12}$&$\Gamma^{(4)}$&
$\tau^{33}$&$\tau^{43}$&$\tau^{53}$&$\tau^{58}$&$\tau^{61}$&$Y$&$Y'$\\
\hline\hline
&& ${\rm Antioctet\; VII},\;\Gamma^{(1,7)} =-1,\;\Gamma^{(6)} = 1,$&&&&&&&&&& \\
&& ${\rm of \; antiquarks}$&&&&&&&&&&\\
\hline\hline
49&$\bar{d}_{L}^{\bar{c3}}$&$\stackrel{03}{[-i]}\stackrel{12}{(+)}|\stackrel{56}{(+)}\stackrel{78}{(+)}
||\stackrel{9 \;10}{(+)}\stackrel{11\;12}{[+]}\stackrel{13\;14}{(-)}$
&-1&1/2&1&0&1/2&0&$1/\sqrt{3}$&-1/6&1/3&-2/3\\
\hline 
50&$\bar{d}_{L}^{\bar{c3}}$&$\stackrel{03}{(+i)}\stackrel{12}{[-]}|\stackrel{56}{(+)}\stackrel{78}{(+)}
||\stackrel{9 \;10}{(+)}\stackrel{11\;12}{[+]}\stackrel{13\;14}{(-)}$
&-1&-1/2&1&0&1/2&0&$1/\sqrt{3}$&-1/6&1/3&-2/3\\
\hline
51&$\bar{u}_{L}^{\bar{c}}$&$\stackrel{03}{[-i]}\stackrel{12}{(+)}|\stackrel{56}{[-]}\stackrel{78}{[-]}
||\stackrel{9 \;10}{(+)}\stackrel{11\;12}{[+]}\stackrel{13\;14}{(-)}$
&-1&1/2&1&0&-1/2&0&$1/\sqrt{3}$&-1/6&-2/3&1/3\\
\hline 
52&$\bar{u}_{L}^{\bar{c3}}$&$\stackrel{03}{(+i)}\stackrel{12}{[-]}|\stackrel{56}{[-]}\stackrel{78}{[-]}
||\stackrel{9 \;10}{(+)}\stackrel{11\;12}{[+]}\stackrel{13\;14}{(-)}$
&-1&-1/2&1&0&-1/2&0&$1/\sqrt{3}$&-1/6&-2/3&1/3\\
\hline
53&$\bar{d}_{R}^{\bar{c3}}$&$\stackrel{03}{(+i)}\stackrel{12}{(+)}|\stackrel{56}{(+)}\stackrel{78}{[-]}
||\stackrel{9 \;10}{(+)}\stackrel{11\;12}{[+]}\stackrel{13\;14}{(-)}$
&1&1/2&-1&1/2&0&0&$1/\sqrt{3}$&-1/6&-1/6&-1/6\\
\hline
54&$\bar{d}_{R}^{\bar{c3}}$&$\stackrel{03}{[-i])}\stackrel{12}{[-]}|\stackrel{56}{(+)}\stackrel{78}{[-]}
||\stackrel{9 \;10}{(+)}\stackrel{11\;12}{[+]}\stackrel{13\;14}{(-)}$
&1&-1/2&-1&1/2&0&0&$1/\sqrt{3}$&-1/6&-1/6&-1/6\\
\hline
55&$\bar{u}_{R}^{\bar{c3}}$&$\stackrel{03}{(+i)}\stackrel{12}{(+)}|\stackrel{56}{[-]}\stackrel{78}{(+)}
||\stackrel{9 \;10}{(+)}\stackrel{11\;12}{[+]}\stackrel{13\;14}{(-)}$
&1&1/2&-1&-1/2&0&0&$1/\sqrt{3}$&-1/6&-1/6&-1/6\\
\hline
56&$\bar{u}_{R}^{\bar{c3}}$&$\stackrel{03}{[-i]}\stackrel{12}{[-]}|\stackrel{56}{[-]}\stackrel{78}{(+)}
||\stackrel{9 \;10}{(+)}\stackrel{11\;12}{[+]}\stackrel{13\;14}{(-)}$
&1&-1/2&-1&-1/2&0&0&$1/\sqrt{3}$&-1/6&-1/6&-1/6\\
\hline\hline 
\end{tabular}
\\[1mm]
\begin{tabular}{|r|c||c||c|c||c|c|c||c|c|c||r|r|}
\hline
i&$$&$|^a\psi_i>$&$\Gamma^{(1,3)}$&$ S^{12}$&$\Gamma^{(4)}$&
$\tau^{33}$&$\tau^{43}$&$\tau^{53}$&$\tau^{58}$&$\tau^{61}$&$Y$&$Y'$\\
\hline\hline
&& ${\rm Antioctet\; VIII},\;\Gamma^{(1,7)} =-1,\;\Gamma^{(6)} = 1,$&&&&&&&&&& \\
&& ${\rm of \; antileptons}$&&&&&&&&&&\\
\hline\hline
57&$\bar{e}_{L}$&$\stackrel{03}{[-i]}\stackrel{12}{(+)}|\stackrel{56}{(+)}\stackrel{78}{(+)}
||\stackrel{9 \;10}{[-]}\stackrel{11\;12}{(-)}\stackrel{13\;14}{(-)}$
&-1&1/2&1&0&1/2&0&$0$&1/2&1&0\\
\hline 
58&$\bar{e}_{L}$&$\stackrel{03}{(+i)}\stackrel{12}{[-]}|\stackrel{56}{(+)}\stackrel{78}{(+)}
||\stackrel{9 \;10}{[-]}\stackrel{11\;12}{(-)}\stackrel{13\;14}{(-)}$
&-1&-1/2&1&0&1/2&0&$0$&1/2&1&0\\
\hline
59&$\bar{\nu}_{L}$&$\stackrel{03}{[-i]}\stackrel{12}{(+)}|\stackrel{56}{[-]}\stackrel{78}{[-]}
||\stackrel{9 \;10}{[-]}\stackrel{11\;12}{(-)}\stackrel{13\;14}{(-)}$
&-1&1/2&1&0&-1/2&0&$0$&1/2&0&1\\
\hline 
60&$\bar{\nu}_{L}$&$\stackrel{03}{[-i]}\stackrel{12}{[-]}|\stackrel{56}{[-]}\stackrel{78}{[-]}
||\stackrel{9 \;10}{[-]}\stackrel{11\;12}{(-)}\stackrel{13\;14}{(-)}$
&-1&-1/2&1&0&-1/2&0&$0$&1/2&0&1\\
\hline
61&$\bar{\nu}_{R}$&$\stackrel{03}{(+i)}\stackrel{12}{(+)}|\stackrel{56}{[-]}\stackrel{78}{(+)}
||\stackrel{9 \;10}{[-]}\stackrel{11\;12}{(-)}\stackrel{13\;14}{(-)}$
&1&1/2&-1&-1/2&0&0&$0$&1/2&1/2&1/2\\
\hline
62&$\bar{\nu}_{R}$&$\stackrel{03}{[-i])}\stackrel{12}{[-]}|\stackrel{56}{[-]}\stackrel{78}{(+)}
||\stackrel{9 \;10}{[-]}\stackrel{11\;12}{(-)}\stackrel{13\;14}{(-)}$
&1&-1/2&-1&-1/2&0&0&$0$&1/2&1/2&1/2\\
\hline
63&$\bar{e}_{R}$&$\stackrel{03}{(+i)}\stackrel{12}{(+)}|\stackrel{56}{(+)}\stackrel{78}{[-]}
||\stackrel{9 \;10}{[-]}\stackrel{11\;12}{(-)}\stackrel{13\;14}{(-)}$
&1&1/2&-1&1/2&0&0&$0$&1/2&1/2&1/2\\
\hline
64&$\bar{e}_{R}$&$\stackrel{03}{[-i]}\stackrel{12}{[-]}|\stackrel{56}{(+)}\stackrel{78}{[-]}
||\stackrel{9 \;10}{[-]}\stackrel{11\;12}{(-)}\stackrel{13\;14}{(-)}$
&1&-1/2&-1&1/2&0&0&$0$&1/2&1/2&1/2\\
\hline\hline
\end{tabular}
\caption{Parts VII and VIII of Table I}
\end{sidewaystable}
In Table I (spread over the Tables 1--4)  the 64-plet of one Weyl spinor of $\SO(1,13)$ is presented. The multiplet contains  spinors - all the quarks 
(Octets I,II,III) and leptons (Octet IV)- and the corresponding ``antispinors''- antiquarks 
(Antioctets V, VI, VII) and antileptons 
(Antioctet VIII) of the Standard model, that is left handed weak charged quarks and leptons and right handed 
weak chargeless antiquarks and antileptons, as well as left handed weak chargeless antiquarks and antileptons
and weak charged right handed antiquarks and antileptons.  It also contains right handed weak chargeless 
neutrinos and left handed weak char\-geless antineutrinos in addition.

The first two states of each of the triplet (that is of Octet I, Octet II, Octet III), 
the antitriplet (that is of Antioctet V, Antioctet VI,
Antioctet VII), the singlet (that is of Octet IV) and the antisinglet (that is of Antioctet VIII) 
form the ($2^{10/2-1}$ =16)-plet of the group $\SO(1,9)$. (The $\SO(4)$ part is only a spectator,
for the chosen multiplet of $\SO(1,9)$ taken to be $\stackrel{56}{(+)}\stackrel{78}{(+)}$). 
Since the starting state is a $\SU(2)$ singlet 
(a weak chargeless state), all 
the states of the chosen multiplet of the group $\SO(1,9)$ are $\SU(2)$ singlets. Since the starting state 
has a right 
handedness with respect to the subgroup $\SO(1,3)$
($\Gamma^{(1,3)}=1$) all the quarks and the leptons are right handed, while all  antiquarks and 
antileptons are left handed. We see that
the generators $S^{ab},$ $a,b = 0,1,2,3,9,10,11,12,13,14$, transform quarks  into other quarks or to
leptons, without changing handedness $\Gamma^{(1,3)}$ of particles (a member of a triplet transforms into other 
members of the triplet or to a singlet without changing $\Gamma^{(1,3)}$),
while they transform quarks ($\SU(3)$ triplets) and leptons ($\SU(3)$ singlets)  
into antiquarks (antitriplets of $\SU(3)$) and antileptons (antisinglets of $\SU(3)$)
of opposite handedness. 
For the group $\SO(1,7)$, however,
left handed weak charged quarks ($\SU(2)$ doublets) or leptons ($\SU(2)$ doublets) transform 
into right handed weak chargeless quarks or leptons (both $\SU(2)$ singlets).

{\em The difference in  properties of representations of the two subgroups -  $\SO(1,9)$} 
(which is a complex group) {\em and $\SO(1,7)$ }(which is a real group)- {\em explains, why
the weak charge breaks parity, while the colour charge does not}. {\em While the group $\SO(1,9)$ has
complex representations }(and so has complex representations also $\SU(3)$ as a subgroup of the group
$\SO(1,9)$), {\em bringing the particle and antiparticle concept into the theory}
(requiring that particles have opposite handedness than antiparticles), {\em the group $\SO(1,7)$, which has real 
representations} (and has accordingly no particle and antiparticle
concept) {\em  has left and right handed particles in the same multiplet, one which is a $\SU(2)$ doublet and
another which is a $\SU(2)$ singlet}. {\em And here is the reason, why the weak charge breaks parity, while the 
colour charge does not: Colour chargeless and colour charged states have both the same handedness, while weak
charged and weak chargeless states have different handedness}.

We also notice that the two  $\unit(1)$'s charges $ \tau^{43}$ and $\tau^{61}$ combined into $Y$ and $Y'$
connect the $\SO(4)$ (and accordingly $\SU(2)$, that is the weak charge) and the $\SO(6)$ (and accordingly 
$\SU(3)$, that is the colour charge) degrees of freedom.

The Cartan subalgebra eigenstates appear in one irreducible representation always in pairs, it is with a plus 
and with a minus sign. Since the trace of all $S^{ab}$, which do not belong to the Cartan subalgebra,
are zero as well, it follows that the trace of 
any superposition of $S^{ab}$'s (and accordingly also of any of $\sum_{ab}\; c^{Ai}{ }_{ab} S^{ab}$) is equal
to zero. But one can notice that {\em the trace of any superposition of $S^{ab}$ within only the 
particle part}  (or only within the antiparticle part) {\em of one Weyl representation of $\SO(1,13)$ alone is 
already equal to
zero}, due to the complex character of the representation.
This fact guaranties the appropriate anomaly cancellation as we shall discussed in section \ref{conclusions-sc3}.

%\subsection{Discrete symmetries}
%\label{Discrete}

%Charge conjugation, which should be the product of all imaginary $\gamma$'s (\gamma^2,\gamma^5,$
% $\gamma^7, \gamma^9, \gamma^11, \gamma^13$ and antilinear operator $K$.
% Reversal of moovement is strange. It should be a product of $\gamma^1, \gamma^3, K$ and $\tau_x$.
% For the hole space does not
% mean a lot. But for a appropriately chosen Yukawas, it works -approximately.
% Parity is also strange. It should be a product of $\gamma^0, \gamma^7$, since it should change
% right handed singlets into into a left handed doublet. We don't have two Weyls, only one.
% It has to be worked on this.

\subsection{Families}
\label{familiessub}

We demonstrated in subsection \ref{left} that one Weyl irreducible representation of the group $\SO(1,13)$ contains
just one family and one antifamily of the Standard model (with right handed weak chargeless neutrino and left handed 
weak chargeless antineutrino in addition). 

We can, however, notice, that other 64-plets can be reached, which have all the properties of the
Weyl multiplet of Table I, simply by changing the starting state, for example, into
\begin{eqnarray}
\stackrel{03}{[+i]}\stackrel{12}{[+]}|\stackrel{56}{(+)}\stackrel{78}{(+)}
||\stackrel{9 \;10}{(+)}\;\stackrel{11\;12}{(-)}\;\stackrel{13\;14}{(-)}
\nonumber\\
\stackrel{03}{(+i)}\stackrel{12}{(+)}|\stackrel{56}{[+]}\stackrel{78}{[+]}
||\stackrel{9 \;10}{(+)}\;\stackrel{11\;12}{(-)}\;\stackrel{13\;14}{(-)}.
\label{family}
\end{eqnarray}
None of these two states can be reached by multiplying the states of the Weyl spinor of subsection \ref{left}
from the left by either $S^{ab}$ or by $\gamma^a$. We namely learned that $\gamma^a$ when multiplying from the
left the operator $\stackrel{ab}{(+)}$ or $\stackrel{ab}{(+i)}$ transforms it to $\stackrel{ab}{[-]}$
or to $\stackrel{ab}{[-i]}$, never to $\stackrel{ab}{[+]}$ or $\stackrel{ab}{[+i]}$. We can prove (see 
ref. \cite{holgernorma2002sc3}) that
any of the states of Eq.(\ref{family}) are orthogonal to all the states of Table I and to each other. 

States of the two additional 64-plets, which follow from the starting two states of Eq.(\ref{family})
by the application of $S^{ab}$ will accordingly be orthogonal to each other and to the states of Table I.

We are proposing that different 64-plets represent different families of quarks and leptons (and antiquarks 
and antileptons).
There are $2^{14/2}$ families in the group $\SO(1,13)$ of Dirac  (two Weyl) spinors. 
To select three  out of these families in order to identify them with the families of the Standard model
a mechanism is needed.
We shall comment on a possible mechanism in section \ref{mechanism}, demonstrating that the same mechanism
offers the Yukawa couplings as well.

\section{A possible mechanism generating families}
\label{mechanism}

In subsection \ref{familiessub} we demonstrated different Weyl representations, all with the same properties with
respect to the group $\SO(1,13)$ and the subgroups of this group interesting for 
the Standard model content of elementary particles. We expect these Weyl representations to be
candidates for describing the three families of quarks and leptons, provided that there is a model behind, 
supporting such an expectation. In this subsection we briefly point out those results of 
refs.\cite{norma92,norma93,norma95,norma01,holgernorma00,pikanormaproceedings}, which show that there are operators,
which transform one family into another. 

In section \ref{Yukawa} we then demonstrate that there is a Lagrange function for spinors in $d=14$
dimensional space-time,
which in four dimensional subspace manifest  the needed Yukawa couplings of the Standard model.

%\subsection{Operators transforming one family into another}
%\label{operators}

In the approach of one of us\cite{norma92,norma93,norma95,norma01,holgernorma00,pikanormaproceedings} 
unifying spins and charges
we pointed out that there are two kinds of operators in Grassmann space (as well as  in space of 
differential forms), which can be identified with the Clifford algebra objects $\gamma^a$ of Eq.(\ref{clif}).
In subsection  \ref{technique} we introduced only one kind of $\gamma^a$'s and express 
spinor states as polynomials of these $\gamma^a$'s, requiring that all the operators
operate on these polynomials from the left. In subsection \ref{left} we then presented  one Weyl spinor
of the group $\SO(1,13)$, reached by multiplying the starting state from the left by $S^{ab}$.

We saw that the multiplication by $\gamma^a$ from the left  changes the operator $\stackrel{ab}{(+)}$ or
the operator $\stackrel{ab}{(+i)}$
into the operator $\stackrel{ab}{[-]}$ or the operator $\stackrel{ab}{[-i]}$, respectively,
never to $\stackrel{ab}{[+]}$
or to $\stackrel{ab}{[+i]}$, respectively. To obtain the two starting states of Eq.(\ref{family}) the 
operator is needed, which would do the later transformation.

In refs.\cite{norma92,norma93,norma95,norma01,holgernorma00} we presented two kinds of operators, 
both fulfilling the Clifford algebra relation\footnote{
The two possible superpositions of a Grassmann coordinate $\theta^a$ (or equivalently of 
a differential form) and the conjugate momentum $p^{\theta}_a := -i \overrightarrow{\frac{\partial}{\partial
\theta^a }} := -i\overrightarrow{\partial}_a$ (or equivalently of the conjugate momentum to the
differential form\cite{holgernorma00}) define two different Clifford algebra objects, namely 
$\tilde{a}^a := i(p^{\theta a}-i\theta^a), \quad \tilde{\tilde{a}}{ }^a :
= -(p^{\theta a}+i\theta^a)$, with the  properties $\{\tilde{a}^a, \tilde{a}^b\} 
= 2\eta^{ab} = \{\tilde{\tilde{a}}{ }^a, \tilde{\tilde{a}}{ }^b\}, \quad \{\tilde{a}^a,
\tilde{\tilde{a}}{ }^b\} = 0$. One can check that while $\tilde{a}^a $ transforms
$(\tilde{a}^a - \tilde{a}^b)$ into $(\eta^{aa} - \tilde{a}^a \tilde{a}^b)$, transforms $ \tilde{\tilde{a}}{ }^a$
$(\tilde{a}^a - \tilde{a}^b)$ into $(\eta^{aa} + \tilde{a}^a \tilde{a}^b)$ and this is what is needed to 
come from one family to another.}, while anticommuting with each other. These two different superpositions of
Grassmann coordinates and the corresponding momentum can in the presenting technique be simulated by the
left and the right multiplication of the Clifford algebra objects.
To come from one family to another we need to define operators, which would transform  for example 
$(\gamma^a - \gamma^b)$ into $(\eta^{aa} + \gamma^a \gamma^b)$. Let us multiply the operator 
$(\gamma^a - \gamma^b)$ by  $\gamma^a$ from the right instead of from the left. We find
\begin{eqnarray}
(\gamma^a - \gamma^b) \gamma^a = (\eta^{aa} + \gamma^a \gamma^b).
\label{rightgamma}
\end{eqnarray}
This is just what we need. 

The operators $\gamma^a$  operating from the left and the operators $\gamma^a$ operating from the right
are obviously not the same operators, although
both fulfill the Clifford algebra relation(\ref{clif}). They namely generate different states 
(orthogonal to each other).
Both are odd Clifford algebra objects. Knowing which kind of transformations the right multiplication
causes on the objects which are polynomials of $\gamma^a$'s, it is meaningful to rename
the operators $\gamma^a$'s when they operate from the right hand side.

{\em We define the second kind of operators} $\tilde{\gamma}^a$ {\em as operators, 
which formally  operate from the left hand side (as
$\gamma^a$ do)
on  states $\stackrel{ab}{(\pm i)}$, $ \stackrel{ab}{(\pm )}$, $\stackrel{ab}{[\pm i]}$ and $\stackrel{ab}{[\pm]}$
transforming these states as $\gamma^a$ would do if being applied from the right hand side}
\begin{eqnarray}
\tilde{\gamma}^a \stackrel{ab}{(\pm i)} := \stackrel{ab}{(\pm i)} \gamma^a = \eta^{aa} \stackrel{ab}{[\pm i]},\quad 
\tilde{\gamma}^a \stackrel{ab}{(\pm )} := \stackrel{ab}{(\pm )} \gamma^a = \eta^{aa} \stackrel{ab}{[\pm]},
\nonumber\\
\tilde{\gamma}^a \stackrel{ab}{[\pm i]} := \stackrel{ab}{[\pm i]}\gamma^a = \stackrel{ab}{(\pm i)}, \quad
\tilde{\gamma}^a \stackrel{ab}{[\pm ]} := \stackrel{ab}{[\pm ]}\gamma^a = \eta^{aa} \stackrel{ab}{(\pm)}.
\label{rightgammastate}
\end{eqnarray}
Then we can write
\begin{eqnarray}
\{\tilde{\gamma}^a, \tilde \gamma^b \}_+ = 2 \eta^{ab} = \{\gamma^a,\gamma^b\}_+,
\nonumber\\
\{ \tilde{\gamma^a}, \gamma^b \}_+ = 0.
\label{rightgammaleft}
\end{eqnarray}

We can define accordingly besides $S^{ab}$ of Eq.(\ref{sabsc3}) also $\tilde{S}^{ab}$
\begin{eqnarray}
\tilde{S}^{ab} = \frac{i}{4} [\tilde{\gamma}^a, \tilde \gamma^b ],
\label{tildesab}
\end{eqnarray}
which fulfill the Lorentz algebra of Eq.(\ref{loralg}), as accordingly also ${\cal S}^{ab} 
= S^{ab} + \tilde{S}^{ab}$ do, while
\begin{eqnarray}
\{\tilde{S}^{ab}, \gamma^c\}_- = 0 = \{ S^{ab}, \tilde{\gamma}^c \}_-.
\label{tildesabsabcom}
\end{eqnarray}

\section{Lagrange function including Yukawa couplings}
\label{Yukawa}

Refereeing to the work of one of us\cite{norma01sc3,pikanormaproceedings} we write the Lagrange density 
function for a Weyl (massless)
spinor in $d(=1+13)$ - dimensional space as
\begin{eqnarray}
{\cal L} &=& \bar{\psi}\gamma^a p_{0a} \psi = \bar{\psi} \gamma^a f^{\mu}_a p_{0\mu}, 
\nonumber\\
\quad {\rm with}\quad
p_{0\mu} &=& p_{\mu} - \frac{1}{2}S^{ab} \omega_{ab\mu} - \frac{1}{2}\tilde{S}^{ab} \tilde{\omega}_{ab\mu}.
\label{lagrange}
\end{eqnarray}
The Lagrange density of Eq.(\ref{lagrange}) is the Lagrange density for a Weyl spinor in 14-dimensional space
in a free-falling system ($\bar{\psi}\gamma^a p_{0a} \psi$) that is with no gravitational (and accordingly,
due to our approach, with no any other gauge) fields and in an external system 
($\bar{\psi} \gamma^a f^{\mu}_a p_{0\mu}$), where $f^{\mu}_a$ are vielbeins
and $\omega_{ab\mu}$ and $\tilde{\omega}_{ab\mu} $ are spin connections, the gauge fields of $S^{ab}$ and
$\tilde{S}^{ab}$, respectively.

The only internal degree of freedom of spinors in $d=14 $ dimensional space - the spin - might in 
four-dimensional subspace appear as the ordinary spin and all the known charges (as presented in Table I).
The gravitational field presented with spin connections and vielbeins might accordingly in four 
dimensional subspace manifest as all the known gauge fields as
well as the Yukawa couplings, if the break of symmetries occurs in an appropriate way. 
To see that let us first rewrite the Lagrange density of Eq.(\ref{lagrange}) in an appropriate way. 
According to section \ref{spinor} we can rewrite 
Eq.(\ref{lagrange}) as follows
\begin{eqnarray}
{\cal L} &=& \bar{\psi}\gamma^{\alpha} (p_{\alpha}- \sum_{A,i}\; g^{A}\tau^{Ai} A^{Ai}_{alpha} \psi) 
\nonumber\\
&+& i\psi^+ S^{0h} S^{k k'} f^{\sigma}_h \omega_{k k' \sigma} \psi + \quad 
i \psi^+ S^{0h} \tilde{S}^{k k'} f^{\sigma}_h \tilde{\omega}_{kk' \sigma} \psi,
\label{lagrangein4}
\end{eqnarray}
with  $\psi$, which does not depend on coordinates $x^{\sigma}, \sigma
=\{5,6, \cdots ,14 \}$. We assume for simplicity in addition  that there is no gravitational field in 
four-dimensional subspace
($f^{\alpha}_m = \delta^{\alpha}_{m}, m=\{0,1,2,3 \}, \alpha =\{0,1,2,3 \},\; \omega_{mn\alpha} =0$).

The second and the third term look like a mass term, since 
$f^{\sigma}_h \tilde{\omega}_{kk' \sigma}$ behaves in $d(=1+3)-$
dimensional subspace like a scalar field, while the operator $S^{0h}, h=7,8$, for example, 
transforms a right handed
weak chargeless spinor into a left handed weak charged spinor, without changing the spin. 
Since masses of quarks of one family differ from masses of leptons in the same family, it is meaningful to
rewrite the term $\psi^+ S^{0h} S^{k k'} f^{\sigma}_h \omega_{k k' \sigma} \psi $ as $- \gamma^0 \gamma^h
\tau^{Ai} A^{Ai}_{\sigma} f^{\sigma}_h$ to point out that hypercharges ($Y$ and $Y'$) are important for 
the appropriate Yukawa couplings. We also note that the term $i \psi^+ S^{0h} \tilde{S}^{k k'} 
f^{\sigma}_h \tilde{\omega}_{kk' \sigma} \psi$ contributes to the Yukawa couplings, which connect
different families. 

One should of course ask oneself whether or not it is at all possible to choose 
spin connections and vielbeins $f^{\sigma}_h \omega_{k k' \sigma}$ and 
$f^{\sigma}_h \tilde{\omega}_{k k' \sigma}$ in a way to reproduce the masses of the three families of 
quarks and leptons and to predict, what are masses of a possible fourth generation in a way to be in agreement
with the experimental data. 
The work on this topic is under consideration. To demonstrate that the approach used in this paper 
does suggest possible relations among Yukawa couplings
and consequently also the mass matrix, we present in the subsection \ref{sec:mass} a possible choice  
of Yukawa couplings, suggested by Eq.(\ref{lagrangein4}), which leads to
four rather than to three generations of quarks and leptons, with the
values for the masses of the fourth generation, which do not contradict the experimental data\cite{okun-sc3}.

\subsection{Mass matrices for four generations of quarks and leptons}
\label{sec:mass}

Let us assume that the break of symmetries suggests that only terms like 
$$\psi^+ S^{0h} \tilde{S}^{k k'} 
f^{\sigma}_h \tilde{\omega}_{kk' \sigma} \psi $$
appear in the Lagrange density, with $h $ and $\sigma \in 5,6,7,8$ and   $k,k'$ either both equal to 
$0,1,2,3$ or both equal to $5,6,7,8$. Then there are only four families which are measurable at low energies,
namely
\begin{eqnarray}
\stackrel{03}{(+i)} \stackrel{12}{(+)} \stackrel{56}{(+)} \stackrel{78}{(+)} \nonumber\\
\stackrel{03}{[+i]} \stackrel{12}{[+]} \stackrel{56}{(+)} \stackrel{78}{(+)} \nonumber\\
\stackrel{03}{(+i)} \stackrel{12}{(+)} \stackrel{56}{[+]} \stackrel{78}{[+]} \nonumber\\
\stackrel{03}{[+i]} \stackrel{12}{[+]} \stackrel{56}{[+]} \stackrel{78}{[+]}.
\label{threefamilies}
\end{eqnarray}
 If we denote by $A_i$ the matrix element for the transition from  a right handed weak chargeless 
 spinor of type $i = u,d,e,\nu$ to the left handed weak charged spinor (these transitions occur within 
 one family and are caused by the second term $- \gamma^0 \gamma^h
\tau^{Ai} A^{Ai}_{\sigma} f^{\sigma}_h$) of Eq.(\ref{lagrangein4} ), by $B_i$ the matrix element 
causing the transition,
 in which $\stackrel{03}{(+i)} \stackrel{12}{(+)}$ changes to $\stackrel{03}{[+i]} \stackrel{12}{[+]}$
 or opposite (such are transitions between the first and the second family or transitions between the 
 third family and the fourth of Eq.(\ref{threefamilies}) caused by 
 $\tilde{S}^{mm'}f^{\sigma}_h \tilde{\omega}_{mm' \sigma}$ with $m,m'=0,1,2,3$ and $h = 5,6,7,8$
 ), by $C_i$  the matrix element causing the transition 
 in which $\stackrel{56}{(i)} \stackrel{78}{(+)}$ changes to $\stackrel{56}{[+]} \stackrel{78}{[+]}$
 or opposite (such are transitions between the first and the third family or transitions between the 
 second and the fourth family of Eq.(\ref{threefamilies}) caused by $ \tilde{S}^{kk'}f^{\sigma}_h
 \tilde{\omega}_{kk' \sigma}$ with $h,k,k'=5,6,7,8$) and by $D_i$  transitions in which all four 
 factors change, that is the transitions,
 in which $\stackrel{03}{(+i)} \stackrel{12}{(+)}$ changes to $\stackrel{03}{[+i]} \stackrel{12}{[+]}$
 or opposite and  $\stackrel{56}{(+i)} \stackrel{78}{(+)}$ changes to $\stackrel{56}{[+]} \stackrel{78}{[+]}$ or
 opposite
 (such are transitions between the first and the fourth family or transitions between the 
 second  and the third family of Eq.(\ref{threefamilies}))  and if we further assume that 
 the elements are real numbers, we find 
 the following mass matrix
\begin{displaymath}
\left( \begin{array}{cccc}
A_a&B_a&C_a&D_a\\
B_a&A_a&D_a&C_a\\
C_a&D_a&A_a&B_a\\
D_a&C_a&B_a&A_a
\end{array} \right),
\label{mass-sc3}
\end{displaymath}
with $a$, which stays for a family members $u,d,\nu,e$.  
 
 To evaluate the matrix elements $A_a,B_a,C_a,D_a$ one should make a precise model, taking into account that
 matrix elements within one family depend on quantum numbers of the members
 of the family, like $Y,Y'$  and others, and accordingly also on the way how symmetries are broken. 
 We noticed in addition that the elements $D_a$ correspond to two step processes and are expected to be smaller. 
 We mention again that we simplified the problem by assuming  that matrix elements are real, while in general they are 
 complex. All these need further study.
 
We can noticed, that  the mass matrices have the symmetry
\begin{displaymath}
\left( \begin{array}{cc}
X&Y\\
Y&X
\end{array} \right),
\label{xy}
\end{displaymath}
which makes the diagonalization of the mass matrix of Eq.(\ref{mass-sc3})
simple. We find
\begin{eqnarray}
\lambda_{a_1} &=& A_a-B_a-C_a+D_a,\nonumber\\
\lambda_{a_2} &=& A_a-B_a+C_a-D_a,\nonumber\\
\lambda_{a_3} &=& A_a+B_a-C_a-D_a,\nonumber\\
\lambda_{a_4} &=& A_a+B_a+C_a+D_a. 
\label{formalabcd}
\end{eqnarray}
We immediately see that a ''democratic'' matrix with $A_a=B_b=C_c=D_d$  (ref.\cite{fritsch})
leads to $\lambda_{a_1}=\lambda_{a_2}=\lambda_{a_3}= 0, \lambda_{a_4} = 4 A_a$. The diagonal matrix leads to
four equal values $\lambda_{a_i} = A_a. $ We expect that break of symmetries of the group $\SO(1,13)$ down
to $\SO(1,3), \SU(3)$ and $\unit(1)$ will lead to something in between.
If we fit $\lambda_{a_i}$ with the masses of families $m_{ai}$, with $a= u,d,\nu,e$ and $i$ is
the number of family,
we find
\begin{eqnarray}
A_a &=& \{(m_{a4}+ m_{a3}) + (m_{a2} + m_{a1})\}/4,\nonumber\\
B_a &=& \{(m_{a4}+ m_{a3}) - (m_{a2} + m_{a1})\}/4,\nonumber\\
C_a &=& \{(m_{a4}- m_{a3}) + (m_{a2} - m_{a1})\}/4,\nonumber\\
D_a &=& \{(m_{a4}- m_{a3}) - (m_{a2} - m_{a1})\}/4.
\label{formalabcdwithm}
\end{eqnarray}
For the masses of quarks and leptons to agree with the experimental ones 
{\it i.e.}  $m_{u_i}/GeV = 0.0004, 1.4, 
180, 285 (215)$ and $m_{d_i}/GeV = 0.009, 0.2, 6.3, 215 (285)$ for quarks, and
for leptons $m_{e_i}/GeV = 0.0005, 0.105, 1.78, 100$ and for 
$m_{\nu_i}/GeV$ let say $1.10^{-11}, 2.10^{-11}, 6.10{-11}$
and $50$, which would agree also with what Okun and coauthors\cite{okun-sc3} have found as possible values 
for masses of the fourth family, 
we find 
\begin{equation}
\begin{array}{cccc}
A_u = 116.601 & B_u = 115.899 & C_u = 26.599 & D_u = 25.901 \\
(A_u' = 99.101 & B_u' = 98.399 & C_u' = 9.099 & D_u' = 8.401)  \\
A_d = 55.377 & B_d = 55.2728 & C_d = 52.223 & D_d = 52.127 \\
(A_d' = 72.877 & B_d' = 72.773 & C_d' = 69.723 & D_d' = 69.627) \\
A_e = 25.471 & B_e = 25.419 & C_e = 24.581 & D_e = 24.529 \\
A_\nu = 12.5 & B_\nu = 12.5 & C_\nu = 12.5  & D_\nu = 12.5.  
\end{array}
\end{equation}
Values in  the parentheses correspond to the values in the parentheses for the masses of quarks.
The mass matrices are for leptons and even for $d$ quarks very close to a ''democratic''
one\cite{fritsch}. One could also notice that for 
quarks $A_a$ are roughly proportional to the charge $Y$. Further studies are needed to 
comment more on mass matrices, suggested by our approach unifying spins and charges.
Some further discussions can be found in ref.\cite{astridragannorma}. There the generalization 
to more than four family is done. The approach unifying spins and charges namely suggests
$2^{2k}$ families, with $k=1,2,3.$ It turns out that the structure of Eq.(\ref{xy}) repeats
whenever the number of families doubles. This symmetry suggestss for $k=2$ for example $2^{2k} =16$ at most
independent parameters. If we put the first row equal to 
$A_a,B_a,C_a,D_a,E_a,F_a,G_a,H_a,I_a,J_a,K_a,L_a,M_a,N_a,O_a,P_a$, we can then easily write down the mass matrices
(see also ref.\cite{astridragannorma}). 
Also the eigenvalues can for any $2^{2k}$, due to the symmetry of mass matrices, 
easily be found. We find for $k=2$ if we write ${\cal A}_a = 
\{(A_a-B_a)-(C_a-D_a)\},\; {\cal B}_a = \{(E_a-F_a)-(G_a-H_a)\},\; {\cal C}_a = \{(I_a-J_a)-(K_a-L_a)\},\;
{\cal D}_a = \{(M_a-N_a)-(O_a-P_a)\},\;$  for the four (lowest) $\lambda_{ai}$ the values of 
Eq.{\ref{formalabcd}} if we exchange $A$ with ${\cal A}$, $B$ with ${\cal B}$, $C$ with ${\cal C}$ and $D$,
${\cal D}$. The other $12$ values (which can easily be expressed in terms of the
$16$ matrix elements, if taking into account the symmetry of the mass matrices) 
can always be found to lie as high as needed.

\section{Possible break of symmetries}
\label{break}

We shall comment on a possible break of symmetries which leads to physics of the Standard model. 
Taking into account that mass protection mechanism occurs only in even dimensional 
spaces\cite{norma01sc3,holgernorma2002,holgernormawhy}, that
spinors and ``antispinors`` should not transform into each others at low energies, that the colour charge 
should be a conserved quantity, as well as other above discussed phenomena, we find as a promising way of 
breaking symmetries\cite{holgernormaren} the following one\footnote{Two of the authors of this paper have 
shown\cite{holgernormaren} that the way of breaking the group $\SO(1,13)$, presented also in this section, leads 
to unification of all the three  coupling constants at high enough energy and that the proton decay does not contradict
the experimental data.}

\[
\begin{array}{c}
\begin{array}{c}
\underbrace{%
\begin{array}{rrcll}
 & & \SO(1,13) \\
 & & \downarrow \\
 & & \SO(1,7) \otimes \SO(6) \\
 & \swarrow & &  \searrow \\
 & \SO(1,7) & & \SU(4)\\
 \swarrow\qquad & & & \qquad\downarrow \\
 \SO(1,3)\otimes \SO(4) & & & \SU(3)\otimes\unit(1) \\
 \downarrow\qquad & & & \qquad\downarrow \\
\SO(1,3)\otimes\SU(2)\otimes\unit(1) & & & \SU(3)\otimes\unit(1)\\
& & & \\
\end{array}} \\
\SO(1,3)\otimes\SU(2)\otimes\unit(1)\otimes\SU(3)\otimes \unit(1)\\
\end{array}\\
\downarrow \\
\SO(1,3)\otimes\SU(2)\otimes\unit(1)\otimes\SU(3)\\
\end{array}
\]
%*** picture***
%
%\begin{eqnarray}
%\quad \quad \quad \quad \SO(1,13)\nonumber\\
%\quad \quad \quad \SO(1,7)\times \SO(6)\nonumber\\
%\quad \quad \SO(1,3)\times \SO(4)\times \SU(3)\times \unit(1)_{\SO(6)}\nonumber\\
%\quad  \SO(1,3)\times \SU(2)\times \unit(1)_{\SO(4)} \times \SU(3)\times \unit(1)_{\SO(6)}\nonumber\\
%\SO(1,3)\times \SU(2)\times Y \times \SU(3)
%\label{scheme}
%\end{eqnarray}

In the first step of breaking $\SO(1,13)$ the ${\cal S}^{ab}\omega_{ab \mu}$ term should break into
$S^{ab}\omega_{ab \mu} + $ $\tilde{S}^{ab}\tilde{\omega}_{ab \mu}$. Further breaks of symmetries then
should take care that particle-antiparticle transitions should not occur any longer, which means that
transitions caused by $S^{ab}$ or $\tilde{S}^{ab}$, with $a= 0,1,2,3,5,6,7,8$ and $b= 9,10,11,12 $
(or opposite, with $a$ and $b$ exchanged), should at lower energies appear with a negligible
probability. The same should appear also for transitions transforming (coloured) quarks into (colourless)
leptons while  
quarks and leptons demonstrate a $\SO(1,7)$ multiplet, with (as we have seen)
left handed weak charged
and right handed weak chargeless spinors. Further break should occur then close to the weak scale, leading
to all the Standard model spinors.

\section{Discussions and conclusions}
\label{conclusions-sc3}

We have presented in this paper some possible answers to the open questions of the Standard model, 
using the approach of one of us.

We see that a left handed $\SO(1,13)$ Weyl multiplet (we have no mirror symmetry in this approach!) 
contains, if represented in a way to demonstrate the
$\SO(1,3)$, $\SU(2)$ and $\unit(1)$'s substructure of the group $\SO(1,13)$, just all the quarks and the leptons
and all the antiquarks and the antileptons of the Standard model, with  
right handed (charged only with
$Y'$) neutrinos and left handed (again charged only with $Y'$) antineutrinos in addition. 

The group $\SO(1,13)$ (with the rank $7$), which unifies the spin and all the known charges, has  complex 
representations. 
Its two regular subgroups  $\SO(1,7)$ (with the rank $ 4$ and possible regular subgroups  
$(\SU(2)\times \SU(2))^2$) and
$\SO(6)$ (with the rank $3$ and possible regular subgroups $\SU(3)$ and $\unit(1)$ - 
neither $\SO(1,13)$ nor $\SO(6)$ have as regular
subgroups groups $(\SU(2)\times \SU(2))^k$) have real and complex representations,
respectively. Complexity of the representations of $\SO(1,13)$ and accordingly of $\SO(6)$ enables the concept
of spinors and ``antispinors``.

We further see that due to real representations of the group $\SO(1,7)$, the
left handed weak charged quarks and leptons together with right handed weak chargeless quarks and leptons 
appear in the Weyl multiplet of $\SO(1,13)$, causing that the weak charge violets the parity. 

A spinor multiplet of the  subgroup $\SO(1,9)$ would instead contain, 
due to the complex character of its representations, 
the colour charged and chargeless spinors of left handedness and the colour anticharged and antichargeless 
``antispinors`` of right handedness and could accordingly not
break the parity. (All spinors of one multiplet of $\SO(1,9)$ have namely the same handedness 
while ``antispinors`` of the same multiplet have the opposite handedness.) 

It is the real and the complex nature of representations of the two subgroups $\SO(1,7)$ and
$\SO(1,9)$, respectively, which is responsible for the fact that weak charge breaks parity 
while the colour charge can not, if the spinor-antispinor concept should stay. Accordingly, also the
colour charge is conserved.

We see that invariants of the Lorentz group appear to be important.
One well known invariant of groups $\SO(1,d-1)$ is $1/2S^{ab}S_{ab}$ which is for spinors equal to 
$1/4 d(d-1)/2$ and it is smaller,
the smaller is d.
We pay attention to the invariant,
which  is the operator of 
handedness\cite{normasuper94,norma01,bojannorma2001,holgernorma0203}.
For an even $d$ it can be defined for any spin as $\Gamma = \alpha \varepsilon_{a_1 a_2 \dots a_d}
S^{a_1 a_2} S^{a_3 a_4} \cdots S^{a_{d-1} a_d} $, with the constant $\alpha $, which can be chosen in a way
that for any spin $\Gamma = \pm 1$ and $\Gamma^2 =1, \Gamma^+ = \Gamma$. 
According to Table I, the value of $\Gamma^{(1,13)} = -1$ and  stays  (of course) unchanged also if only a
part of the group is concerned. We also see that $\Gamma^{(1,7)}$ is for all the spinors equal to 1 and
for all the ``antispinors`` equal to -1. One also sees that when the subgroup $\SU(2)$ (the weak charge group) is broken,
that is when $Q= (Y + \tau^{33})$ (the electromagnetic charge) only is the well defined quantity besides the 
colour charge ($\tau^{53}, \tau^{58}$), the handedness 
$\Gamma^{(1,3)}$  is broken, leading to the superposition of states $u_L$ and $u_R$, for example, 
since both states have
the same colour properties ($\tau^{53}$, $\tau^{58}$) and the electromegnetic charge $Q.$

We saw  in subsection 
%\ref{left} 
that the trace of all the Cartan subalgebra operators within half a 
Weyl representation of $\SO(1,13)$, namely the particle part alone (as well as the antiparticle part alone),
is equal
to zero. This is true also for all the other generators of $\SO(1,13)$ and is accordingly true for all the linear
superpositions of $S^{ab}$. Due to this fact the anomaly cancellation for spinors is guaranteed.

A break of a symmetry leads to smaller subgroups (of smaller ranks and smaller representations, 
the sum of ranks of regular subgroups being just equal to the rank of the group). 
The smallest spinor representations besides a scalar one \footnote{Scalar representations are trivial, 
leading in the proposed approach to scalars without any charge.}
would occur for  the break of the type $(\SU(2)\times \SU(2))^k$ (since all the eigenvalues of the Cartan subalgebra
would be $\pm 1/2$ and accordingly also the eigenvalues of the superpositions of $S^{ab}$ which lead to
invariant subgroups, would then be $\pm 1/2$ for doublets and $0$  for singlets), if the group would be of such a type. 
For the group $\SO(1,13)$ this is not the case. Accordingly, instead of only $\SU(2)$ (and $\unit(1)$'s from $\SU(2)$) 
also $\SU(3)$ appears (as well as $\unit(1)$ from $\SO(6)$) 
(enabling the concept of spinors
and ``antispinors``, as already mentioned). These are (small) representations of spinors in the agreement
with the concept of the Standard model. 

In the proposed approach unifying spins and charges the concept of families naturally appears, 
reached from the 
starting family by the operators $\tilde{S}^{ab} = -i/4 [\tilde{\gamma}^a \tilde{\gamma}^b - 
\tilde{\gamma}^b \tilde{\gamma}^a]$. We present in this paper the definition as well as the 
meaning of the Clifford odd objects
$\tilde{\gamma}^a$'s, which anticommute with ordinary $\gamma^a$'s. The number of families at low 
energies appear to depend on the interaction.

We propose a Lagrange density for spinors in $d =14$ dimensional space 
\begin{eqnarray}
\psi^+ \gamma^0 \gamma^a f^{\mu}_a\; (p_{\mu} -\frac{1}{2} {\cal S}^{bc}\; \omega_{bc\mu}\;)\; \psi.
\label{lagrange14}
\end{eqnarray}
It contains the 
(gauged) gravitational field with spin connections $\omega_{bc\mu}$ and vielbeins $f^{\mu}_a$ only, 
which then manifest at low energies
as all the known gauge fields  $A^{Ai}_{\alpha}$ 
\begin{eqnarray}
\psi^+ \gamma^0 \gamma^m f^{\alpha}_m\; 
(p_{\alpha} - \sum_{{Ai}}\; \tau^{Ai} A^{Ai}_{\alpha}) \psi
\label{lagrangetau}
\end{eqnarray}
as well as the Yukawa couplings 
\begin{eqnarray}
\psi^+ \gamma^0 \gamma^h S^{h'h''} \omega_{h'h''\sigma} f^{\sigma}_h \psi,
\label{lagrangeyukawas}
\end{eqnarray}
and
\begin{eqnarray}
\psi^+ \gamma^0 \gamma^h \tilde{S}^{h'h''} \tilde{\omega}_{h'h''\sigma} f^{\sigma}_h \psi
\label{lagrangeyukawast}
\end{eqnarray}
with no index $h,h',h''$ from $\{0,1,2,3\}$), with operators of the type $S^{h h'}$, 
which transform right handed $\SU(2)$ singlets into left handed $\SU(2)$ doublets within
one family. For  operators  of the type $\tilde{S}^{h h'}$  
cause transitions betweem states of the same properties with respect to the $S^{ab}$ 
belonging to different families. Both types of terms manifest
accordingly as mass terms quarks and leptons (and antiquarks and antileptons). 
It is the interactions in higher dimensions which 
look like a Higgs causing the Yukawa 
couplings
in a four-dimensional subspace.

The mass matrices suggested by the approach have the dimension $4 \times 4$ at least (or the multiple
of $2^{2k} \times 2^{2k},$ with $k=1,2,3$) and (if assuming that the matrix elements are real)
the symmetry of Eq.(\ref{xy}), which starts at $2 \times 2$ (so that the 
number of different matrix elements is allways equal to  the rank of the matrix). 
Accordingly the eigenvalues can  easily be found and also fitted to the experimental data in a way, that
they do agree with the experimental data. The approach suggests an even number of families, with the
fourth family in agreement with the ref.\cite{okun-sc3}.

We presented\cite{holgernormaren} a possible scheme of breaking symmetry $\SO(1,13)$, which manifests the
above mentioned properties of representations. 
We studyed the arguments for  the three families at low energies and the fourth family, which does not
contradict the experimental data\cite{okun-sc3}. The study is not yet finished\cite{astridragannorma}. 
We have so far learned from this study that the break of symmetry, 
presented in this 
paper, seems to be the appropriate one, reproducing the experimental mass matrices,
as well as the probabilities for the weak transitions.

The mass protection mechanism works in even dimensional spaces 
\cite{norma01sc3,holgernorma2002,holgernorma0203} only\footnote{In even dimensional 
spaces the Dirac equation for massive spinors
has solutions only if left and right handed representations are involved, while in odd dimensional
spaces within one either left or right handed representation the solution exists.}, provided that only one Weyl 
spinor (one irreducible representation) is assumed. Since we assume one Weyl spinor representation of $\SO(1,13)$,
spinors in $d=14$ are massless. And as we have said, it is the interaction in $d=(1+13)$, which 
manifests in $d=(1+3)$ subspace
as a mass term.

One has to ask oneself why a Weyl spinor in $d= (1+13)$? Why the break of symmetry of the group 
$\SO(1,13)$ occurs at all, leading then to small charges and to further break of symmetries in which,
however the baryon number and accordingly the colour charge is conserved?
How accurately can within the  approach of unifying spins and charges the 
parameters of the Standard model and physics behind  be predicted?

There should be several answers to these questions, connected with the properties of the internal space
\footnote{In refs.\cite{holgernormawhy,holgernorma2002} several reasons, why Nature has made a choice of one time and three 
space coordinates are presenting, 
pointing out that there are internal degrees of freedom (spins) which forced the Nature to manifest mainly
$(1+3)$ dimensions.} and also the ordinary part of the space, like topology properties, differential geometry properties,
dynamical properties and also the many body properties.

\section*{Acknowledgements} We would like to express the thanks to Ministry of education, Science and sport for the
grant, Desy for nice hospitality of one of us {HBN}. 
It is a pleasure to thank all participants of the fourth workshop What comes beyond the Standard model at Bled, July
2002 for fruitful discussions, in particular to Holger Bech Nielsen, whose thoughts and comments were extremely
fruitful.

% Astri 
\title*{A Tight Packing Problem}
\author{A. Kleppe}
\institute{%
Bjornvn. 52, 07730 Oslo, Norway}

\titlerunning{A Tight Packing Problem}
\authorrunning{A. Kleppe}
\maketitle

%****************************************************************************
\begin{abstract}
In order to determine the abelian coupling in the context of the Multiple
Point Principle, we seek a tight packing in a $N_{gen}$=3-dimensional space
where the U(1) coupling is absorbed in the action metric. A face-centered
cubic lattice
was originally assumed, a tighter packing is however obtained for an
identification lattice corresponding to a 3-dimensional tesselation, using
rhombic dodecahedra. This suggests a description in terms of a Han-Nambu-like
system of charges corresponding to the $2^3$-1=7 linearly
independent basis vectors of a $Z_2^3$ projective space.
\end{abstract}
%****************************************************************************
\section{Introduction}

In the course of this article, a specific geometrical object is defined,
namely an almost regular three-dimensional lattice corresponding to a
3-dimensional tesselation (space-filling system) of rhombic dodecahedra. The
motivation is the need within the scheme of Anti-Grand Unification and
the Multiple Point Principle, for a tight packed three-dimensional lattice
which basically
represents the abelian sector of the Standard Model group, i.e. $U(1)^3$.
We look for a non-orthogonal lattice, since on such a lattice each site has
more neigbouring sites than on an
orthogonal lattice. The lattice directions of these different sites are
interpreted as corresponding to different (vaccum) phases, and we want as many
different phases as possible. The search for a lattice where each site has
a maximal number of neighbouring sites can be formulated as a tight packing
problem.

The Anti-Grand Unification scheme and the Multiple Point Principle have
many interesting properties. Some very appealing features are that the
Anti-Grand Unification scheme provides a prediction
of the number of generations, and the Multiple Point Principle
supplies a finetuning mechanism for the free parameters in the Standard Model.
According to The Principle of Multiple Point Criticality there is
a multiple point in the action parameter space, corresponding
to a situation where the vacuum
is maximally degenerate, and a maximal number of vacuum phases come together.
The observed coupling values are then explained as the critical multiple point
values that the couplings take when the vacuum is maximally degenerate.

In the pursuit of the critical multiple point, we look for a tightest packed
three-dimensional lattice (corresponding to three generations). Dense packing
means that each lattice site is
surrounded by as many nearest neighbours as possible, in our scheme at
(approximately) the same, critical distance.
The densest regular three-dimenional lattices are the face-centered
cubic lattice and the hexagonal lattice. The face-centered cubic lattice
with twelve ``critical'' neighbours of each site has already been studied,
and it was found that this lattice corresponds to a total of twelve phases
coming together at the critical point. Six of the twelve phases coming
together at the critical point correspond to confined one-dimensional
subgroups of $U(1)^3$. In this article the goal is to find an alternative
lattice, in order to maximize the number of phases corresponding to these
confined one-dimensional subgroups.

We find that while on the face-centered cubic lattice there are six such
phases,
on an almost regular three-dimensional lattice, dual to the face-centered
cubic lattice, seven of the phases that come together at the critical point
correspond to confined one-dimensional subgroups.\\

\section{Anti-grand unification and the Multiple Point Principle}

A flaw of the Standard Model is that it gives no prediction
either of the generation replication pattern, or of the number of generations.
All the  generations are really treated on the same footing.

Another shortcoming of the Standard Model is the large number of
free parameters. There are on the one hand parameters that are not
mass related, and on the other hand the mass parameters associated
with the weak symmetry breaking, i.e. the Higgs quartic and quadratic
self couplings which determine the Higgs mass and vacuum expectation
value; and the Yukawa couplings, which determine the quark masses
and mixing angles. The introduction of the Higgs field and the weak
spontaneous symmetry breaking thus implies a substantial increase
in the number of free parameters.

In the Anti-Grand Unification scheme \cite{AGUT}, it is
assumed that the Standard Model gauge group is derived from the group
$(SMG)^{N_{gen}}$ where ${N_{gen}}$ is the number of generations, and
\begin{equation}
SMG \equiv S(U(2){\rm{x}}U(3))
\end{equation}

In the Standard Model, running coupling constants for
$SU(2)$ and $SU(3)$ at Planck scale deviate by approximately 3 from the
critical values for $SMG$ lattice theory (i.e. from the low energy values of
the coupling constants extrapolated to the Planck scale).
This can be explained if we assume that there are three generations, and the
Standard Model group $SMG$ comes about at Planck scale, by the
breakdown of $(SMG)^3$ to the diagonal subgroup.
With $N_{gen}=3$, 
$$(SMG)^{N_{gen}}=SMG_1 \otimes SMG_2
\otimes SMG_3=\{(g_1,g_2,g_3)|g_j \in SMG_j\},$$ 
and the special case
where $g_1=g_2=g_3$ corresponds to the diagonal subgroup,
$$(SMG^{N_{gen}=3})_{diagonal}\equiv \{(g,g,g)|g \in SMG \}.$$
The $1/g^2$ for the diagonal subgroup couplings being approximately
three times larger than for the single $SU(2)$ and $SU(3)$ subgroups, suggests
that these inverses of the squared coupling are exactly three times larger
for the non-abelian subgroups. This constitutes a prediction of three
generations.

The Multiple Point Principle \cite{Holdon1} can be perceived
as a finetuning mechanism accounting for the free parameters
of the Standard Model. The idea is that as certain extensive parameters
are globally fixed, intensive parameters are finetuned. The extensive
parameters are furthermore fixed at values such that the universe
is supposed to contain a combination of different degenerate vacua, i.e. more
than one phase.
The philosophy is that the lattice artifact phases in a lattice gauge
theory really have physical meaning (assuming that space-time is fundamentally
discrete and we really have a physical 4-dimensional lattice, acting as a
non-arbitrary regulator).

With first order transitions between these different phases, the extensive
qu\-antities can be fixed at values that are not situated at certain points,
but within finite intervals. Several coexisting phases constitute constraints
on the conjugate intensive parameters, like the gauge couplings. The situation
is similar to an equilibrium system in an insulated container
where water exists in its three phases. This system has fixed
extensive parameter values, like the volume and the energy. It is a very
stable system, since there is a whole range of
parameter values for the average energy and average volume per
molecule, and the three phases continue to coexist because the
heat of fusion and the heat of sublimation and vaporization are finite.

Since knowledge of the number of extensive quantities implies a knowledge of
the number of the various phases, the global fixation of extensive parameters
breaks locality.
This $mild$ type of nonlocality however implies no contradictions with
data, since it
exists everywhere in space-time, and can thus be incorporated in the
coupling constants, which are universal intensive parameters.

In this scheme it is not only assumed that the vacuum exists in several
phases. According to The Principle of Multiple Point Criticality, Nature
moreover seeks out a point in the action parameter space where a maximal
number of phases come together - phases in a lattice gauge theory
of the Standard Model gauge group $(SMG)^3$.
At this $multiple$ $point$ the vacuum is maximally degenerate, and
the running gauge coupling constants assume "multiple
point critical values" at Planck scale.

In this way, the Multiple Point Principle supplies an explanation for the
finetuning of the parameter values.

The prediction that there are three generations also depends
the Multiple Point Critical Principle, which predicts that the couplings
are driven
to one and the same multiple point value ${g_{MP}}$ for all three generations,
that is, for the diagonal subgroups of $SU(3)$ and $SU(2)$,
\begin{equation}
\frac{1}{g^2_{diag}} = \frac{1}{g^2_1} +\frac{1}{g^2_2}+
\frac{1}{g^2_3}+\ =\frac{3}{g^2_{MP}}
\end{equation}
In the abelian case the situation is more complicated.
For $U(1)$ the deviation of the coupling in going from the multiple point
of $(SMG)^3$ to the diagonal subgroup, is not three but $\sim 6$. So instead of
$U(1)^3$ we would have $\sim U(1)^6$. It is thus necessary to consider the
abelian sector separately.\\

\section{Phase transitions}

A first order phase transition is characterized by
a sharp distinction between the phases, in the sense that the division between
phases is so clear that we do not need much stuff to determine what
phase we are in, at first order phase transitions.
In a second order phase transition we need more stuff to determine
which phase we are in, that is, we need to go to the long wavelength limit.

In lattice gauge theory, phases are defined in a more abstract way than
in the case of water and ice.
A lattice is a system of $P$ elements $x,y,z,..$ which have a reflexive,
transitive and antisymmetric relation, usually represented
as lattice sites connected by lattice links.
A gauge field on a lattice assigns a gauge group element
$U({\tiny{\bullet}}$-${\tiny{\bullet}})$
to each link of
the lattice. These links in their turn constitute the lattice plaquettes.
A gauge invariant action on the lattice must be formulated in terms of lattice
gauge variables, which not apriori satisfy the Bianchi
identities, $\sum_{\Box} U(\Box)$ = 0, where
$ U(\Box)$=$U(l_1)U(l_2)U(l_3)U(l_4)$,
\begin{center}\begin{picture}(220,80)(-40,0)
\Line(7,9)(90,9)
\Line(25,0)(25,70)
\Line(75,0)(75,70)
\Line(7,59)(90,59)
\Vertex(25,59){2}
\Vertex(75,59){2}
\Vertex(25,9){2}
\Vertex(75,9){2}
\Text(50,0)[]{$l_1$}
\Text(83,34)[]{$l_2$}
\Text(50,67)[]{$l_3$}
\Text(19,34)[]{$l_4$}
\end{picture}\end{center}
In $D>2$, the Bianchi identities which do not apply to plaquette variables,
however do apply to the volumes enclosed
by the plaquettes, like  the cubes in a 3-dimensional cubic lattice.

There are several vacuum phases, which are determined by the transformation
properties of the vacuum under gauge transformations corresponding to
a set of gauge functions that are constant or linear in the space-time
coordinates.
Phenomenologically different phases can be distinguished by their
different fluctuation patterns,
the vacuum phases are thus determined by the different quantum
fluctuations of space-time.
The strength of quantum fluctuations is governed by the coupling
constants, so for small coupling constants the fluctuations
are small, and for large coupling constants they are large.
With the running coupling constants increasing with energy, at the
fundamental lattice scale the fluctuations are thus very large.

A "confinement-like" phase corresponds to short range correlations, and
a "Coulomb-like" phase corresponds to infinite range correlations. The ordinary
electromagnetic potential $A_{\mu}$ is in the Coulomb phase, and
photons, having infinite range, correspond to infinite correlations.
On the lattice, the Bianchi identities can in the confinement-like phase
be neglected (to a rough approximation), meaning that the fluctuations of one
plaquette variable is approximately independent of the fluctuations of the
other plaquettes that enclose a lattice cell.
In the Coulomb phase, on the other hand, the Bianchi identities are
non-negligeable, leading to a correlation of plaquette variable fluctuations
over long distances, i.e. distances of at least several lattice
constants.

Each possible combination of gauge field phases corresponds to a specific
vacuum, so when you have a gauge field you consider what phase it is in.
What phases are possible depends on the gauge group.
For gauge groups with many subgroups, like $(SMG)^3$, the degrees of freedom
corresponding to the different subgroups can have qualitatively different
fluctuation patterns, corresponding to different regions of the action
parameter space.
At the multiple point the phase transitions are first order, in the sense that
the phases are distinguishable at the lattice scale.
The determination of gauge couplings by means of first order phase transitions
however has the problem of non-universality, so in that sense second order
phase transitions would be preferable for giving well defined numbers for the
transition points. The problem is then that there is
no scale for second order phase transtions. At the multiple point of
a phase diagram for a gauge lattice theory, we thus assume that the
lattice phases are separated by first order phase transtions.
Different short distance physics can be represented by different distributions
of group elements along different subgroups of the gauge group in
different parts of the action parameter space.
The focusing on the short distance aspect does however not mean that
the long distance behaviour is not influenced by the passage from one phase
to another.

The abelian and non-abelian subgroups of the anti-grand unification group
$(SMG)^3$ can be treated separately,
because in the approximation where we identify the $U(1)_j$ in $U(1)^3$ with
$SMG_j/(SU(2){\rm{x}}SU(3))_j$ ($j=1,2,3$), the interaction between abelian and
non-abelian subgroups may be discarded, and $U(1)^3$, $SU(2)^3$ and
$SU(3)^3$ can be treated separately.

The necessecity for the different treatment of the abelian and non-abelian
sectors is obvious when dealing with the action corresponding to $(SMG)^3$.
It is possible to express the action corresponding to the Cartesian product
$(SMG)^{N_{gen}}$ as a sum of contributions from each factor group,
${\cal{S}}=\sum_j {\cal{S}}_j$.
This means that the confining phases must correspond to factorizable invariant
subgroups, implying that the phase diagram for $(SMG)^{N_{gen}}$ can be
determined from the phase diagrams of the individual factors (a group $G$ with
the group operation $\cdot$ is factorizable if it has two proper nontrivial
subgroups $H$ and $F$ such that $G=H \cdot F$).
That the action is restricted to being additive however means that the phases
corresponding to confinement along non-factorizable subgroups are missing.
For the non-abelian subgroups there are rather few such phases, but for the
abelian subgroups there is an infinity of partially confining phases of
$U(1)^3$ missing. The action ${\cal{S}}=\sum_j {\cal{S}}_j$ thus gives a
good approximation for the non-abelian subgroups, but not for the abelian.
In order to find the multiple point, it is therefore necessary to
consider the whole group $(SMG)^3$, or, to a good approximation, $U(1)^3$.

The non-factorizable subgroups of $U(1)^3$ occur as diagonal-like
subgroups of all possible Cartesian products with 2 or 3 repeated factors,
which can be the subgroup $U(1)$ or the discrete groups $Z_N$ (Appendix III).
$U(1)^3$ is a compact factor group of the covering group ${\bf{R}}^3$,
by going to $U(1)^3$ we identify the 3-dimensional $identification$ $lattice$
$L$ of elements of ${\bf{R}}^3$ with the unit element, the 3-dimensionality of
the identification lattice thus representing the assumed number of generations.
The idea is to use the Multiple Point Principle Criticality to find the
identification lattice $L$ which will bring a maximal number of
phases together.

The elements of the group $U(1)$ can be represented as $\{ e^{igA^{\mu}} \}
\sim \{ e^{i\theta} \}$,
so $U(1)$ is compactified as $\theta$ runs around
the unit circle. In that sense, the sites $0$, $2 \pi$, $4\pi$,...are
identified with each other,
\begin{center}\begin{picture}(285,25)(0,0)
\LongArrow(0,9)(270,9)
\Text(10,2)[]{0}
\Text(50,3)[]{$2 \pi$}
\Text(90,3)[]{$4 \pi$}
\Text(130,3)[]{$6 \pi$}
\Text(170,3)[]{$8 \pi$}
\Text(280,9)[]{$\theta$}
\Vertex(10,9){2}
\Vertex(50,9){2}
\Vertex(90,9){2}
\Vertex(130,9){2}
\Vertex(170,9){2}
\Text(190,4)[]{$\cdots$}
\end{picture}\end{center}
{\vspace{2mm}}
With two $U(1)$ groups we have
\begin{center}\begin{picture}(315,296)(0,9)
\LongArrow(20,145)(290,145)
\LongArrow(145,12)(145,284)
\Text(185,140)[]{$2 \pi$}
\Text(225,140)[]{$4 \pi$}
\Text(250,140)[]{$\cdots$}
\Text(314,145)[]{$U(1)_1$}
\Text(147,296)[]{$U(1)_2$}
\Vertex(185,145){2}
\Vertex(225,145){2}
\Vertex(25,145){2}
\Vertex(105,145){2}
\Vertex(65,145){2}
\Vertex(145,185){2}
\Vertex(145,225){2}
\Vertex(145,225){2}
\Vertex(145,145){2}
\Vertex(145,105){2}
\Vertex(145,65){2}
\end{picture}\end{center}
Here the sites $2 \pi$, $4 \pi$,.. correspond to charges or some
other quantized quantity. This system can be ``dynamized'' by absorbing the
$U(1)$ coupling in the metric
$g_{\mu \nu}$, which is the metric on the space of charges, i.e. the
metric used in forming the inner product of two $\theta$-variables.
When the coupling is thus absorbed
the distance between two charges is well defined, as the metric is
included in the formalism.

The sites on the axes are identified with the origin - this is the
compactification. As the couplings are included in the definition of the
distance, the distance between two charges will vary according to the charge
values of the two charges between which the distance goes.
So from the point of view of $U(1)$, all the sites
\begin{center}\begin{picture}(285,30)(0,0)
\LongArrow(7,9)(270,9)
\Vertex(50,9){2}
\Vertex(90,9){2}
\Vertex(130,9){2}
\Vertex(170,9){2}
\end{picture}\end{center}
are the same site, while from the point of view of the charges,
they are different.
The sites can also be interpreted as monopoles, i.e. sinks or sources for
magnetic fields.
In any lattice gauge theory we get "effective monopoles", because
certain flux combinations of the different abelian magnetic field can
disappear into a lattice cube, as if there were a monopole. This monopole
is a lattice artifact, but behaves exactly like a fundamental monopole.
In the Coulomb-like phase vacuum contains very few monopoles, while when a
subgroup $U(1)_j$ ($j=1,2,3$) is confined, there is
statistically an abundance of lattice cubes for which the monopole charge
$2\pi$$(n_1,n_2,n_3)$
is $\pm 2\pi(1,0,0)$, a few with charges $\pm 2\pi(2,0,0)$ or
$\pm 2\pi(3,0,0)$, and also some monopoles with $n_k \neq 0$, $k \neq j$.

When there is confinement along $U(1)_j$, the distance between nearest
neighbours on the $U(1)_j$ identification lattice is less than the critical
distance.
If we assume that the critical distance along one lattice direction is
independent of the spacing between the neighbours in other lattice directions,
the transition in
one subgroup does not really influence the fluctuation pattern for the other
subgroups.
In this approximation the multiple point criticality corresponds to
having a critical distance between neighbouring identification lattice points
in all directions, as critical distances correspond to critical couplings.

The identification lattice - a 3-dimensional lattice of elements of
${\bf{R^3}}$ - is a generalization of the
identification mod $2\pi$. If ${\bf{R^3}}$ has a well-defined
inner product, a unique action can be defined on the lattice.
The action is then described by the geometry of the identitification lattice
in the sense that the different phases can be reached by changes in the
action, corresponding to deviations from the critical distance between nearest
lattice neighbours.
There is one phase defined for each pair $({\cal{G}}_j,{\tilde{{\cal{G}}_j}})$
of subgroups of $SMG$, where ${\cal{G}}_j \in SMG$ and
${\tilde{{\cal{G}}_j}}$ is an invariant subgroup of ${\cal{G}}_j$.
The action must be formulated in such a way that it encompasses all the phases.

Let ${\cal{G}}_1$ = $U(1)$ and ${\tilde{{\cal{G}}_1}}=Z_2$.
A simple lattice action for $U(1)$ is ${\cal{S}}_{\Box}$ = $\beta \cos \theta$,
\begin{center}\begin{picture}(275,200)(0,0)
\Text(2,129)[]{$\beta \cos \theta$}
\Text(220,10)[]{$\theta$}
\Text(214,70)[]{${\bf{\it{{2{\hspace{1mm}} phases}}}}$}
\Text(97,0)[]{$\beta_{crit}$}
\Line(97,7)(97,13)
\Text(50,45)[]{Confinement}
\Text(135,45)[]{Coulomb}
\Curve{(1,10)(87,15)}
\Curve{(89,16)(87,15)}
\Curve{(89,16)(91,17)}
\Curve{(91,17)(94,19)}
\Curve{(94,19)(95,40)}
\Curve{(95,40)(96,50)}
\Curve{(96,50)(97,60)}
\Curve{(97,60)(98,70)}
\Curve{(100,105)(98,70)}
\Curve{(100,105)(101,106)(102,109)}
\Curve{(102,109)(104,111)}
\Curve{(104,111)(105,113)}
\Curve{(107,114)(105,113)}
\Curve{(107,114)(109,115)}
\Curve{(109,115)(195,117)}
\LongArrow(1,10)(1,120)
\LongArrow(1,10)(210,10)
\end{picture}\end{center}
{\vspace{15mm}}
The phase transition between the confinement and Coulomb phases is a function
of $\beta$, that is, of $\beta_{crit}$.
With one more term in the Lagrangian, i.e.
${\cal{S}}_{\Box}$ = $\beta \cos \theta+\gamma \cos 2\theta$, the critical
point is modified,
\begin{center}\begin{picture}(250,200)(0,0)
\Text(13,205)[]{$\gamma$}
\Text(210,10)[]{$\beta$}
\LongArrow(13,10)(13,195)
\LongArrow(13,10)(200,10)
\Curve{(13,150)(70,120)(85,100)(95,80)(101,60)}
\Curve{(101,60)(102,40)}
\Curve{(104,10)(102,40)}
\Curve{(80,110)(77,150)}
\Curve{(81,190)(77,150)}
\DashLine(13,110)(80,110){2}
\DashLine(80,10)(80,110){2}
\Text(1,110)[]{$\gamma_{crit}$}
\Text(40,180)[]{$Z_2$ alone}
\Text(40,169)[]{confined}
\Text(43,80)[]{total}
\Text(47,69)[]{confinement}
\Text(150,75)[]{Coulomb}
\Text(80,0)[]{$\beta_{crit}$}
\Text(190,140)[]{${\bf{\it{{3{\hspace{1mm}} phases}}}}$}
\end{picture}\end{center}
{\vspace{10mm}}
With yet another term, the critical point is again modified and
yet more phases meet.\\

\section{Abelian and non-abelian subgroups}

The abelian and non-abelian sectors are separated also in that
the abelian and non-abelian couplings need different
considerations. The reason is that in non-abelian groups the commutator
relations constitute constraints that fix the normalization of the gauge
couplings, while in the abelian case there are no such constraints.
In the non-abelian case the Yang-Mills fields are themselves
charged and can therefore be used to define an "inner convention" of the
gauge charges to be established.
In this case the Lie algebra commutator
relations are nonlinear and thus not invariant under rescalings of the
gauge potential. If such rescalings were allowed, the gauge
couplings would be deprived of physical significance.

In the abelian case, however, such rescalings of the gauge potential $are$
possible. For the $U(1)$ there is therefore no natural unit of charge.
So the weak hypercharge fine structure constant is normalizable
only by reference to some quantum of charge - but which quantum of charge?

There is the notion that a fundamental quantum of charge should in some sense
be "small", this is the philosophy of the Han-Nambu scheme\cite{HN1}.
The question is how to define "smallness".
The non-abelian representations of the Standard Model gauge group
realized in nature are small in the sense that in the non-abelian sector
of Standard Model there is the trivial representation and the lowest
dimensional representation after the trivial.
In the abelian sector it is more unclear what small means. Abelian
subgroups always have 1-dimensional irreducible representations, so dimension
tells nothing.
Therefore, replace the demand for ``smallness'' by the demand that
the number of ``Han-Nambu''charges of the fundamental Weyl components taking
the (``small'') values $-1,0,1$ should be maximal.
The reason for the notation is that this approach resembles
an early quark model by Han and Nambu \cite{HN}, where a group of eight
integrally charged quarks was proposed, quarks
that also carry a colour-like quantum number. In this group two
of the charges have charge $Q=1$, and two had $Q=-1$, the other four are
electrically neutral.

\section{Tight packing}

On the lattice a confining subgroup is found along a lattice direction where
the distance between two nearest neighbours is smaller than the distance
corresponding to a critical coupling value.
The critical coupling for a given subgroup generally depends on
which phases are realized for the remaining $U(1)$ degrees of freedom.
The critical distance in one direction is however approximately independent
of the distance between neighbouring lattice sites in other
directions, approximate MPP criticality is thus achieved when the distance
between neighbouring identification lattice sites is critical in all
directions. This corresponds to a maximal number of 1-dimensional subgroups,
or to having the tightest possible packing of identification lattice points.

This tight packing problem is closely related to the problem of finding
the tightest packing of equal non-overlapping spheres in $N$-dimensional space,
that is, the problem of packing spheres in such a
way that one sphere "kisses" as many other spheres as possible.
In 2 dimensions there are two periodic packings for identical circles:
square lattice and hexagonal lattice, in 1940 the hexagonal lattice was proven
to be the densest of all possible plane packings \cite{Ftoth}.
The lattice plaquettes are the fundamental regions of the lattice, and
the packing density or filling factor for a lattice is defined as
\begin{equation}
\rho_P=\frac{Volume{\hspace{1mm}}of{\hspace{1mm}}one{\hspace{1mm}}sphere}
{Volume{\hspace{1mm}}of{\hspace{1mm}}fundamental{\hspace{1mm}}region}
\end{equation}
For example, in the hexagonal 2-dimensional lattice the lattice plaquettes are
rhombuses, and (with lattice constant $1$) $\rho_P = (\pi 1/4)
/( \sqrt{3}/2)$=$\pi/2\sqrt{3} \sim$ 0.907, and each circle kisses 6 other
circles,
\begin{center}\begin{picture}(220,125)(0,0)
\LongArrow(7,9)(220,9)
\LongArrow(7,9)(7,125)
\SetScale{1.5}
\DashLine(65,60)(55,42){2}
\DashLine(85,60)(75,42){2}
\DashLine(55,42)(75,42){2}
\DashLine(65,60)(85,60){2}
\Oval(65,60)(10,10)(0)
\Oval(75,42)(10,10)(0)
\Oval(55,42)(10,10)(0)
\Oval(85,60)(10,10)(0)
\Oval(95,42)(10,10)(0)
\Oval(65,25)(10,10)(0)
\Oval(85,25)(10,10)(0)
\Vertex(25,25){2}
\Vertex(45,25){2}
\Vertex(65,25){2}
\Vertex(85,25){2}
\Vertex(105,25){2}
\Vertex(125,25){2}
\Vertex(35,42){2}
\Vertex(55,42){2}
\Vertex(75,42){2}
\Vertex(95,42){2}
\Vertex(115,42){2}
\Vertex(135,42){2}
\Vertex(25,60){2}
\Vertex(45,60){2}
\Vertex(65,60){2}
\Vertex(85,60){2}
\Vertex(105,60){2}
\Vertex(125,60){2}
\Vertex(35,77){2}
\Vertex(55,77){2}
\Vertex(75,77){2}
\Vertex(95,77){2}
\Vertex(115,77){2}
\Vertex(135,77){2}
\DashLine(65,60)(55,42){2}
\DashLine(85,60)(75,42){2}
\DashLine(55,42)(75,42){2}
\DashLine(65,60)(85,60){2}
\end{picture}\end{center}
In the square lattice the fundamental regions are squares,
\begin{center}\begin{picture}(220,132)(0,6)
\LongArrow(7,9)(230,9)
\LongArrow(7,9)(7,132)
\SetScale{1.4}
\DashLine(65,45)(65,65){2}
\DashLine(85,45)(85,65){2}
\DashLine(85,45)(65,45){2}
\DashLine(65,65)(85,65){2}
\Oval(65,45)(10,10)(0)
\Oval(85,45)(10,10)(0)
\Oval(65,65)(10,10)(0)
\Oval(85,65)(10,10)(0)
\Vertex(25,25){2}
\Vertex(45,25){2}
\Vertex(65,25){2}
\Vertex(85,25){2}
\Vertex(105,25){2}
\Vertex(125,25){2}
\Vertex(25,25){2}
\Vertex(45,25){2}
\Vertex(65,25){2}
\Vertex(85,25){2}
\Vertex(105,25){2}
\Vertex(125,25){2}
\Vertex(25,65){2}
\Vertex(45,65){2}
\Vertex(65,65){2}
\Vertex(85,65){2}
\Vertex(105,65){2}
\Vertex(125,65){2}
\Vertex(25,85){2}
\Vertex(45,85){2}
\Vertex(65,85){2}
\Vertex(85,85){2}
\Vertex(105,85){2}
\Vertex(125,85){2}
\Vertex(25,45){2}
\Vertex(45,45){2}
\Vertex(65,45){2}
\Vertex(85,45){2}
\Vertex(105,45){2}
\Vertex(125,45){2}
\end{picture}\end{center}
the kissing number is 4, and density is $\rho_P = (\pi 1/4)/1$ =
$\pi/4 \sim$ 0.785.
So in two dimensions the hexagonal lattice gives the densest packing,
and also the biggest kissing number.

In 3 dimensions there are three periodic packings for identical spheres:
cubic, face-centered cubic (fcc), and hexagonal, where the fcc and the
hexagonal packing have the same filling factor.
There is  also random tight
packing \cite{Jager}, which has $\rho_P$ $\sim$ 0. 64.

In studying the tight packing problem, one can use root lattices
(Appendix I). In 1 dimension, the only tight packing is the root lattice
$A_1$ generated by the vector $(-1,1)$,
\begin{center}\begin{picture}(285,25)(0,0)
\Line(10,9)(240,9)
\Vertex(210,9){3}
\Vertex(50,9){3}
\Vertex(90,9){3}
\Vertex(130,9){3}
\Vertex(170,9){3}
\Text(270,9)[]{$A_1$}
\end{picture}\end{center}
Assuming that our spheres have radius $1/\sqrt{2}$, the minimal
distance between distinct packing points is 2.

In 2 dimensions the only tight packing is the root lattice
$A_2$, which is generated by stacking displaced $A_1$'s,
with a distance between the layers of at least
$\sqrt{2-1/2}=\sqrt{3/2}$,
\begin{center}\begin{picture}(285,100)(0,0)
\Line(10,9)(240,9)
\Vertex(210,9){3}
\Vertex(50,9){3}
\Vertex(90,9){3}
\Vertex(130,9){3}
\Vertex(170,9){3}
\Text(270,9)[]{$A_1$}
\Line(10,44)(240,44)
\Vertex(30,44){3}
\Vertex(70,44){3}
\Vertex(110,44){3}
\Vertex(150,44){3}
\Vertex(190,44){3}
\Text(270,44)[]{$A_1^{'}$}
\Line(10,79)(240,79)
\Vertex(210,79){3}
\Vertex(50,79){3}
\Vertex(90,79){3}
\Vertex(130,79){3}
\Vertex(170,79){3}
\Text(270,79)[]{$A_1^{"}$}
\end{picture}\end{center}
Likewise, tight packing in 3 dimensions is generated by stacking
copies of $A_2$, with a separation between the layers of at least
$\sqrt{2-2/3}=\sqrt{4/3}$.
A sphere in a 3-dimensional can be surrounded in essentially two
different ways.
First, there are two layers of spheres stacked in such a way that
each sphere in the top layer is in contact with three spheres in the
bottom layer. Then a third layer is added in two different ways.
If the spheres in the third layer are placed over the holes in the
first layer not occupied by the second layer, we get a face-centered
cubic close packing, where the contact points of the twelve neighbours of each
sphere form the vertices of a cuboctahedron.
Fruits in fruit stands are often stacked in this fcc fashion.

If the spheres in the third layer are instead
placed directly over the spheres in the first layer, the result is
a hexagonal close packing, where the contact points of the twelve neighbours
form the corners of an anti-cuboctahedron, obtained by
twisting the cuboctahedron about one of its four equatorial planes.

In 1611 Kepler hypothesized that fcc tight packing is the densest
possible 3-dimensional packing, with $\rho_P = \pi\sqrt{18}$ $\sim$ 0.74.
And in 1831 it was proven by Gauss that the face-centered cubic $lattice$
is the densest lattice packing in 3 dimensions.

But lattice packing is not the same as sphere packing. A regular lattice is the
most regular arrangement of spheres, and can be constructed by repeating a
single lattice "crystal". In that sense, the maximum lattice density is
a finite problem.
In sphere-packing, on the other hand, the position of every sphere is a
variable, and it takes infinitely many spheres to fill up all of mathematical
space.
One thus has to reduce the problem to a finite part of space, with
a finite amount of spheres. In the process of proving Kepler's
conjecture, it was therefore suggested to use Voronoi cells in the analysis of
the problem. A Voronoi cell is the set of points that are closer to one
particular sphere than any other. For example, the Voronoi cell of a
2-dimensional disc is the set of points in the plane which are as
close or closer to the centre of that disc than to the centre of any other
disc.

The Voronoi cells corresponding to a 2-dimensional hexagonal lattice
are hexagons, and the Voronoi cells of the 3-dimensional
face-centered cubic lattice are rhombic dodecahedra. Using an extension
of the notation of cells, Kepler's conjecture was proven \cite{Hales}
in the 1990-ies. It has also been proven that there are packings of congruent
ellipsoids with density considerably greater than $\pi\sqrt{18}$. \\

\section{A tighter packing}

In a 3-dimensional lattice with mutually orthogonal directions, there
are 8 phases that can be reached, with confinement of eight subgroups.
With a fcc lattice corresponding to the cuboctahedron, a larger number
of phases can come together, confining a total of twelve subgroups, six of
which are the one-dimensional subgroups of $U(1)^3$. The
fcc lattice is relevant for all the continuous subgroups of $U(1)^3$, we
are however also interested in the phases confined
with respect to discrete abelian subgroups in contact with the multiple
point.

So instead of a fcc identification lattice, we introduce
a Han-Nambu like system of charges, corresponding to the $2^3-1=7$
linearly independent basis vectors of a $Z^3_2$ projective space \cite{HD}.
The seven vectors with the coordinates
\begin{equation}\label{wec}
(1,0,0), (0,1,0), (0,0,1), (1,1,0), (1,0,1), (0,1,1), (1,1,1)
\end{equation}
constitute, together with the vector $(0,0,0)$, a representation of
the discrete group $Z_2^3$.

Since the seven one-dimensional subgroups of $U(1)^3$ should all be
critical for the same action metric, these vectors should all have
the same length, set to unity. As they represent $Z_2^3$, this is
however not possible. The next best thing is to have these vectors
as close as possible to unity. This can be accomplished by determining
a metric which minimizes the amount of deviation from unity for all seven
vectors.

Therefore, define a metric on the space of charges such that a critical
distance is determined by phase transitions that make confinement in
the direction along the charges.
The $Z_2^3$ symmetry constrains the number of adjustable parameters
in the metric to two, giving the metric tensor

\begin{equation}\label{ball}
\eta_{kl} = \left(\begin{array}{rcl}
                  a   &  b{\hspace{4mm}}b\nonumber\\
                  b   &  a{\hspace{4mm}}b\nonumber\\
                  b   &  b{\hspace{4mm}}a
                      \end{array}
                \right)
\end{equation}
{\vspace{4mm}}
In terms of $\eta_{kl}$ the squared length of the vectors
$(1,0,0), (0,1,0), (0,0,1)$ is $a$, e.g.
\begin{equation}\label{valp}
(1,0,0) \left(\begin{array}{rcl}
                  a   &  b{\hspace{4mm}}b\nonumber\\
                  b   &  a{\hspace{4mm}}b\nonumber\\
                  b   &  b{\hspace{4mm}}a
                      \end{array}
                \right)
\left(\begin{array}{rcl}
                  1\\
                  0\\
                  0
       \end{array}
\right)=a
\end{equation}
{\vspace{4mm}}

\noindent The squared deviation length from unity for these vectors is then
$(a-1)^2$. For the vectors $(1,1,0), (1,0,1), (0,1,1)$ it is $2(a+b)$, and the
squared deviation length from unity for these vectors is
$[2(a+b)-1]^2$. The vector $(1,1,1)$ has the length $3(a+2b)$, with the
squared deviation from unity $[3(a+2b)-1]^2$.
The total squared deviation length for all seven vectors is then
\begin{equation}
D=3(a-1)^2+3[2(a+b)-1]^2+[3(a+2b)-1]^2=
24 a^2+48 b^2 +60 ab - 24 a - 24 b +7
\end{equation}
Minimizing the total squared deviation length with respect to $a$ and $b$
gives $\partial D/ \partial a= 4 a+ 5b -2 = 0$ and
$\partial D/ \partial b= 5a + 8b -2 = 0$, giving $a=6/7$ and $b=-2/7$.
This gives the metric tensor

\begin{equation}\label{namo}
\eta_{kl}=\frac{2}{7} \left(\begin{array}{rcl}
                  3   &{\hspace{2mm}}  -1{\hspace{4mm}}-1\nonumber\\
                 -1   &{\hspace{6mm}}3{\hspace{5mm}}-1\nonumber\\
                 -1   &{\hspace{1mm}}  -1{\hspace{8mm}}3
                      \end{array}
                \right)
\end{equation}
{\vspace{5mm}}

\noindent The vectors $((1,0,0),
(0,1,0), (0,0,1))$, $((1,1,0), (1,0,1), (0,1,1))$ and
$(1,1,1)$ thus have the lengths $\sqrt{6/7}$, $\sqrt{8/7}$ and $\sqrt{6/7}$,
respectively.

They span a rhombic dodecahedron which is made out of twelve rhombuses shaped
in accordance with the
Golden Section., i.e. the two diagonals of each rhombus satisfies
$x=\sqrt{2}y$.
The rhombic dodecahedron has fourteen vertices. Three faces meet at eight of
these vertices, and four faces meet at six of the vertices. On this lattice,
seven of the phases that come together at the critical point correspond
to confined one-dimensional subgroups of $U(1)^3$, i.e. one more than for the
facec-centered cubic lattice,
\begin{center}\begin{picture}(285,220)(0,0)
\LongArrow(100,180)(100,210)
\LongArrow(180,100)(240,100)
\Line(2,100)(20,100)
\DashLine(20,100)(180,100){2}
\DashLine(100,180)(100,20){2}
\Line(100,20)(100,0)

\LongArrow(41,60)(10,38)
\DashLine(41,60)(100,100){2}

\Vertex(100,180){2}
\Vertex(100,20){2}
\Vertex(180,100){2}
\Vertex(20,100){2}
\Vertex(47,144){2}
\Vertex(80,80){2}
\Vertex(130,60){2}
\Vertex(130,117){2}
\Vertex(155,155){2}
\Vertex(162,70){2}
\Vertex(49,40){2}

\Line(180,100)(130,60)
\Line(100,20)(130,60)
\Line(80,80)(130,60)
\Line(20,100)(47,144)
\Line(100,180)(47,144)
\Line(80,80)(47,144)
\Line(80,80)(130,117)
\Line(100,20)(49,40)
\Line(49,40)(80,80)
\Line(49,40)(20,100)
\Line(100,180)(155,155)
\Line(100,180)(130,117)
\Line(155,155)(180,100)
\Line(180,100)(162,70)
\Line(162,70)(100,20)
\Line(180,100)(130,117)
\end{picture}\end{center}
The rhombic dodecahedron is dual to the cuboctahedron, and thus the
Voronoi cell of the fcc lattice. The rhombic dodecahedron is moreover a
3-dimensional projection of the 4-dimensional hypercube.
An object $O$ in a D-dimensional space can be projected onto an object $O^{'}$
in a space of (D-1) dimensions, in many different ways. For an object $O$ with
well-defined symmetries, we are however interested in projections $O^{'}$ that
retain some of these symmetries.

The cube has an octahedral symmetry, with three axes of 4-fold rotational
symmetry which join centres of
opposite faces, four axes of 3-fold rotational symmetry which join
diagonally opposite vertices, six 2-fold rotation axes joining the
midpoints of opposite edges.
One of the 2-dimensional projections of the 3-dimensional
cube is the square,
with its 4-fold rotational symmetry, another 2-dimensional projection
of the cube is the hexagon, with its $\pi/3$ rotational symmetry,
\begin{center}\begin{picture}(285,180)(0,0)
\Line(90,25)(25,63)
\Line(25,138)(25,63)
\Line(25,138)(90,100)
\Line(90,25)(155,63)
\Line(155,63)(155,138)
\Line(90,25)(25,63)
\Line(25,138)(25,63)
\Line(25,138)(90,100)
\Line(90,25)(155,63)
\Line(155,63)(155,138)
\DashLine(90,100)(25,63){3}
\DashLine(90,100)(155,63){3}
\Line(90,25)(90,100)
\Line(155,138)(90,100)
\Line(155,138)(90,175)
\Line(25,138)(90,175)
\Line(90,25)(90,100)
\Line(155,138)(90,100)
\Line(155,138)(90,175)
\Line(25,138)(90,175)
\DashLine(90,100)(90,175){3}
\end{picture}\end{center}
The cube is space filling, i.e. a 3-dimensional tesselation.
Its symmetric 2-dimen\-sional projections, the square and the hexagon,
correspondingly fill the plane.
In its turn, the cube is a 3-dimensional projection of the
4-dimensional hypercube, and the hypercube fills 4-dimensional space.
The cube and the rhombic dodecahedron are (symmetric) 3-dimensional
projections of the 4-dimensional hypercube, and they both fill 3-dimensional
space.

In this way, the tesselations in spaces of different dimensions are related.
To get a notion of 4-dimensionality, consider how to go from one
known dimension to the next: first, let 1 dimension be represented
by a 1-dimensional line:
\begin{center}\begin{picture}(285,25)(0,0)
\Line(100,10)(25,10)
\end{picture}\end{center}
the go from 1 to 2 dimensions by connecting two 1-dimensional lines to
make a 2-dimensional square
\begin{center}\begin{picture}(285,100)(0,0)
\Line(100,10)(25,10)
\Line(100,85)(25,85)
\DashLine(100,85)(100,10){4}
\DashLine(25,85)(25,10){4}
\end{picture}\end{center}
Then repeat the procedure in going from 2 to 3 dimensions, connect
two 2-dimen\-sional squares to make a 3-dimensional cube:
\begin{center}\begin{picture}(285,116)(0,0)
\Line(100,10)(25,10)
\Line(100,85)(25,85)
\DashLine(25,85)(25,10){4}
\DashLine(100,85)(100,10){4}

\Line(120,30)(45,30)
\Line(120,105)(45,105)

\Line(120,105)(100,85)
\Line(120,30)(100,10)

\Line(100,10)(120,30)
\Line(45,105)(25,85)

\DashLine(120,30)(120,105){4}
\DashLine(45,30)(45,105){4}

\Line(25,10)(45,30)
\end{picture}\end{center}
Now to go to 4 dimensions do the same thing: connect two 3-dimensional
cubes to make a 4-dimensional hypercube
\begin{center}\begin{picture}(285,150)(-20,60)
\SetScale{1.3}
\Line(5,120)(5,70)
\Line(55,70)(5,70)
\Line(55,70)(55,120)
\Line(97,150)(55,120)
\Line(97,150)(47,150)
\Line(5,120)(47,150)
\Line(5,120)(55,120)
\Line(97,150)(97,100)
\Line(55,70)(97,100)

\DashLine(5,120)(24,95){4}
\DashLine(5,70)(24,45){4}
\DashLine(55,120)(74,95){4}
\DashLine(55,70)(74,45){4}
\DashLine(97,150)(116,125){4}
\DashLine(97,150)(116,125){4}
\DashLine(47,150)(66,125){4}
\DashLine(97,100)(116,75){4}
\Line(24,95)(24,45)
\Line(74,45)(24,45)
\Line(74,45)(74,95)
\Line(116,125)(74,95)
\Line(116,125)(66,125)
\Line(24,95)(66,125)
\Line(24,95)(74,95)
\Line(116,125)(116,75)
\Line(74,45)(116,75)
\end{picture}\end{center}
\section{Conclusion}
A tight packing in a $N_{gen}$=3-dimensional space
where the U(1) coupling is absorbed in the action metric was found,
for a system of charges corresponding to the $2^3$-1=7 linearly
independent basis vectors of a $Z_2^3$ projective space.
This lattice, which corresponds to a 3-dimensional tesselation using
rhombic dodecahedra, implies that seven of the phases that come together
at the critical point correspond to confined one-dimensional subgroups.

\section{APPENDIX}

\ \ \ \ \    

{\bf{I. Root lattices}}

\noindent A Lie algebra ${\cal{L}}$ is a flat vector space V over some field (usually
the real or complex numbers), together with the Lie bracket $[,]$, which is a
binary operation, $[,]: V$x$V$ $\rightarrow$ $V$. It is bilinear, satifies
the Jacobi identity, and $[v,v]$ = 0, $v \in V$. A Lie algebra is usually
represented as a vector space of square matrices over the real/complex numbers.
A simple Lie algebra contains no non-trivial ideals, and a semisimple Lie
algebra is the direct sum of simple Lie algebras.
An abelian subalgebra of a Lie algebra has vanishing commutators, and is
often called a torus, the Cartan subalgebra is a maximal torus.
Semisimple Lie algebras are classified through the representations of their
Cartan subalgebras, that is, by their $root$ system.

In a collection of diagonal matrices which span a subspace of
the represenattion space, the diagonal elements constitute weights.
The Cartan subalgebra ${\cal{H}}$ is abelian, and can be put into
diagonal form.
For example, in the standard representation of the special linear Lie algebra
$sl_3({\cal{C}})$ on ${\cal{C}}^3$, the matrices
\begin{equation}\label{mat}
{\bf{H_1}} = \left(\begin{array}{rcl}
                  1   & {\hspace{4mm}} 0{\hspace{7mm}}0\nonumber\\
                  0   & -1{\hspace{7mm}}0\nonumber\\
                  0   &  {\hspace{4mm}}0{\hspace{7mm}}0\nonumber
                      \end{array}
                \right),
{\bf{H_2}} = \left(\begin{array}{rcl}
                  0   & {\hspace{4mm}} 0{\hspace{7mm}}0\nonumber\\
                  0   & -1{\hspace{7mm}}0\nonumber\\
                  0   &  {\hspace{4mm}}0{\hspace{7mm}}1\nonumber
                      \end{array}
                \right)
\end{equation}
span the Cartan algebra. There are then three weights, $w_1=h_{11}$,
$w_2=h_{22}$ $w_3=h_{33}$, corresponding to the decomposition of
${\cal{C}}^3$ into its eigenspaces, ${\cal{C}}^3 =
<{\bf{w_1}}>\oplus<{\bf{w_2}}>\oplus<{\bf{w_3}}>$. The eigenvectors
${\bf{w_j}}$ are the weight vectors, and
the corresponding spaces $<{\bf{w_j}}>$ are the weight spaces.

The roots of a semisimple Lie algebra are the weights of the adjoint
representation, they generate the discrete root lattice in the
dual of the representation space. In their turn, the weights form a weight
lattice which contains the root lattice.
For example, the special Lie algebra
$sl_2{\cal{C}}$ of $2{\rm{x}}2$ with trace zero, has the basis
\begin{equation}\label{B1}
{\bf{B_1}}=\left(\begin{array}{rcl}
                  1   &  0\nonumber\\
                  0   &  -1
                      \end{array}
                \right),
{\bf{B_2}}=\left(\begin{array}{rcl}
                  0   &  1\nonumber\\
                  0   &  0
                      \end{array}
                \right),
{\bf{B_3}}=\left(\begin{array}{rcl}
                  0   &  0\nonumber\\
                  1   &  0
                      \end{array}
                \right),
\end{equation}
and the adjoint representation is given by
$[{\bf{B_1}},{\bf{B_2}}]=2{\bf{B_2}}$ and
$[{\bf{B_1}},{\bf{B_3}}]=-2{\bf{B_3}}$, so there are two roots,
$2$, $-2$.

A finite reflection group is a finite group of linear transformations
of $R^n$, which can be represented by a Dynkin diagram.
That is, the group is generated by reflections through
unit vectors that are all at angles $\pi/n$ from each other. These vectors are
represented by dots in the diagram, so a Dynkin diagram like
\begin{center}\begin{picture}(285,25)(0,0)
\Line(10,9)(210,9)
\Oval(130,9)(2,2)(0)
\Oval(50,9)(2,2)(0)
\Oval(90,9)(2,2)(0)
\Oval(170,9)(2,2)(0)
\end{picture}\end{center}
corresponds to a reflection group having one generator for each dot, such that
$r^2=1$ for each generator ($\sim$ reflections).
Not all Dynkin diagrams correspond to finite groups, but
those denoted by $A_n,B_n, C_n, D_n, E_6, E_7, E_8, F_4, G_2$, do, and
they moreover correspond to lattice symmetries.
The dots act as a basis of the lattice and the links keep track of the
angles between the basis vectors, which are the roots.

Lie algebras of compact simple Lie groups are also classified by Dynkin
diagrams. Then $A_n$ is identified as the Lie algebra corresponding to
$sl_{n+1}(C)$, i.e. (n+1)x(n+1) traceless matrices. The compact real form
of $sl_n(C)$ is $su_n$, and the corresponding compact Lie group is
$SU(n)$.\\
\\

{\bf{II. Projective geometry}}\\
Projective geometry enables us to view geometry as a whole. It is based on
the concept of projective transformation. If
the points on a line $a$ are projected into the points on a line $b$,
which in their turn are projected into the points on a line $c$, etc.
the last line being $l$, then
each point on $a$ corresponds to a definite point $L$ on $l$.
This corresponds to a projective transformation
\begin{equation}
\{A\} {\bar{\wedge}} \{K\}
\end{equation}
The set of all points on a line $l$ is called the $range$ (or $pencil$
ir $row$) of points. The correspondence of not the whole range, but of
individual points, is written
\begin{equation}
A_1A_2A_3... {\bar{\wedge}} K_1K_2K_3...
\end{equation}

Theorem: Any three points of a straight line can be projected into any
three points of a straight line, i.e. $A_1A_2A_3 {\bar{\wedge}}
K_1K_2K_3$ provided that $A_1A_2A_3$ as well as $K_1K_2K_3$ are distinct and
collinear.

Theorem: Any projective transformation of a line is fully determined
by the fate of the three points on the line. So if $A_j \rightarrow K_j$
for $j=1,2,3$, then for each other point $A_4$ on $a$, there is a
uniquely determined point $K_4$ on $k$ to which $A_4$ will go.

Parallell lines are dealt with in a special way; by introducing
points at infinity. To each straight line there is attributed a point
at infinity, and each line parallell to $a$ meets $a$ in the point
at infinity, of $a$.
The points at infinity of the straight lines of a plane constitute a straight
line at infinity, of this plane.
All the points at infinity of the 3D space constitute the plane at infinity.
\\
\\

{\bf{III.The discrete subgroups $Z_N$}}\\
$Z_2$ = the integers modulo 2;
a mod n = b means that a = np + b, so b is the remainder when a is divided by n.
For example, 8 mod2 = 0, 9 mod2 = 1.
This means that $Z_2$ divides the integers ${\cal{Z}}$ into the two equivalence
classes of even and odd integers.
In the general case, $Z_N$ is generated by a homomorphism of the integers
${\cal{Z}}$ to the factor group $(0,1,...,N)+N{\cal{Z}}$ of the integers
${\cal{Z}}$. So $Z_2 = (0,1)=\{a | a \in(0,1)\}$,
\begin{center}\begin{picture}(220,30)(0,0)
\Line(7,9)(200,9)
\Vertex(25,9){2}
\Vertex(75,9){2}
\Text(25,0)[]{0}
\Text(75,0)[]{1}
\end{picture}\end{center}
is a very discrete group, with addition as group operation:
\begin{eqnarray}
&&0+1=1\nonumber\\
&&0+0=0\nonumber\\
&&1+1=2=0(mod2) \nonumber
\end{eqnarray}
From $Z_2$ we can compose $Z_2^2=(0,1)\otimes (0,1) =
\{(a,b) | a,b \in(0,1)\}$,
\begin{center}\begin{picture}(220,100)(0,0)
\Line(7,9)(170,9)
\Line (25,4)(25,100)
\Line(75,9)(75,59)
\Line(25,59)(75,59)

\Vertex(25,59){2}
\Vertex(75,59){2}
\Vertex(25,9){2}
\Vertex(75,9){2}
\Text(23,0)[]{(0,0)}
\Text(77,0)[]{(1,0)}
\Text(77,67)[]{(1,1)}
\Text(23,67)[]{(0,1)}
\end{picture}\end{center}
{\vspace{6mm}}
and $Z_2^3=(0,1) \otimes (0,1)\otimes(0,1)=
\{(a,b,c) | a,b,c \in(0,1)\}$,
\begin{center}\begin{picture}(220,130)(0,0)
\Line(0,20)(5,20)
\DashLine(5,20)(75,20){2}
\Line(75,20)(120,20)
\Line(25,0)(25,5)
\DashLine(25,5)(25,70){2}
\DashLine(25,20)(5,5){2}
\Line(25,70)(25,120)
\Line(75,20)(75,70)
\Line(25,70)(75,70)

\Line(5,5)(5,55)
\Line(5,5)(55,5)
\Line(55,5)(55,55)

\Line(55,55)(75,70)
\Line(55,5)(75,20)
\Line(5,55)(55,55)
\Line(5,55)(25,70)
\Vertex(5,5){2}
\Vertex(25,70){2}
\Vertex(75,70){2}
\Vertex(25,20){2}
\Vertex(75,20){2}
\Vertex(5,55){2}
\Vertex(55,55){2}
\Vertex(55,5){2}
\end{picture}\end{center}
{\vspace{8mm}}
$SO(3)=SU(2)/Z_2= \{ rZ_2, r \in SU(2) \}$
amounts to identifying the elements.
The elements of $SU(2)$ are $e^{ia^j\sigma_j}$ where $a^j \in [0, 2 \pi]$
for a given normalization.
In the Lie algebra, $0$ corresponds to the identity group element, since
$e^{i0}={\bf{1}}$, and $2 \pi$ corresponds to $-{\bf{1}}$, and
$({\bf{1}},-{\bf{1}})$ is the center of $SU(2)$, i.e. the maximum set
of elements that commute with all other elements in the group.

When creating explicit representation of $SU(2)/Z_2$, we use
a representation of $Z_2$ which has the same group operation as $SU(2)$,
that is, matrix multiplication. So with
\begin{eqnarray}\label{ztwo}
Z_2 = \{ \left(\begin{array}{rcl}
                  1   &{\hspace{3mm}}  0\nonumber\\
                  0   &{\hspace{3mm}}  1\nonumber
                     \end{array}
                \right),
                \left(\begin{array}{rcl}
                 -1   &  0\nonumber\\
                  0   &  -1\nonumber
                     \end{array}
                \right) \},
SU(2)/Z_2 = \left[ g\left\{ \left(\begin{array}{rcl}
                  1   &{\hspace{3mm}}  0\nonumber\\
                  0   &{\hspace{3mm}}  1\nonumber
                     \end{array}
                \right),
                \left(\begin{array}{rcl}
                 -1   &  0\nonumber\\
                  0   &  -1\nonumber
                     \end{array}
                \right) \right\}| g \in SU(2)\right]
\end{eqnarray}

\noindent We can view $Z_2^N$ projective geometrically:
All the non-vanishing elements of $Z_2^N$ are basically rays in $Z_2^N$ because in the $Z_2$-field there is only one non-vanishing
element, so a ray has only one element.
If $P_d(Z_2)$ is the projective space of $d$ dimensions over the
field $Z_2$, $P_d(Z_2)=\{rays \in Z_2^{d+1}\}$. For example
\begin{eqnarray}
P_2(Z_2)=\{rays \in  Z_2^3\}\nonumber
\end{eqnarray}
The number of elements in $Z_2^3$ = $2^3=8$. The number of rays is
equal to the number of non-vanishing elements in $Z_2^3$, i.e. $8-1=7$.
Thus the number of points in the projective plane over $Z_2^3$ is 7.

%% Don
\newtheorem{thm1}{Theorem}

\title*{%
Why so Few Particle Species? Why Only Fermions \&  Bosons? 
Why  U(1), SU(2) \& SU(3) Symmetries?\\  (A Speculative Model)\thanks{%
Editors' note: This contribution was intended for the Holger Bech Nielsen's
Festschrift (Vol. 1 of this Proceedings), but was received late, so
we include it here.}}
\author{D.L. Bennett and A. Kleppe}
\institute{%
Brookes Institute for Advanced Studies,
B\o gevej 6, 2900 Hellerup, Denmark}  

\authorrunning{D.L. Bennett and A. Kleppe}
\titlerunning{Why so Few Particle Species}
\maketitle

\section{Holger Bech Nielsen}

Happy Birthday dear Holger and congratulations with your 60 years day. Thank you
for many exciting years of work together. Because this is a contribution to a 
``festschrift'' for you, we shall let you escape this once from being a 
co-author (which has the advantage that you are not strictly accountable
for the contents).

\section{Introduction}

Elementary particles come as fermions and bosons. Maybe there are even fewer
categories than these two, maybe we could perceive fermions and bosons as
different modi of one sole type of particle. 

Or is it the other way round? While the standard picture of the physics at the
fundamental scale is one of oneness and simplicity, in the Random Dynamics 
scheme \cite{randyn1,randyn2,frogniel}, the assumption is that at the Planck level,
or at an even more fundamental level, there is a multitude of individual 
particles. Because of the lack of information that characterizes our low energy
world, these are perceived by us as the few species of identical fermions 
or bosons 
described by particle physics.

Here we outline how this complex world of non-identical particles at the
fundamental level ends up as a few different species of fermions or bosons
having symmetries that are subgroups of the Standard Model group at 
energies accessible to the poor physicist. After first 
introducing large exchange forces, we see that particles, 
by assumption all different fundamentally, are effectively all identical
at low energies. Some diversity reemerges when we examine the consequences
of the generic structure of Young diagrams and the so-called HOMO-LUMO gap 
effect. Column endings of Young tableaux function as different isolated
Fermi surfaces and the action of
the HOMO-LUMO gap effect "bunches" these Fermi surfaces together into what
we perceive as $SU(N)$ multiplets.

\section{Toy Model}

To illustrate the idea, let us consisder a universe consisting of $n=2$ different particles.
Assume furthermore that there are two single particle states 
(wave packets/orbits) $\Psi_1$ and $\Psi_2$.
Then there are 2! possible 2-particle states

\begin{equation} \Psi_1(1)\Psi_2(2)  \mbox{   and    }  \Psi_1(2)\Psi_2(1) \end{equation}

which for later convenience are assumed to be normalized to unity.\\
Now assume that the Hamiltonian $\cal{H}$ contains  a very strong exchange term

\begin{equation} K_{Pl}\left( \begin{array}{cc}0&1\\1&0 \end{array} \right) \hat{=}\; \hat{\cal{O}}_{exch} \end{equation} 

where $K_{Pl}$ is a large coefficient of the order of the Planck mass.

Initially assume that the single particle states  $\Psi_1$ and $\Psi_2$ are close together in phase space
so that the exchange force term in the Hamiltonian dominates.

Solving for eigenstates of the exchange operator $\hat{\cal{O}}$:

\begin{equation} \det \left( \begin{array}{cc} -\lambda&K_{Pl}\\K_{Pl}&-\lambda \end{array}\right) = 0 \end{equation}

yields

\begin{equation} \lambda = \pm K_{Pl}. \end{equation}

Only one of these  eigenstates will be accessible to us ``poor physicists'',
living at our (low) energy scale.
Which one depends on the sign of $K_{Pl}$.

Now let us project the set $\{\Psi_1(1)\Psi_2(2), \Psi_1(2)\Psi_2(1)\}$ of 2-body states onto the
irreducible representations of the symmetric group $S_2$. There are  
two irreducible representations corresponding to the Young 
tableaux $S_{ij}$ and
$A_{ij}$ (see Figure~\ref{one}) 
that correspond respectively to the symmetric and antisymmetric 
eigenstates of the 
exchange operator $\hat{\cal{O}}$:

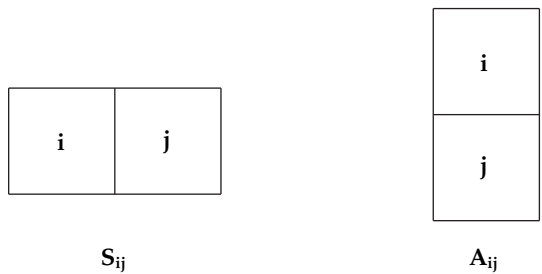
\begin{figure}  
\begin{center}\begin{picture}(285,140)(0,0)
\Line(40,40)(120,40)
\Line(40,80)(120,80)
\Line(80,80)(80,40)
\Line(40,80)(40,40)
\Line(120,80)(120,40)
\Text(80,15)[]{${\bf{S_{ij}}}$}
\Text(60,60)[]{${\bf{i}}$}
\Text(100,60)[]{${\bf{j}}$}

\Line(200,110)(200,30)
\Line(240,110)(240,30)
\Line(240,110)(200,110)
\Line(200,70)(240,70)
\Line(200,30)(240,30)
\Text(220,90)[]{${\bf{i}}$}
\Text(220,50)[]{${\bf{j}}$}
\Text(220,15)[]{${\bf{A_{ij}}}$}

%\Text(152,0)[]{${\bf{[Fig{\hspace{2mm}}1]}}$}
\end{picture}\end{center}
\caption{\label{one} The Young tableaux corresponding to the symmetric
and anti-symmetric irreducible representations of $S_2$.}
\end{figure}

\begin{equation} S_{12}=\frac{1}{\sqrt{2}}(\Psi_1(1)\Psi_2(2),\Psi_1(2)\Psi_2(1))\left( \begin{array}{c}1\\1 
\end{array}
\right) = \frac{1}{\sqrt{2}}\left(\begin{array}{c}1\\1\end{array} \right)\label{exchs}\end{equation} 

\begin{equation} A_{12}= \frac{1}{\sqrt{2}}(\Psi_1(1)\Psi_2(2),\Psi_1(2)\Psi_2(1))\left( \begin{array}{c}{\hspace{3mm}}1\\-1 \end{array}
\right)=\frac{1}{\sqrt{2}}\left(\begin{array}{c}{\hspace{3mm}}1\\-1\end{array} \right)\label{excha}\end{equation} 

These are written in the basis 

\begin{equation} \left(\begin{array}{c}1\\0\end{array}\right) = (\Psi_1(1)\Psi_2(2),\Psi_1(2)\Psi_2(1)) 
 \left(\begin{array}{c}1\\0\end{array}\right)
\label{meas1}\end{equation}

\begin{equation} \left(\begin{array}{c}0\\1\end{array}\right) = (\Psi_1(1)\Psi_2(2),\Psi_1(2)\Psi_2(1)) 
 \left(\begin{array}{c}0\\1\end{array}\right)
\label{meas2}\end{equation}

corresponding respectively to the eigenstates of the system when a measurement reveals the system to 
be in the state 
$\Psi_1(1)\Psi_2(2)  \mbox{   or the state   }  \Psi_1(2)\Psi_2(1)$.
In the basis of the eigenstates of the exchange operator $\hat{\cal{O}}$ (i.e., (\ref{exchs}) and  (\ref{excha}) 
above), the eigenstates (\ref{meas1}) and (\ref{meas2}) are

\begin{equation}  \textstyle\left(\begin{array}{c}1\\0\end{array}\right)=
\frac{1}{2}\left(\begin{array}{c}1\\1\end{array}\right)+\frac{1}{2}\left(\begin{array}{c}{\hspace{3mm}}1\\-1\end{array}\right)
\end{equation}
\begin{equation}  \left(\begin{array}{c}0\\1\end{array}\right)=
\frac{1}{2}\left(\begin{array}{c}1\\1\end{array}\right)-\frac{1}{2}\left(\begin{array}{c}{\hspace{3mm}}1\\-1\end{array}\right)
\end{equation}

Consider now the uncertainty in the exchange force energy if we, 
for example, want to know that particle 1 is in the  state 
$\Psi_1$ and particle 2 is in the state $\Psi_2$ corresponding to the eigenstate
$\left(\begin{array}{c}1\\0\end{array}\right)$:

\begin{equation} \Delta \hat{\cal{O}}_{exch} = K_{Pl}
\left( (1,0)\left(\begin{array}{cc}0&1\\1&0\end{array}\right)^2\left(\begin{array}{c}1\\0\end{array}\right)-
\left[(1,0)\left(\begin{array}{cc}0&1\\1&0\end{array}\right)\left(\begin{array}{c}1\\0\end{array}\right)\right]^2
\right)^{\frac{1}{2}}=K_{Pl}.
\end{equation}
It is seen that if we want to determine that the system is in the state 
$\Psi_1(1)\Psi_2(2)$, the cost in energy
is of the order of the Planck mass. ``Poor physicists'' could therefore never 
afford to establish the
individual identity of the particles 1 and 2 (even though they are by 
assumption non-identical  at the fundamental scale which is here taken 
to be of the order of the Planck mass). The ``poor physicist'' can 
therefore regard the two particles as being $identical$ (and,
moreover, in this simple example, as fermions $or$  bosons).

Initially we have considered the effect of the assumed strong exchange force 
when the single particle states
$\Psi_1$ and $\Psi_2$ essentially $coincide$ in phase space. Hence there is an
approximate $\it{symmetry}$ under $\hat{\cal{O}}_{exch}\propto\sigma_x$; the 
Hamiltonian  $\cal{H}$ must therefore essentially  commute
with $\sigma_x$. The contribution to $\cal{H}$ from e.g. $\sigma_z$ must 
accordingly vanish. Denote by {\bf A} the position in phase space at which the
single particle states $\Psi_1$ and $\Psi_2$
coincide. It has been argued that the strong exchange force singles out one of 
the two irreducible representations
$S_{ij}$ or $A_{ij}$ (corresponding to bosons or fermions) that is accessible 
to ``poor physicists''.

Imagine now displacing $\Psi_1$ and $\Psi_2$ to non-coinciding positions in 
phase space so that the
invariance under $\hat{\cal{O}}_{exch}\propto\sigma_x$ is no longer expected. 
Let us subsequently
imagine that we again move the states $\Psi_1$ and $\Psi_2$ so that these  
once more become coincident at
some new position {\bf B} in phase space. Now we ask whether we can expect 
the system to ``remember''
upon reunion at {\bf B} that it started out at {\bf A} as fermionic 
(or bosonic)? The answer is, generally speaking, yes.

When the single particle states $\Psi_1$and $\Psi_2$ are separated, it is 
really not meaningful to ask
whether particles 1 and 2 are fermions or bosons, but generally the 
``reunited'' system at
{\bf B} will have the same statistics (fermi or bose) as it had at {\bf A}.

The argument for this goes as follows. Think of displacing the system from {\bf A} to {\bf B} in such
a way that single particle states $\Psi_1$ and $\Psi_2$ remain essentially coincident along the way.
Then, unless the eigenvalues of $\hat{\cal{O}}_{exch}$ become degenerate somewhere along the way from {\bf A} to
{\bf B}, the exchange force maintains the same statistics for the entire journey. In particular,
the ``path'' followed  in phase space from {\bf A} to {\bf B} is inconsequential;
whenever  $\Psi_1$ and $\Psi_2$ again become coincident
in phase space, the exchange force works in the same and the statistics (fermi or bose) is ``conserved''
$\it{unless}$ a degeneracy surface/line is encountered along the way.

\section{What is a generic Young Diagram?}
Assuming that we have a system of $n$ (different) particles, we ask what a 
typical Young diagram looks like. To this end, consider a Young diagram with $C$ columns and label by $\lambda_i$ the number
of rows in the $i$th column. Then the total number of boxes in the Young diagram is just

\begin{equation} \sum_{i=1}^C \lambda_i \hat{=} \lambda \end{equation}

Now consider a canonical ensemble of Young diagrams. For a Young diagram with $\lambda$ boxes we assign a weight
$\exp(-\xi \lambda)$ where $\xi$ is determined so that

\begin{equation} n=<\lambda>=\frac{\sum_{all\;Young\;d.} \lambda e^{-\xi \lambda}}{\sum_{all\;Young\;d.}e^{-\xi \lambda}}. \end{equation}

It follows that the distribution of the different column lengths 
$\lambda_i$ are proportional to $\exp(-\xi \lambda)$.
Now rewrite $\lambda=\sum_{i=1}^C \lambda_i$ as

\begin{equation} \lambda=(\lambda_1-\lambda_2)\cdot 1+(\lambda_2-\lambda_3)\cdot 2+(\lambda_3-\lambda_4)\cdot 3+\cdots
=\sum_{i=1}^C (\lambda_i-\lambda_{i+1})\cdot i \end{equation}

where $\lambda_{C+1}=0$.

Now calculate the average difference in the length of  (i.e., number of rows in) two adjacent columns 
$\lambda_i$ and $\lambda_{i+1}$:

\begin{equation} <\lambda_i-\lambda_{i+1}>=
\frac{\sum_{all\;Young\;d.}(\lambda_i-\lambda_{i+1}) e^{-\xi \lambda}}{\sum_{all\;Young\;d.}e^{-\xi \lambda}} \end{equation}

Use now the approximation in which $<\lambda_i-\lambda_{i+1}>$ is taken as being equal to the value of
$\lambda_i-\lambda_{i+1}$ that maximizes  $<\lambda_i-\lambda_{i+1}>$:

\begin{equation} \frac{d<\lambda_i-\lambda_{i+1}>}{d(\lambda_i-\lambda_{i+1})}  =
\frac{\sum_{all\;Young\;d.} (1-\xi(\lambda_i-\lambda_{i+1})\cdot i) e^{-\xi \lambda}}{\sum_{all\;Young\;d.}e^{-\xi \lambda}}
=0 \end{equation}
or

\begin{equation}  <\lambda_i-\lambda_{i+1}>  \approx (\lambda_i-\lambda_{i+1} )_{max}=\frac{1}{\xi\cdot i} \end{equation}

%For ``poor physicists'', column endings correspond effectively to fermions close to a Fermi surface
%located at the column ending. This is because he can only afford hole excitati%ons that are near such an effective Fermi surface.

\begin{figure}
\begin{center}\begin{picture}(250,240)(0,0)
\Line(80,6)(100,6)
\Line(100,110)(120,110)
\Line(120,180)(140,180)
\Line(80,220)(140,220)
\Line(80,220)(80,6)
\Line(100,110)(100,6)
\Line(120,180)(120,110)
\Line(140,220)(140,180)
\Text(91,0)[]{$\lambda_1$}
\Text(112,103)[]{$\lambda_2$}
\Text(133,173)[]{$\lambda_3$}
\Text(75,57)[]{$\displaystyle\left \{ \begin{array}{c}\\ \\ \\ \\ \\ \\ \\ \end{array}\right. $}
\Text(75,147)[]{$\displaystyle\left \{ \begin{array}{c}\\ \\ \\ \\ \\ \end{array}\right. $}
\Text(45,57)[]{$\sim 1/\xi$}
\Text(45,147)[]{$\sim 1/2\xi$}
\Text(45,203)[]{$\sim 1/3\xi$}
\Text(75,203)[]{$\displaystyle\left \{ \begin{array}{c} \\ \\ \end{array}\right. $}
\end{picture}\end{center}
\caption{\label{two} The figure is intended to suggest that, 
in a generic
Young diagram, column endings at $\lambda_1$, $\lambda_2, \cdots $ are
widely separated and that just one column (of plaquettes) ends at a 
given $\lambda_i$.}

%\Text(250,0)[]{${\bf{[Fig{\hspace{2mm}}2]}}$}
\end{figure}
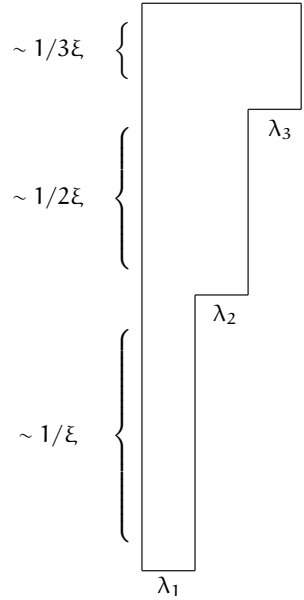

We see that for a generic Young diagram, only one column ends at a given row 
number and the distance between column endings is typically large (see 
Figure~\ref{two}).

For ``poor physicists'', each column ending corresponds to a 
different fermion species 
associated with the effective  Fermi surface located at the column ending. 
This is because he can 
only afford hole excitations that are near the termination of a column 
and not so
deep down in the Fermi sea as to be near the next column ending. 
So because the ``poor physicist'' is restricted to only working 
very close to a column ending, he
is prevented energetically from doing a comparison of particles 
associated with (generically)
widely separated column endings (i.e., effective Fermi surfaces) 
that would establish
them as being of the same species. 

In summary, we have first argued (Section 3) that the effect of strong exchange forces is to render fundamentally different particles  effectively identical at low energies. Subsequently (Section 4) we have argued that the generic structure of a Young diagram effectively insures a few different species corresponding to the widely separated column terminations of a typical Young diagram.

\section{The HOMO-LUMO gap}

Inspired by the Jahn-Teller Effect observed in nuclear physics and the
HOMO-LUMO\footnote{HOMO stands for Highest Occupied Molecular Orbital and
LUMO stands for Lowest Unoccupied Molecular Orbital} gap 
observed in the energy levels of molecular orbitals,
we put forth a generalization suggestively referred to as  the HOMO-LUMO
Gap Effect. Our generalization states that for any system of fermions, the
energy interface between occupied and unoccupied energy levels 
(i.e., the Fermi surface) tends to occur at an energy $E_{FS}$ 
for which the density of states 
$\rho(E_{FS})$ is low. This
depletion of the energy level density at the Fermi surface has the effect
of pushing unoccupied
energy levels to higher energies (at no ``cost'' in energy) concurrent with
pushing occupied levels to lower energies (thereby yielding a lower
energy for the system). Our claim is that there is a universal tendency
for fermion systems to have available degrees adjusted in such a way that
a HOMO-LUMO gap is formed.

\begin{figure}
\begin{center}\begin{picture}(405,245)(0,0)
\Text(55,80)[]{$\leftarrow$$magic{\hspace{1mm}} number$}
\Text(55,150)[]{$\leftarrow$$magic{\hspace{1mm}} number$}

\LongArrow(148,115)(205,115)
\Text(180,103)[]{$Homo-Lumo$}

\Text(140,160)[]{empty}
\Text(140,70)[]{filled}

\Line(4,200)(75,200)
\Line(4,190)(75,190)
\Line(4,180)(75,180)
\Line(4,170)(75,170)

\Line(4,130)(75,130)
\Line(4,120)(75,120)
\Line(4,110)(75,110)
\Line(4,100)(75,100)

\Vertex(30,50){2}
\Vertex(30,60){2}
\Vertex(30,100){2}
\Vertex(30,40){2}
\Vertex(30,30){2}
\Vertex(30,110){2}

\Line(4,30)(75,30)
\Line(4,40)(75,40)
\Line(4,50)(75,50)
\Line(4,60)(75,60)

\Text(340,175)[]{empty}
\Text(340,55)[]{filled}
\Text(320,175)[]{$\displaystyle\left \} \begin{array}{c} \\ \\ \\ \\  \end{array}\right. $}
\Text(320,55)[]{$\displaystyle\left \} \begin{array}{c} \\ \\ \\ \\  \end{array}\right. $}

\Text(113,72)[]{$\displaystyle\left \} \begin{array}{c} \\ \\ \\ \\ \\ \\  \end{array}\right. $}
\Text(113,160)[]{$\displaystyle\left \} \begin{array}{c} \\ \\ \\ \\ \\ \\  \end{array}\right. $}
\Line(224,200)(295,200)
\Line(224,190)(295,190)
\Line(224,180)(295,180)
\Line(224,170)(295,170)
\Line(224,160)(295,160)
\Line(224,150)(295,150)

\Vertex(240,50){2}
\Vertex(240,60){2}
\Vertex(240,70){2}
\Vertex(240,40){2}
\Vertex(240,30){2}
\Vertex(240,80){2}

\Line(224,80)(295,80)
\Line(224,70)(295,70)
\Line(224,30)(295,30)
\Line(224,40)(295,40)
\Line(224,50)(295,50)
\Line(224,60)(295,60)
\end{picture}\end{center}

\begin{center}\begin{picture}(405,115)(0,0)
\Text(60,40)[]{Sphere}
\Text(280,50)[]{Ellipsoid}

\Oval(60,100)(45,45)(0)
\LongArrow(143,100)(200,100)

\Oval(280,100)(25,65)(0)
%%\Text(200,10)[]{${\bf{[Fig{\hspace{2mm}}3]}}$}
\end{picture}\end{center}

\caption{\label{three} Highest occupied states are pushed down in 
energy (and
lowest unoccupied states are push up in energy) concurrent with a breaking of 
spherical symmetry when the filling of states stops within a bunch of states
that otherwise would be degenerate under the symmetry.}
\end{figure}
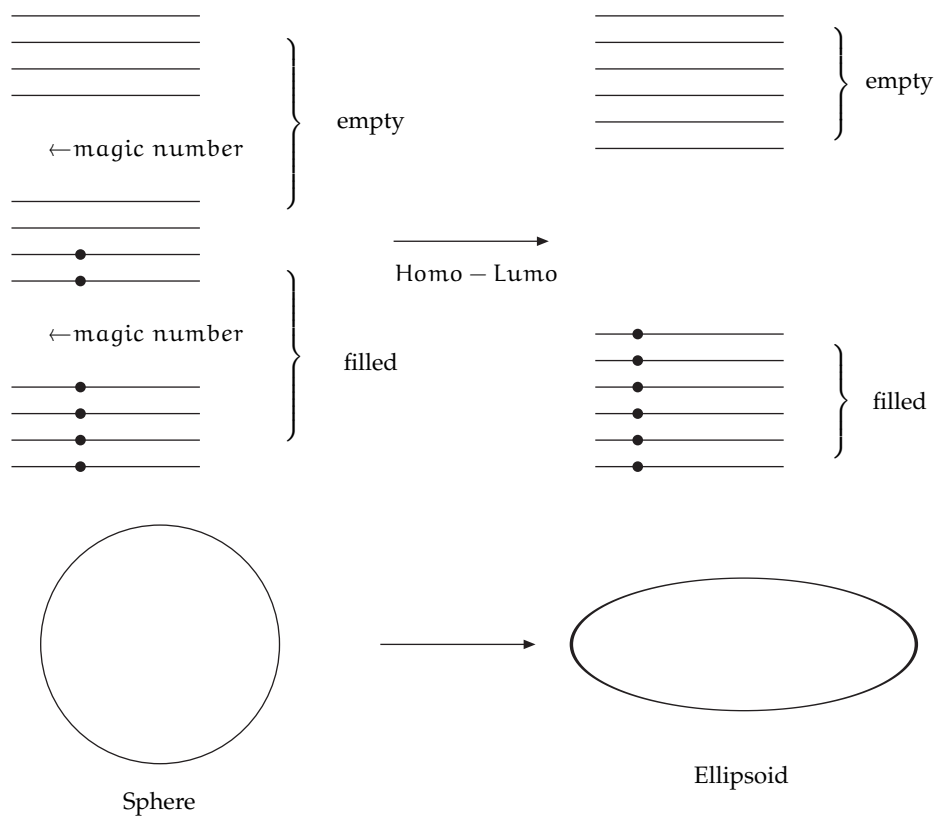

The Jahn-Teller Effect is observed in nuclei for which the highest occupied
nucleon single particle state falls between two magic numbers (see
Figure~\ref{three}). For the filled
states of a nucleus, there are energy gaps at magic numbers that separate
bunches (shells) of levels degenerate with respect to some symmetry.
Especially the rotational symmetry (spherical symmetry) of the nucleus
is important for the level bunching. So if the Fermi surface for
neutrons or protons falls between a pair of  magic numbers, i.e.,  within a
shell, a HOMO-LUMO gap is observed between these magic numbers at the Fermi
surface. The appearence of this extra gap 
(i.e., a gap  not at a magic number) is 
accompanied by the adjustment of available degrees of freedom in such a
way that  the rotational symmetry of the
nucleus is spontaneously broken and the level degeneracy hereby lifted. 
  
\begin{figure}

\begin{center}\begin{picture}(400,170)(0,0)
\Text(68,170)[]{A symmetric molecule}
\Text(68,25)[]{${\bf{Degenerate{\hspace{1mm}} orbits}}$}
\Text(260,25)[]{${\bf{Degeneracy{\hspace{1mm}}lifted}}$}
\Text(260,170)[]{A skew molecule}
\DashLine(200,80)(280,110){2}
\Line(200,80)(250,60)
\Line(250,60)(280,110)
\Line(250,60)(260,140)
\Line(200,80)(260,140)
\Line(260,140)(280,110)
\Vertex(200,80){3}
\Vertex(250,60){3}
\Vertex(280,110){3}
\Vertex(260,140){3}
\LongArrow(140,105)(170,105)
\Line(10,80)(60,50)
\Line(60,50)(110,80)
\Line(60,50)(68,135)
\Line(10,80)(68,135)
\Line(110,80)(68,135)
\Vertex(68,135){3}
\Vertex(10,80){3}
\Vertex(110,80){3}
\Vertex(60,50){3}
\DashLine(10,80)(110,80){2}
%\Text(160,0)[]{${\bf{[Fig{\hspace{2mm}}4]}}$}
\end{picture}\end{center}
\caption{\label{four} In a skewed molecule (e.g., having unequal 
bond lengths)
a symmetry that otherwise would lead to degenerate states at the Fermi surface
is spontaneously broken in such a way that occupied states are pushed to
lower energies and unoccupied state to higher energies. This is the HOMO-LUMO
Gap Effect.} 
\end{figure}
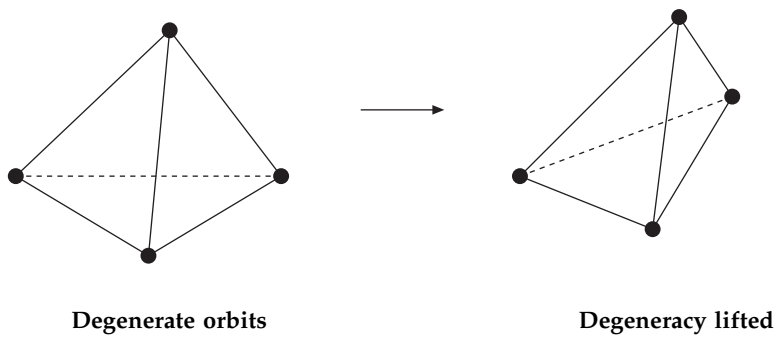

The scenario for having a HOMO-LUMO gap in the spectrum of molecular orbitals 
parallels that for the Jahn-Teller Effect: the molecule is deformed in such
a way that a molecular symmetry is broken (see Figure~\ref{four})  
with the result that 
states that
would have been degenerate under the symmetry are split by a HOMO-LUMO gap.
This happens if 
the filling of states stops within a grouping of several orbitals that would
otherwise be degenerate under some (typically) non-Abelian discrete symmetry
(e.g., permutation symmetry). When this happens,  
the Fermi surface is just here in the midst of several would-be 
degenerate states
and a HOMO-LUMO gap is created by spontaneously
breaking the symmetry responsible for the degeneracy
provided there are available degrees of freedom that can
be adjusted to bring about the symmetry breaking (e.g., equal bonds lengths 
become unequal leading to a skewed molecule - see Figure~\ref{four}.

The HOMO-LUMO Gap Effect - our generalization of the Jahn-Teller Effect in nuclei and the HOMO-LUMO gap observed in molecular
spectra - is stated in the language of fermions: the HOMO-LUMO GAP Effect tends to deplete the density
of fermion states at the Fermi surface by the adjustment 
of available  (background) variables. But the HOMO-LUMO
gap effect is even more general than this insofar as there is no need for the restriction to fermions. There is a
analogous effect for bosons that we could call the 
HOMO-LUMO Gap$^{-1}$ Effect or the HOMO-LUMO Pile-up Effect inasmuch as,
for bosons, there is an accumulation or pile-up of bosons at the 
HOMO-LUMO interface. The name HOMO-LUMO Effect might be 
a more encompassing name. For fermions we would mean a {\it depletion}  of 
states at the HOMO-LUMO interface.  For bosons we would mean
a pile-up or accumulation of states at the HOMO-LUMO interface. 
An interface between differently filled orbits for bosons, not usually 
considered, has been proposed by Ninomiija and
H.B. Nielsen. Their viewpoint is that for negative energy orbits, 
a boson is lacking and thus has occupation number $-1$. So 
the ``pile-up'' comes at the transition between positive energy orbits 
with 0 bosons in the ground state and negative energy orbits occupied 
by $-1$ bosons. Such an effect is used in another contribution to these 
proceeding by C. Froggatt and H.B. Nielsen.  
In what follows, we shall
restrict ourselves to the HOMO-LUMO Effect for fermions, i.e., 
state density depletion at the HOMO-LUMO interface. 

Up until now we have first indicated how particles assumed to be fundamentally
non-identical at the Planck scale 
are, in the presence of large exchange forces,
discerned as being identical at our low-energy level.
In the second step however, we argue that such identical particles 
are categorized into different
types or flavours since what we observe are different species of fermions at 
each of the widely separated column endings 
of a generic Young tableaux.

We want now to claim in the third step that column endings of a 
Young tableau not too widely separated in energy tend to get 
bunched together by the HOMO-LUMO gap 
effect. That is, it may be  energetically favorable to push the Fermi 
surfaces at several (e.g., 2 or 3) adjacent column endings  together 
so that they share a common HOMO-LUMO  gap instead of the 
adjustment of available background variables that would be required in order to
create a HOMO-LUMO gap 
at each column 
ending. This bunching together of the Fermi surfaces at several 
adjacent column endings can mean that fermions,  which in the absence of the 
HOMO-LUMO  gap effect, would be perceived as different species, 
are now in some sense
perceived as different isospin states of the some $SU(N)$ multiplet where N 
is the number of column endings sharing a HOMO-LUMO gap.  
It is more likely that 
several columns  end at the same dip if there are relatively 
few dips in the state density 
as a function of energy. There is a balance between the energy cost of
adjusting available background variables so as to make 
more dips and  adjustments that would press
several column endings of different energy into the same dip.

%Such multiplets  
%(i.e., irredicible representations of $U(N)$)  
%are generated by the irreducible representations of the permutation group 
%$S_N$.  
 
%The reasoning goes as follows.
%The endings of the Young tableaux that we considered in section 3 are at a
%large distance from each other.
%In a realistic Random Dynamics world the number of Young tableau plaquettes
%from one of these Homo-Lumo gaps to the next, is of order of the number of
%fermions in a $\mu_{Planck}$ cut-off Dirac sea for the Universe.

%The Homo-Lumo gap effect suggests that the last particle energy in a column,
%i.e. where it ends, should be near a dip in $\rho(E)$.

%But then 
%In such cases one might really need the $U(N)$ description,
%namely when $N$ column-endings end up in the same dip.
%Can we estimate how often that happens, and how big the $N$
%typically is?  

\section{Getting $SU(N)$ Symmetry}

We will now calculate the number of standard Young tableaux allowed for a given filling 
of states in
a configuration with $n$ hole/(excited)particle pairs  at the effective Fermi surface 
located at the terminus of 
$N$ columns of equal length in the ground state  Young tableaux imagined for 
all particles of the Universe. We then
show that this number of allowed standard Young tableaux is the 
same as the number of $SU(N)$ singlets
that can be obtained by contracting $n$ particles with $N$ 
colors with $n$ anti-particles with $N$ colors.
This allows us to postulate a $SU(N)$ symmetry for the particles occupying states near the terminus of
$N$ columns of a Young tableau of equal length.  

\subsection{Counting the number of standard Young tableaux allowed for a given filling of states in
an excited configuration}

\begin{figure}
\centering
\includegraphics[width=12cm]{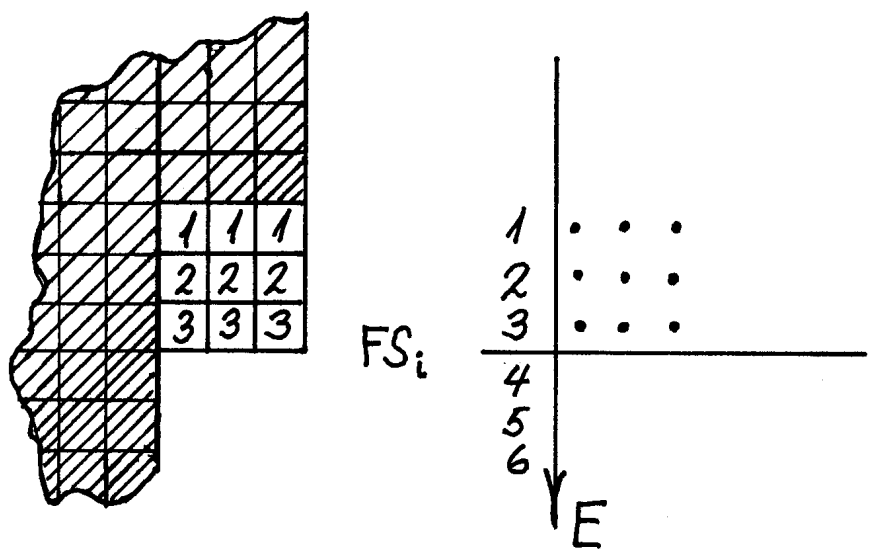}
\caption{\label{groundstate} In the figure on the left, we use as an
example the case of $N_i=3$ columns that have the same length.
This is one of perhaps many 
occurences of $N_j$ columns having the same length in the ground state Young 
tableau for the 
Universe. The  $N_j$ 
columns of the same length
define an effective Fermi surfaces $FS_j$ at the terminius of these $N_j$ 
columns. $N_j$ is expected to be small,  
perhaps 2 or 3,  because column endings are generically widely
separated but can be bunched into small clusters because of the 
HOMO-LUMO gap effect. Only the states
above the $N$ coincident column endings need be considered and of these, 
only those near the column endings. The unimportant 
part of the Young tableau is
cross-hatched in the figure.
In the Figure we consider a Fermi sea that is ``3 states deep''. 
Using a deeper Fermi sea has
no effect on the ensuing arguments.}
\end{figure}

Consisder the prototype situation in which $N$ columns of a Young tableaux have the same length (i.e.,
end at the same effective ``Fermi surface''$FS_i$ say)
We can imagine that $N$ column endings get bunched
together at a local minimum value $\rho(E_{FS_i})$ 
of the density of states  $\rho (E)$ due to
the HOMO-LUMO Gap Effect. The situation is depicted in 
Figure~\ref{groundstate} for $N=3$. The cross-hatched area suggests that
the Young tableau continues above and to the left of the 9 plaquettes
that are numbered by three different states. The poor
physicist need only be concerned with the states near the effective Fermi
surface $FS_i$ defined by the $N_i=3$ columns that all end at $FS_i$. 
That we consider in the example of Figure~\ref{groundstate} 
a Fermi sea that is three states deep is not crucial to the arguments that follow. We could as well have considered a deeper sea. The (cross-hatched) states
to the left of the numbered plaquettes are typically very far below the
next effective Fermi surface. As the poor physicist cannot afford to create a
hole so deep down in the Fermi sea measured relative to the next effective 
Fermi surface,
he can forget about the cross-hatched states to the left of the (accessible)
numbered states in the figure.

The situation of Figure~\ref{groundstate} is imagined to be the 
$i$th occurence of perhaps a 
whole set of similar occurences $\{FS_j,N_j\}$ where $N_j$
columns of equal length end at an effective Fermi surface 
$FS_j$. Each of these $FS_j$ corresponds to a particle species $j$ for
which the poor physicist has access to excitations near $FS_j$ but not
to excitations that would allow him to bring particles from $FS_i$ to $FS_j$
($i \neq j$). Being able to do this would reveal to him that the $i$th and 
$j$th particles were really the same (by the arguments of section 3). 

We have already argued that the generic
stucture of Young diagrams indicates that
the $FS_j$ are expected to be energetically widely separated and that the 
$N_j=1$ is expected. Only
upon invoking the HOMO-LUMO Gap Effect can we argue that some $N_j$ can be 
expected to be slightly larger
than 1 (e.g., 2 or 3). 

In Figure~\ref{groundstate}  the integers 1,2,3 label the different single
partice states in order of increasing energy. This is the ground state and corresponds to the
``clean'' Dirac sea (i.e., no holes and no  excitations of single particle states lying above the Fermi surface $FS_i$). 
The state 1 is occupied by 3 particles.
The same is true for states 2 and 3 as seen in the part of the total  
Young tableau of Figure~\ref{groundstate}
that is labelled. The labelling is in accord with that for a standard Young tableau: a {\it standard Young Tabeaux}
has labels that never decrease in in going from left to right in any row and labels that increase steadily in value
in going from the top to the bottom of any column. A {\it Young frame} is a tabeau with labels removed; i.e.,
a Young frame consists of rows (and also columns of course) of plaquettes without labels in them. A
Young frame and Young diagram are the same thing and we have used these two terms interchangeably.

\begin{figure}
\centering
\includegraphics[width=12cm]{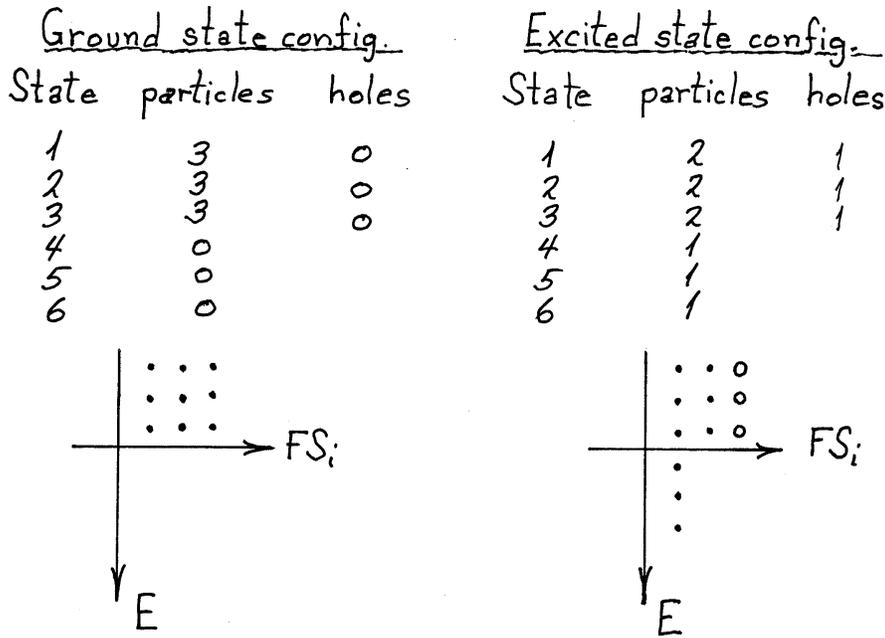}
\caption{\label{filling} In Figure~\ref{filling}a (on the left) 
the ground state filling is shown. There are $N=3$ particles i state $1$,
$N=3$ particles in state $2$ and $N=3$ particles in state $3$. 
$N$ is the number of (equal length) columns that 
all end at the effective Fermi surface labelled ``$FS_i$. 
In Figure~\ref{filling}b (to the right) we show the $n=3$ excited state
configuration that we want to consider: there are $n=3$ different excited states each occupied by one particle and
the $n=3$  Fermi sea states 1,2 and 3 are each occupied by one hole.}
\end{figure}

Now let us consider an excited configuration corresponding to 
the creation of $n$ holes and the occupation by $n$ 
particles of single particle states lying above the Fermi surface $FS_i$.
Let us use $n=3$ as an example. An excited configration  is specified by 
telling how many of the $n$ holes are in each Dirac Sea state and how 
many of the 
$n$ excited particles are in each state above the Fermi Sea $FS_i$. 
 For
brevity we refer to such a specification as a given ``filling'' 
corresponding to an excited configuration. For $n=3$ an example
of an excited configuration ``filling'' is as in Figure~\ref{filling}b. 
There are of course also many other possible
excited configuraion ``fillings'' for $n=3$ hole/excited-particle pairs. 
In Figure~\ref{filling}a we show the (unique) ground state ``filling''. 
The number of holes in  a state in the excited configuration ``filling'' 
is  obtained by comparing
the number of particles in that state in the ground state 
``filling'' and in the excited configuration ``filling''.  

%For a given ``filling'' 
%(recall we use the word ``filling''
%to denote a specification of an excited configuration 
%as, e.g., in Figure~\ref{filling})
 
\begin{figure}
\centering
\includegraphics[width=10cm]{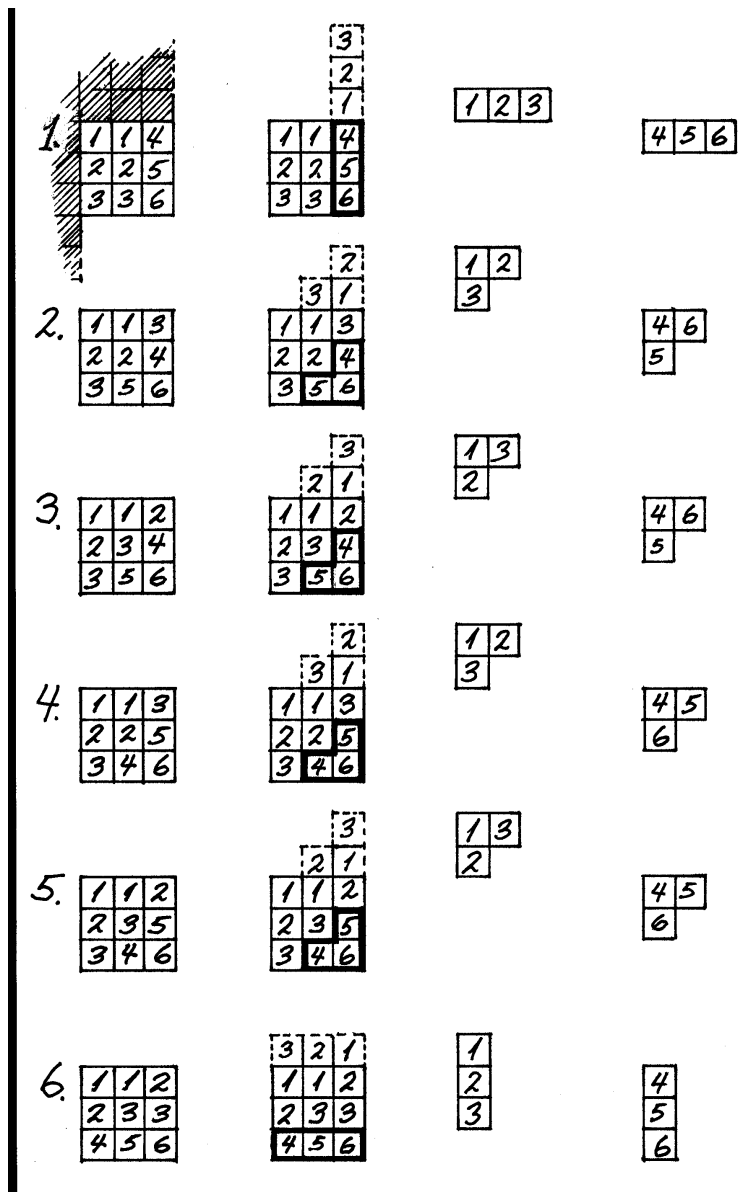}
\caption{\label{tableaux} Seen in the  left column are the  
f-allowed standard Young tableaux (numbered 1 to 6) 
for the ``filling'' corresponding to the configuration with $n=3$ 
holes and $n=3$ excited particles
at the terminus of $N=3$ columns of equal length as shown in
Figure~\ref{filling} In the next column, the (non-standard) 
tablaux for the $n=3$ ``holes and $n=3$ ``excited particles''
are identified respectively by broken-line frames drawn above the 
f-allowed standard Young tableau in question  and a 
bold-line frame within the f-allowed standard Young tableau in question. 
In the third and fourth columns, the (non-standard) respectively ``hole'',
and   
``excited particle'' tableaux identified in the second column are rendered 
in standard
Young tableau form.}
\end{figure}

We want now to construct the set of
standard Young tabeaux consistent with the given ``filling'' 
We shall refer to these
as ``f-allowed standard Young tableaux. We continue to 
use the filling of Figure~\ref{filling} with $n=3$ as an example.  
The possible f-allowed standard Young tableaux for the filling 
of Figure~\ref{filling} are depicted  
in the left column of Figure~\ref{tableaux}. We see that there 
are 6 f-allowed standard Young tableaux for the ``filling''
of Figure~\ref{filling}. These f-allowed standard Young tableaux 
are numbered from 1 to 6 for future reference.

The aim, using our example with $N=3$ and $n=3$, is, for each 
f-allowed standard Young tableaux in the allowed set 
(corresponding to the given ``filling''), to illustrate a correspondence
between such an f-allowed standard 
Young tableaux and a pair of standard Young tableaux (each consisting
of $n=3$ plaquettes) 
where one member of the pair is for the $n$ holes created in the given
``filling'' (i.e., Figure~\ref{filling} in which $n=3$) and the other 
member of the pair
is for the excited particles of the ``filling'' (i.e., the $n=3$ particles
occupying states above the Fermi sea $FS_i$).
Let us refer to these pair members as the ``hole''standard Young tableau and
``excited particle''standard Young tableau respectively. So in the sequel, 
we shall be referring to standard Young
tableaux of three types: {\it f-allowed} standard Young tableaux, 
{\it ``hole''} standard Young tableaux and 
{\it ``excited particle} standard Young tableaux. We now proceed to identify 
the ``hole/(excited)particle'' pair of standard Young tableaux corresponding 
to each f-allowed standard 
Young tableau.

In the next column of Figure~\ref{tableaux}, the f-allowed 
standard Young tableau 
are drawn again but this time we, for each f-allowed standard Young tableau, 
identify an 
(as yet non-standard) ``excited particle'' tableau that is enclosed by 
a bold line frame and a (as yet non-standard) 
``hole'' tableau that is drawn with broken lines above the relevant 
f-allowed standard Young tableau. 
%of the set  
%standard Young tableaux allowed by the ``filling'' specified in 
%Figure~\ref{filling}. 
Each column of the ``hole'' tableau is labelled by the
missing Dirac sea states in that column relative to the ground state 
``filling''. We use the rule that the label of
the missing state of largest negative energy is put in the bottom 
plaquette of the broken-line column, the next to 
largest negative energy missing
state in the next to bottom plaquette, etc. All the ``hole'' 
and all the ``excited particle'' tableaux consist of
$n=3$  plaquettes since the excited configuration of Figure~\ref{filling} 
that we are using as an examnpe has
$n=3$ excited particles and $n=3$ holes. 

Having now identified the (as yet non-standard) ``hole'' and 
``excited particle''  tableau pair  corresponding 
to each allowed standard Young Tableau,
we put these non-standard ``hole'' and ``excited particle'' 
tableax into the standard Young tableau form as follows. 
For the ``hole'' non-standard tableaux,
we reflect the (labelled) plaquettes about the ``SW to NE'' 
diagonal. This yields
the desired  ``hole'' standard Young tableaux as seen in column 3 of 
Figure~\ref{tableaux}. For the ``excited particle''
non-standard tableaux, we reflect the (labelled) plaquettes 
%%%%%%%%%%%%%%%%%%%%%
about
the ``SW to NE'' diagonal and reverse the order of the labels to get 
the standard Young tabeaux form for the ``excited particle'' 
tableaux (shown in Column 4 of
Figure~\ref{tableaux}).

So far we have -  for each f-allowed standard form Young tableau 
belonging to the set of f-allowed standard Young tableaux
for the given ``filling'' -  found a recipe for being able to see that there is a one-to-one correspondence between
each f-allowed standard Young tableau and a pair of standard $n=3$ Young tableaux where the pair consists
of a standard ``hole'' Young tableau and a standard ``excited particle'' Young tableau. These are shown in Columns
3 and 4 respectively og Figure~\ref{tableaux}. 

We now want to use this one-to-one correspondence in conjunction with well known knowledge about the total number of
standard Young tableaux havng $n$ plaquettes ($n=3$ in our example) to tell us the total number of f-allowed standard Young
tableax allowed by a given ``filling''. But first let us finish the special case of our example with $N=n=3$. 

Observe now that there are 3 different Young frames occuring  in Columns 3 and 4 of Figure~\ref{tableaux}. One
type - call it Young frame 1 - has the $n=3$ plaquettes all in one row. Another type - call it Young frame 2 - 
has two plaquettes in the first row and one plaquette in the second row. The third frame type - call it Young
frame 3 - has all $n=3$ plaquettes in the same column. 
  
The f-allowed standard Young tableau ``number 1'' in the first column of Figure~\ref{tableaux}  
 is labelled by one standard Young tableau for the ``holes'' 
and one standard Young tableau  for the ``excited particles'' both of which are of Young frame type 1 and 
corresponds by convention to the fully symmetrized (singlet) representation of the symmetric group $S_3$ (all
plaquettes in a single row). We denote by the symbol $f_1$ the number of standard Young tableaux
that have the Young frame 1. For the fully symmetrized Young frame we have $f_1=1$ (this is also true for any $n$). 
The f-allowed standard Young tableau ``number 6'' in the first column of Figure~\ref{tableaux} 
is labelled in an analogous fashion by a ``hole'' standard Young tableau and a 
 ``excited particle'' standard Young tableau both of Young frame type 3 corresponding to the ``fully antisymmetrized'' 
(singlet) representation of the symmetric group $S_3$ (all plaquettes in a single column).  We denote by the symbol 
$f_3$ the number of standard Young tableaux
that have the Young frame 3. For the fully antisymmetrized Young frame we have $f_3=1$ (this is also true for any $n$).

Note now that the ``hole'' standard Young tableaux and the ``excited particle'' standard Young tableaux identified
with the f-allowed standard Young tableaux numbers 2,3,4 and 5 in the left column of Figure~\ref{tableaux} all have the type 2 
Young frame - 
namely the Young frame which corresponds to the mixed symmetry case (with two plaquettes in the first row
and one plaquette in the second row). Denote by the symbol $f_2$ the number of standard Young tableaux having 
the type 2  Young frame. For both the ``hole'' and the ``excited particle'' standard Young tableax we see that
$f_2=2$ correponding to the two doublet representations of the 
group $S_3$. Each of the f-allowed standard Young 
tableaux 2,3,4 and 5 in Figure~\ref{tableaux} is in a one-to-one
correspondence with a pair of  standard Young tableaux
namely one of the $f_{2,``hole''}=2$ possible ``hole'' standard Young tableaux and one of the $f_{2,``excited\;part.''}=2$
possible ``excited particle'' standard Young tableaux. For the ``hole'' and the ``excited particle'' standard Young 
tableaux of Young frame type 2, this scheme provides a total of 
$f_{2,``hole''}\cdot f_{2,``excited\;part.''}$ = $2^2$ labels that have 
a one-to-one correspondence to the
four f-allowed standard Young tableaux with numbers 2,3,4 and 5 in Figure~\ref{tableaux}.  

In our example with $n=3$ plaquettes, the total number of labels provided by the two sets of standard Young tableaux
(one set for ``holes'' and one set for ``excited particles'') is

\begin{equation} f_1^2+f_2^2+f_3^2=1^2+2^2+1^2=6=3!\end{equation} 

Due to the one-to-one correspondence between f-allowed standard Young tableaux for the given ``filling''
and the 3! labels that we get  using the ``hole'' and ``excited particle'' standard Young tableaux, we
conclude that the number of standard Young tablaeux allowed by the given ``filling'' is also $n!=3!$. 

We now use a couple of theorems\cite{ber} from group theory about Young frames and standard Young tableaux
to establish the general validity of the result obtained in our example with $n=3$.

\begin{thm1}
The number of standard Young tableaux which belongs to the Young frame with row lengths (measured in plaquettes)
$m_1, m_2, \cdots , m_r$  $ (m_1+m_2+ \cdots +m_r=n)$ with $n$ being the total number of plaquettes is

\begin{equation}
f=n!\frac{\prod_{l<k}(l_i-l_k)}{l_1!l_2!\cdots l_r!} 
\end{equation}

where $l_1=m_1+r-1,l_2=m_2+r-2,\cdots ,l_r=m_r$
\end{thm1}

\begin{thm1}
If the index $j$ enumerates the different Young frames that all have $n$ plaquettes, then
the total number of standard Young tableaux corresponding to all Young frames  made up of $n$ plaquettes
is given by

\begin{equation} 
\sum_j f^2_j=n!
\end{equation}
\end{thm1}
Using these theorems in conjunction with the one-to-one 
correspondence between f-allowed standard Young tableaux
for a $n$ hole/(excited)particle configuration and all the $\sum_j f^2_j=n!$ possible labellings that are possible by
using all possible standard Young tableaux having $n$ plaquettes, we conclude 
that there are $n!$ allowed
standard Young tableaux for a configuration with $n$ excited particles (and $n$ holes). In the event that
$N<n$, some of the standard Young tableaux are not realizable (i.e., those containing row lengths greater
than $N$ in which case the number of allowed standard Young tableaux is less than $n!$).

\subsection{Counting the number of $SU(N)$ singlets that can be obtained by 
contracting $n$ particles and $n$ anti-particles
each having $N$ colors}

Consider now $n$ particles $p_a$ and $n$ anti-particles $\overline{p^a}$ where the index $a$ labels $N$
possible colors. We ask how many $SU(N)$ singlets are obtainable 
by contracting the $n$ particles with the
the $n$ anti-particles. It is rather easy to see that (for $n \leq N$) the number of singlets that can be
produced in this way is $n!$.

\subsection{Postulating SU(N) symmetry} 
 
The number of $SU(N)$ singlets that result from contracting $n$ particles that can have
$N$ different ``colors'' with $n$ anti-particles that can have $N$ different ``(anti) colors'' is 
$n!$ (for $n\leq N$).
The number of f-allowed standard Young tableaux for the excitation of $n$ particles
to states lying above an effective  Fermi surface at the terminus of $N$ column of equal length
somewhere in a  Young tableau for ``the Universe'' is also equal to $n!$ (for $n\leq N$). 

It is left as an exercise for the reader to show that an  equality 
also holds for $N<n$
albeit for a number less than $n!$.

We therefore
postulate that the particles near the effective Fermi surface at the terminus  of $N$ columns 
of equal length fall into $SU(N)$  multiplets and have therefore $SU(N)$ symmetry. Because
such particles are totally confined (i.e. only realized as colorless singlets), our postulate
of $SU(N)$ symmetry can never be challenged: singlets always rotate into singlets. We can postulate $SU(N)$ symmetry
because color is globally zero (a singlet). Note that since we have singlets, or rather only
one state of the theory for each representation, the color rotation symmetry does not predict
any (physical) degeneracies.

\section{Conclusion}

The basic idea is that when the poor physicist only sees a 
certain very small number of the (linear superpositions of) states obtained 
from each other by permutation of the genuinely individual (at Planck scale) 
particles, he will most likely
interprete the particles that he observes as a 
poor physicist as being identical. If he only sees one 
single state for each ``filling'' in an excited configuration, 
he can consistently interprete the particles as being either 
simple bosons or simple fermions.

But if the number of f-allowed states per ``filling'' becomes larger than one,
the picture of what a poor physicist sees becomes slightly more sophisticated.
We have found that, in the case of the Young frame with $N$ columns 
of the same length all ending at an effective Fermi surface, one gets
the right number of states by pretending that the (fermions say) are
\underline{$N$}-plets under an $SU(N)$ symmetry with the restriction that {\em
only singlets are allowed} (a kind of confinement
postulate). The poor physicist will encounter no contradiction in assuming such
an $SU(N)$ symmetry because, with only singlet states allowed in the $SU(N)$ 
symmetry, the assumption will have no predictive power. Any 
Hamiltonian is allowed. 

In the third section we assume a Planck scale universe that, 
in the spirit of Random Dynamics, consists of particles
that are all different (individual). By further assuming that the 
Hamiltonian contains very strong
particle exchange forces, we show in the context of a very simple model 
(i.e., a $n$=2 particle universe), that these particles
are perceived as being either identical fermions
or identical bosons  at energies accessible to experiment provided that, in some cases (when more than
one Young frame column have the same length)  we include
a ``color description'' to take into account the degrees of freedom due to a low energy vestige
of the fundamental individuality at the Planck scale.   

Again inspired by Random Dynamics, we examine in the fourth section the generic structure
of a Young tableau for an $n$-particle universe. The characteristic feature is that only
one column of a typical Young tableau ends at a given energy and that adjacent (single) column 
endings are widely separated in energy. At low energies and in the case of fermions, these column endings
function as energetically isolated Fermi surfaces. Inasmuch as the only hole excitations accessible
at low energies must lie very near these isolated Fermi surfaces, the ``poor physicist''
perceives the fermions at each surface as a different species. So even for fermions
that are effectively identical as a result of the exchange 
force mechanism described in section 3,
the energetically poor physicist can not afford to make the comparison needed to establish this 
because he does not have access to 
hole excitations so deep within the Fermi sea so as to be near an adjacent Fermi surface (i.e.,
an adjacent column ending). 
 
In section 5 we consider the way in which the HOMO-LUMO gap effect can  be expected to modify
the spectrum of Fermi surfaces corresponding to the  generic Young diagram for a $n$-particle
universe as described in section 4. The HOMO-LUMO gap effect, which is seen in e.g.
the Jahn-Teller effect, is stated in a more general context (e.g., that of Random Dynamics) as the 
energetically favourable adjustment of ``background'' variables that results in the depletion 
of energy level density at a Fermi surface. This has the effect that some small number $N$,  2 or 3 
perhaps, of not too widely separated column endings get pushed together at one effective Fermi surface
(i.e., $N$ columns of a total Young tableau of the Universe assume equal lengths).

In section 5 we argue that particles in states near the $N$  column endings of the same length can be claimed to
be $SU(N)$ multiplets. The claim of having $SU(N)$ symmetry 
cannot be challenged because the $SU(N)$ symmetry is realized globally as 
(confined) $SU(N)$ singlets. This opens the
door for future work that may reveal that this globally colorless $SU(N)$ symmetry can  turn out to be
manifested locally as non-trivial $SU(N)$ representations with the possibility of interactions through
e.g. locally manifested $SU(N)$ gauge fields.

\title*{About Number of Families}
\author{D. Lukman$^1$, A. Kleppe$^2$ and N.S. Manko\v c Bor\v stnik$^{1,3}$}
\institute{%
${}^1$ Primorska Institute for Natural Sciences and Technology, 
C. Mare\v zganskega upora 2, Koper 6000, Slovenia\\
${}^2$ Bjornvn. 52, 07730 Oslo, Norway\\
${}^3$ Department of Physics, University of
Ljubljana, Jadranska 19, 1111 Ljubljana}

\titlerunning{About Number of Families}
\authorrunning{D. Lukman, A. Kleppe and N.S. Manko\v c Bor\v stnik}
\maketitle

\begin{abstract}
Among open questions, on which the Standard electroweak model gives no answers, is the appearance 
of families of spinors, their numbers and the spinors masses. 
We argue for more than three families, following the approach of one 
of us\cite{norma93fam,norma2001,pikaholgernorma2002} and the ref.\cite{okun}.
\end{abstract}

\section{Introductions}
\label{introductionfam}

The Standard electroweak model gives no explanation for either the number of spinor families (fermions) or for spinor
masses (masses of quarks and leptons).
According to the Standard electroweak model the families of spinors manifest anomaly freedom 
(as also the families of antispinors do) as well as
gauge invariance. Each family consists of at first massless left handed weak charged
doublets and right handed weak chargeless singlets
\begin{eqnarray}\label{qq1}
\left[\begin{array}{rcl}
              &\nu_e& \nonumber\\
                &e&
       \end{array}\nonumber
\right]_L{\hspace{3mm}} {\hspace{3mm}} 
\left[\begin{array}{rcl}
              &u& \nonumber\\
              &d& \nonumber
       \end{array}\nonumber
\right]_L {\hspace{3mm}}
\begin{array}{rcl}
              &(\nu_R) & \nonumber\\
                &e_R&
       \end{array}\nonumber
{\hspace{3mm}}{\hspace{3mm}} 
\begin{array}{rcl}
              &u_R& \nonumber\\
              &d_R& \nonumber
       \end{array}\nonumber
.
\end{eqnarray}
(We put the right handed neutrino into parentheses, since the Standard model does not assume its existence. 
One of the reasons for that is that a right handed neutrino would have zero
as value for all the  charges
of the Standard model. If it exists, it could  interact only through  the gravitational
interaction.) Spinors acquire masses due to Yukawa couplings (this is done by an assumption), 
which is a kind of 
an interaction between weak charge
doublets of  Higgs scalars (with respect to the group $SO(1,3)$) and massless spinors. All the current data
are in agreement with the assumption of only 
three families. In ref.\cite{okun}, however, the authors are analyzing possible masses of a fourth family,
which do not contradict the experimental data.

We present  bellow the experimental values for the masses of the three known families of spinors 
as well as the  masses
of a possible fourth family suggested in ref.\cite{okun}) as values, which do not contradict 
the present experimental data.
\begin{equation}\label{values}
\begin{array}{rcl}
&& m_{u_i} = 0.004{\hspace{1mm}} GeV,{\hspace{1mm}} 1.4{\hspace{1mm}} 
GeV,{\hspace{1mm}} 180{\hspace{1mm}} GeV,{\hspace{1mm}} 285 (215)
{\hspace{1mm}} GeV\nonumber\\                  
&& m_{d_i} = 0.009{\hspace{1mm}} GeV,{\hspace{1mm}} 0.2{\hspace{1mm}} 
GeV,{\hspace{1mm}} 6.3{\hspace{1mm}} GeV,{\hspace{1mm}} 
215 (285){\hspace{1mm}} GeV\nonumber\\                  
&& m_{e_i} = 0.0005{\hspace{1mm}} GeV,{\hspace{1mm}} 0.105{\hspace{1mm}} 
GeV,{\hspace{1mm}} 1.78{\hspace{1mm}} GeV,{\hspace{1mm}} 
100{\hspace{1mm}}GeV, \nonumber\\                
{\hspace{1mm}}{\hspace{1mm}}
&& m_{\nu_i} = 10^{-11} GeV,{\hspace{1mm}} 10^{-9} GeV,{\hspace{1mm}}
10^{-8} GeV,{\hspace{1mm}} 50{\hspace{1mm}} GeV.
\end{array}
\end{equation}
The two values in parentheses represent two possibilities for the masses of the fourth generation, 
suggested by ref.\cite{okun}, both being in agreement with the experimental data. 
Values for the neutrino
masses of three generations are just guessed in agreement with the experimental data, since the experimental
data suggest the square of mass differences among families and not really the masses.

There are many proposals in the literature\cite{babu,jarl,gk,gm} for fermions, which
could occur in addition to the known three generations. The most natural and accordingly the most attractive is
to our understanding the idea for the fourth replication of just the three known families, 
with high enough masses, ''allowed'' by 
the ref. (\cite{okun}) and  small enough mixing matrix elements connecting a possible fourth family with the known
three.

Following the approach of one of 
us\cite{norma93fam,norma2001,pikaholgernorma2002}, we discuss in this contribution the results of the  
mechanism, which gives  the
families of quarks and leptons, as well as  the Yukawa couplings. The details of the approach can
be found in ref.\cite{pikaholgernorma2002}. We argue for four generations of quarks and leptons.

\section{Approach unifying spins and charges}
\label{spinsandcharges}

The approach of one of us\cite{norma93fam,norma2001,pikaholgernorma2002}, unifying spins and charges,
is offering mechanisms for both: the appearance of families and accordingly also for the number of families,
as well as for interactions which manifest as the Yukawa couplings.
The approach assumes the unified internal space of not only all the charges but also of the spins within
the group $SO(1,13)$.
All spinors and antispinors of one family of the Standard model appear in this approach as 
just one Weyl spinor of the group $SO(1,13)$ (with right handed weak chargeless neutrino and left handed
weak chargeless antineutrino included). 

Analyzing the
properties of the Weyl spinor with respect to the subgroups $SO(1,3)$, $SU(3),$ $SU(2)$ and the two $U(1)$'s
of the group $SO(1,13)$
(the rank of $SO(1,13)$ is $7$ as it is also the sum of the ranks of $SO(1,3), SU(3), SU(2)$ and
the two $U(1)$),
one easily sees\cite{pikaholgernorma2002} that there are left handed spinors with respect to the
subgroup $SO(1,3)$, which are the $SU(2)$ doublets and either singlets (leptons) or triplets (quarks) 
with respect to $SU(3)$ 
and right handed (with respect to $SO(1,3)$) $SU(2)$ singlets, which again are either singlets or
triplets with respect to $SU(3)$ within one Weyl spinor (one fundamental irreducible representation).
Within the same Weyl spinor one
finds also the right handed  (with respect to $SO(1,3)$) $SU(2)$ doublets which are antitriplets or 
antisinglets with respect to $SU(3)$ and left handed (with respect to $SO(1,3)$) $SU(2)$ singlets, which
are again antitriplets and antisinglets with respect to $SU(3)$. The Weyl multiplet of $SO(1,13)$ has namely
$2^{d/2-1}$ members, which is $2^6=64$ (that is $16$ spinors with spin up and down and $16$ antispinors with spin up
and down), which means that the family includes also the right handed weak chargeless spinors called right
handed neutrinos and the antifamily includes the left handed weak chargeless antineutrino, the antiparticle
of the right handed neutrino. Since there are two $U(1)$ charges in $SO(1,13)$ besides $SO(1,3)$ spins and $SU(3)$
and $SU(2)$ charges, the right handed neutrino and the left handed antineutrino carry a nonzero (one of the two)
hyper charge.

In the approach unifying spins and charges there are two kinds of generators of the Lorentz 
transformations and accordingly of the corresponding spin connections\cite{pikaholgernorma2002}, 
which cause transitions
between right handed weak chargeless singlets and left handed weak charged doublets. One kind of generators
makes transitions within 
a family, another kind makes transitions among the families, both together manifesting accordingly the properties of
the Yukawa couplings and the Higgs doublets. In this contribution we shall not explain how the Lagrange
density for spinors in $d=1+13$ might split in four dimensional subspace into the Lagrange density with
the usual covariant derivative manifesting the interaction with the known gauge fields and the
part manifesting the mass matrices. The reader can find the explanation in ref.\cite{pikaholgernorma2002}. 
We shall only summarize the results, representing them in  mass matrices. We shall further assume that the elements
are real numbers in order to study a general behavior of families.

The approach unifying spins and charges suggests even number of families, the smallest number is equal to four
\begin{eqnarray}\label{qq2}
\left[\begin{array}{rcl}
              &\nu_i& \nonumber\\
                &e_i&
       \end{array}\nonumber
\right]_L{\hspace{3mm}} {\hspace{3mm}} 
\left[\begin{array}{rcl}
              &u_i& \nonumber\\
              &d_i& \nonumber
       \end{array}\nonumber
\right]_L {\hspace{3mm}}
\begin{array}{rcl}
              & \nu_{iR}& \nonumber\\
                &e_{iR}&
       \end{array}\nonumber
{\hspace{3mm}}{\hspace{3mm}} 
\begin{array}{rcl}
              &u_{iR}& \nonumber\\
              &d_{iR}& \nonumber
       \end{array}\nonumber
,
\end{eqnarray}
with $i \in \{1,2,3,4\}$, which is the family number. The approach unifying spins and 
charges suggests, if we assume real matrix elements (we shall do that for simplicity-just to
study general features of mass matrices),  the following 
structure of the Yukawa couplings\cite{pikaholgernorma2002}
\begin{equation}\label{matfam}
{M_4} = \left(\begin{array}{rcl}
                  A_a   &  B_a{\hspace{4mm}}C_a{\hspace{4mm}}D_a\nonumber\\
                  B_a   &  A_a{\hspace{4mm}}D_a{\hspace{4mm}}C_a\nonumber\\
                  C_a   &  D_a{\hspace{4mm}}A_a{\hspace{4mm}}B_a\nonumber\\
                  D_a   &  C_a{\hspace{4mm}}B_a{\hspace{4mm}}A_a
                      \end{array}
                \right),
\end{equation}
with $a$ which stays for $u,d,\nu,e.$
The matrix has the symmetry structure
\begin{equation}%%\label{m2}
\left(\begin{array}{rcl}
                  X   &  Y\nonumber\\
                  Y   &  X\nonumber\\
                      \end{array}
                \right).
\label{2times2}				
\end{equation}
Diagonalizing this mass matrix we obtain the eigenvalues 
\begin{eqnarray}
\lambda_{a_1} &=& (A_a-B_a)-(C_a-D_a),\nonumber\\
\lambda_{a_2} &=& (A_a-B_a)+(C_a-D_a),\nonumber\\
\lambda_{a_3} &=& (A_a+B_a)-(C_a-D_a),\nonumber\\
\lambda_{a_4} &=& (A_a+B_a)+(C_a+D_a),
\label{lambda4}
\end{eqnarray}
the masses, which should be equalized  with the experimental values $m_{a_i}$ of Eq.(\ref{values}),
with $a_i$, which stays for the members $u_i,d_i,\nu_i,e_i$ of the $i$-th family.

One can express  $A_a,B_a,C_a,D_a$ in terms of masses as follows
\begin{eqnarray}
A_a&=&\{(m_{a_4} + m_{a_3}) + (m_{a_2} + m_{a_1})\}/4 \nonumber\\
B_a&=&\{(m_{a_4} + m_{a_3}) - (m_{a_2} + m_{a_1})\}/4 \nonumber\\
C_a&=&\{(m_{a_4} - m_{a_3}) + (m_{a_2} - m_{a_1})\}/4 \nonumber\\
D_a&=&\{(m_{a_4} - m_{a_3}) - (m_{a_2} - m_{a_1})\}/4. 
\label{abcd}
\end{eqnarray}
Taking into account Eq.({\ref{values}})
we find
\begin{equation}
\begin{array}{cccc}
A_u = 116.6 & B_u = 115.899 & C_u = 26.599 & D_u = 25.901 \\
(A_u' = 99.101 & B_u' = 98.399 & C_u' = 9.099 & D_u' = 8.401)  \\
A_d = 55.377 & B_d = 55.2728 & C_d = 52.223 & D_d = 52.127 \\
(A_d' = 72.877 & B_d' = 72.773 & C_d' = 69.723 & D_d' = 69.627) \\
A_e = 25.471 & B_e = 25.419 & C_e = 24.581 & D_e = 24.529 \\
A_\nu = 12.5 & B_\nu = 12.5 & C_\nu = 12.5  & D_\nu = 12.5,  
\end{array}
\end{equation}
with all the values in $GeV.$
The values in parentheses correspond to two possibilities for the allowed masses of a possible fourth 
family.

The ''democratic'' matrix with $A_a = B_a = C_a = D_a$ would lead, as we can immediately see
from Eq.(\ref{lambda4}), to $\lambda_{a_1} = \lambda_{a_2}
=\lambda_{a_3}= 0, \;\lambda_{a_4} = 4A_a\;$. All the matrix elements besides the diagonal one equal to zero means
$\lambda_{a_i}= A_a, \; i=1,2,3,4.\;$ Since the masses of the fourth family are for all the quarks and the leptons
pretty higher than of the other three families, we expect that the mass matrices are closer to the ''democratic'' one
than to the diagonal one, what indeed is happening\footnote{Ref.\cite{fritzsch} introduces the ''democratic''
matrix in the three family case. 
}. One also can notice that if $C=D=0$ the two by diagonal matrices require the doubling of the eigenvalues 
$\lambda_{a_i}$ that is $\lambda_{a_1}= \lambda_{a_2} = (A-B),\; \lambda_{a_3}=\lambda_{a_4} = (A+B)$.

To understand better what more than four families can mean
 (in the proposed approach of one of us  unifying spins and charges
the number of families is equal to $4^k$) let us look at the
case with eight families, again assuming that matrix elements are real numbers. The approach of one of us suggests
the matrices
\begin{equation}\label{m8}
{M}_8 = \left(\begin{array}{rcl}
A_a & B_a{\hspace{4mm}}C_a{\hspace{4mm}}D_a{\hspace{4mm}}E_a{\hspace{4mm}}F_a{\hspace{4mm}}G_a{\hspace{4mm}}H_a\nonumber\\
B_a & A_a{\hspace{4mm}}D_a{\hspace{4mm}}C_a{\hspace{4mm}}F_a{\hspace{4mm}}E_a{\hspace{4mm}}H_a{\hspace{4mm}}G_a\nonumber\\
C_a & D_a{\hspace{4mm}}A_a{\hspace{4mm}}B_a{\hspace{4mm}}G_a{\hspace{4mm}}H_a{\hspace{4mm}}E_a{\hspace{4mm}}F_a\nonumber\\
D_a & C_a{\hspace{4mm}}B_a{\hspace{4mm}}A_a{\hspace{4mm}}H_a{\hspace{4mm}}G_a{\hspace{4mm}}F_a{\hspace{4mm}}E_a\nonumber\\
E_a & F_a{\hspace{4mm}}G_a{\hspace{4mm}}H_a{\hspace{4mm}}A_a{\hspace{4mm}}B_a{\hspace{4mm}}C_a{\hspace{4mm}}D_a\nonumber\\
F_a & E_a{\hspace{4mm}}H_a{\hspace{4mm}}G_a{\hspace{4mm}}B_a{\hspace{4mm}}A_a{\hspace{4mm}}D_a{\hspace{4mm}}C_a\nonumber\\
G_a & H_a{\hspace{4mm}}E_a{\hspace{4mm}}F_a{\hspace{4mm}}C_a{\hspace{4mm}}D_a{\hspace{4mm}}A_a{\hspace{4mm}}B_a\nonumber\\
H_a & G_a{\hspace{4mm}}F_a{\hspace{4mm}}E_a{\hspace{4mm}}D_a{\hspace{4mm}}C_a{\hspace{4mm}}B_a{\hspace{4mm}}A_a
                      \end{array}
                \right).
\end{equation}
Again $a$ stays for the members of a family $u,d,\nu,e$.
These matrices have the symmetry structure of Eq.(\ref{2times2}), as was already the symmetry structure of
the case with fourth family. Now each of block $X$ and $Y$ itself has  the  two by two structure of Eq.(\ref{2times2}).
One then accordingly easily finds the corresponding eigenvalues
\begin{eqnarray}
\lambda_{a_1}&=&\{(A_a-B_a)-(C_a-D_a)\}-\{(E_a-F_a)-(G_a-H_a)\}\nonumber\\
\lambda_{a_2}&=&\{(A_a-B_a)-(C_a-D_a)\}+\{(E_a-F_a)-(G_a-H_a)\}\nonumber\\
\lambda_{a_3}&=&\{(A_a-B_a)+(C_a-D_a)\}-\{(E_a-F_a)+(G_a-H_a)\}\nonumber\\
\lambda_{a_4}&=&\{(A_a-B_a)+(C_a-D_a)\}+\{(E_a-F_a)+(G_a-H_a)\}\nonumber\\
\lambda_{a_5}&=&\{(A_a+B_a)-(C_a+D_a)\}-\{(E_a+F_a)-(G_a+H_a)\}\nonumber\\
\lambda_{a_6}&=&\{(A_a+B_a)-(C_a+D_a)\}+\{(E_a+F_a)-(G_a+H_a)\}\nonumber\\
\lambda_{a_7}&=&\{(A_a+B_a)+(C_a+D_a)\}-\{(E_a+F_a)+(G_a+H_a)\}\nonumber\\
\lambda_{a_8}&=&\{(A_a+B_a)+(C_a+D_a)\}+\{(E_a+F_a)+(G_a+H_a)\}.
\label{lambda8}
\end{eqnarray}
If we now call ${\cal A}_a = (A_a -B_a),\; {\cal B}_a = (C_a -D_a),\;{\cal C}_a = (E_a -F_a),
\;{\cal D}_a = (G_a -H_a),\;$ then the values of Eq.(\ref{abcd}) for $A,B,C,D$ have to be identified with
${\cal A}, {\cal B},{\cal C},{\cal D}\;$, correspondingly and fitted accordingly with the masses of Eq.(\ref{values}).
One easily sees that then the last four values  for $\lambda_{a_i}$ can allways be chosen to  be as large
as needed in order that the four additional families can not be experimentaly seen.

This kind of doubling the number of families can be continued. Due to (by the approach of unifying spins and charges)
suggested symmetry of the mass matrix (Eq.(\ref{2times2}), this symmetry keeps when ever we double the number of families)
the eigenvalues $\lambda_{a_i}$ can easily be found and the lowest four families then 
identified with the four types of values of Eq.(\ref{values}) in a way that all the rest of the eigenvalues acquire
(very) high masses an can not be seen at present energies. In the next step of doubling ${\cal A}_a = 
\{(A_a-B_a)-(C_a-D_a)\},\; {\cal B}_a = \{(E_a-F_a)-(G_a-H_a)\},\; {\cal C}_a = \{(I_a-J_a)-(K_a-L_a)\},\;
{\cal D}_a = \{(M_a-N_a)-(O_a-P_a)\},\;$ with $I,J,K,L,M,N,O,P\;$ the matrix elements of the doubled structure and with
the starting symmetry all the time.

\section{Conclusions and discussions}
\label{conclusions}

Following the approach of one of us\cite{norma93fam,norma2001,pikaholgernorma2002} and the ref.\cite{okun}
we argue in this contribution for an even number of families, in particular {\em for four families, 
with the fourth family
having all the properties, besides the masses, equal to the properties of the known three families.}

The approach (unifying spins and charges) is offering a mechanism for not only the appearance of families, 
which should be an even number, but also for the appearance of the Yukawa couplings and Higgs: It is a spin
connection in higher dimensions than four ( that is the gravitational interaction) which in 
four dimensional subspace might  manifest
as correct mass terms for (otherwise massless) spinors. This approach has in one $SO(1,13)$ multiplet (one Weyl 
spinor) just all the quarks and the leptons (the left handed weak charged doublets and the right handed
weak chargeless singlets, which are either colour charge triplets - quarks - or colour chargeless singlets-leptons),
as well as all the antiquarks and all the antileptons,  with right handed weak chargeless neutrinos and
the left handed weak chargeless antineutrinos included.

To study properties of mass matrices, we assume that matrix elements, suggested by the approach, are real numbers.
Due to the symmetry of the mass matrix (Eq.(\ref{2times2})), suggested by the approach, and due to the assumption of 
real numbers one then easily finds the eigenvalues of the mass matrices and demonstrate that eigenvalues can agree
with the experimental data and the ref.\cite{okun}. It also turns out that the mass matrix, if fitted to the
experimental data, demonstrate the structure, which is pretty close to the ''democratic'' matrix.

Since the approach is offering indeed $(2^{2k})$, with $k=1,2,3$, families (with the  interaction in higher 
dimensions and 
accordingly with the masses of families, which  depend on the way how the 
 group $SO(1,13)$ breaks and also on the corresponding running coupling constants) we look for the
 eigenvalues of matrices of the dimension $(2^{2k}) \times (2^{2k})$. We find that there are the differences of the
 mass matrix elements, which determine the masses of the four lowest families, while the rest of families lie
 at much higher values.
 
We intend to study further properties of the mass matrices, suggested by the approach unifying spins and charges.
We shall evaluate the mass matrices to predict  differences in mass matrices
among the members of one family, complexity of  matrix elements and others.

\section{Acknowledgement} Two of the authors would like to thank the Ministry of Education, Science and Sport
of Slovenia for funding the research as well as the Workshop. 
All the authors would like to thank the participants at the workshop for really stimulating discussions.

%% SO(1,13)
\title*{%
Coupling Constant Unification in Spin-Charge Unifying Model
Agreeing With Proton Decay Measurements}
\author{N. Manko\v c Bor\v stnik$^{1,2}$ and H. B. Nielsen$^3$}
\institute{%
${}^1$ Department of Physics, University of
Ljubljana, Jadranska 19, 1111 Ljubljana\\
${}^2$ Primorska Institute for Natural Sciences and Technology, 
C. Mare\v zganskega upora 2, Koper 6000, Slovenia\\
${}^3$ Department of Physics, Niels Bohr Institute,
Blegdamsvej 17, Copenhagen, DK-2100}

\authorrunning{N. Manko\v c Bor\v stnik and H. B. Nielsen}
\titlerunning{Coupling Constant Unification}
\maketitle

\begin{abstract} 
We investigate if there could be any contradiction between the measured gauge coupling constants
and the relations between them following from the way of breaking down stepwise the SO(10) GUT to
$SO(4)\times SO(6)$ suggested by a certain spin-charge unifying model proposed by one of us. 
This break down way 
only gives an inequality prediction, telling on which side of SU(5) unification the couplings should fall.
It pushes the unification scale up to $10^{17}\; GeV$ with respect to the best fitting of the $SU(5)$ ($SU(5)$),
thus providing a - very weak! - support for the model mentioned. The proton decay time is accordingly pushed up, 
now with  a model not having ordinary SUSY. 
\end{abstract}
%\pacs{
%04.50.+h, 11.10.Kk,11.30.-j,12.10.-g
%}

\section{Introduction}\label{introduction-so113}

The $SU(5)$ grand unified theory (GUT) is rather predictive concerning the three fine structure 
coupling constants in as far as the three invariant Lie subalgebras of the Standard Model suggest that all
the three corresponding coupling constants get separated at the same scale.
The SUSY-SU(5) coupling predictions of the very unification has much been celebrated. 
Non-susy $SU(5)$-GUT has already since long
\cite{nsgut} been in disagreement with experimentally determined gauge coupling 
constants. Since thus $SU(5)$-G.U.T. is in a bad shape so must be $SO(10)$ broken down along the $SU(5)$-route,
i.e. with $SU(5)$ as an intermediate step. However, if you break down (say) $SO(10)$-G.U.T. along a non-$SU(5)$
route you can easily get so many parameters that you can avoid predictions independent of further 
modelling and thus $SO(10)$ should not be considered killed\cite{SO10rev}.

One obvious way to break down $SO(10)$ is to break it into $SO(4)\times SO(6)$ which as far as the 
Lie algebra is concerned is equivalent to $SU(2)\times SU(2) \times SU(4)$ which can then be further 
broken down by breaking the $SU(4)$ down to $SU(3)\times U(1)$, and one of the $SU(2)$'s to $U(1)$.
Finally the weak hypercharge of the Standard Model can then be identified with a linear combination
of the two $U(1)$, the rest -essentially a $B-L$ charge -  being broken. 

This breaking pattern is closely related to the breaking pattern suggested in connection with the
spin-charge-unifying model put forward by one of us \cite{norma92so,norma93so,norma01so}.
In fact  
this model starts from a $1+13$ dimensional space-time and it contains in principle $SO(10)$ as a subgroup
of its $SO(1,13)$ and thus has an underlying $SO(10)$ symmetry that is even gauged in as far as the
whole $SO(1,13)$ is supposed to be gauged - really due to gravity in the high dimensional space time.
The first breaking - the one at the highest energies - is supposed in this model to be into
$SO(1,7)\times SO(6)$ where the $SO(1,7)$ is supposed to contain the Lorentz group in the 
finally four dimensions also. The part of this $SO(1,7)$ which is purely ordinary gauge transformation
- i.e. not involving any relation to the $1+3$ part of space-time dimensions - is an $SO(4)$ subgroup.
This model has spin-charge-unification in the sense that for instance the whole $SO(1,13)$ is supposed 
to include what at the end shall show up as the Lorentz group in the observed $1+3$ dimensions with all the
known charges. 
Concerning the part of the ``inner space'' meaning spin - which is the main concern of this model - 
this spin got unified with the charges in the sense that the Lorentz group is united with 
the gauge transformation group into the $SO(1,13)$.

\[
\begin{array}{c}

\begin{array}{c}
\underbrace{%
\begin{array}{rrcll}
 & & \SO(1,13) \\
 & & \downarrow \\
 & & \SO(1,7) \otimes \SO(6) \\
 & \swarrow & &  \searrow \\
 & \SO(1,7) & & \SU(4)\\
 \swarrow\qquad & & & \qquad\downarrow \\
 \SO(1,3)\otimes \SO(4) & & & \SU(3)\otimes\unit(1) \\
 \downarrow\qquad & & & \qquad\downarrow \\
\SO(1,3)\otimes\SU(2)\otimes\unit(1) & & & \SU(3)\otimes\unit(1)\\
& & & \\
\end{array}} \\
\SO(1,3)\otimes\SU(2)\otimes\unit(1)\otimes\SU(3)\otimes \unit(1)\\
\end{array}\\
\downarrow \\
\SO(1,3)\otimes\SU(2)\otimes\unit(1)\otimes\SU(3)\\
\end{array}
\]
%*** picture***
%
%\begin{eqnarray}
%\quad \quad \quad \quad SO(1,13)\nonumber\\
%\quad \quad \quad SO(1,7)\times SO(6)\nonumber\\
%\quad \quad SO(1,3)\times SO(4)\times SU(3)\times U(1)_{SO(6)}\nonumber\\
%\quad  SO(1,3)\times SU(2)\times U(1)_{SO(4)} \times SU(3)\times U(1)_{SO(6)}\nonumber\\
%SO(1,3)\times SU(2)\times Y \times SU(3)
%\label{scheme}
%\end{eqnarray}
%
It turns out that in this model, unifying spins and charges\cite{norma01so,pikanormaholger2002},
  in one  $SO(1,7)$ 
spinor multiplet, a left handed $SU(2)$ doublets together with
a right handed $SU(2)$ singlets occur, just as it is needed as an input for the Standard Model.

Apart from the appearance also of the Lorentz group $SO(1,3)$ and some further generators mixing 
the latter with the gauge group $SO(10)$, this model is really much like the $SO(10)$ model, being broken to the 
$ SO(4)\times SO(6)$ way. It is the purpose of the present article to investigate if 
relations among the coupling constants in the Standard Model  - $\alpha_1$, $\alpha_2$, and $\alpha_3$ -
found experimentally as $\alpha_{elm}$, $\alpha_W$, and $\alpha_S$ could possibly be consistent with such a 
break down under the restrictions 
implied by the mentioned model of the spin-charge unification. 
As a test we also present in this (non SUSY ) model the proton decay time, for which it comes out that it causes
no problem.

\section{Group $SO(1,13)$ and subgroups $SO(4)\times SU(4)$ breaking to Standard Model group and $ U(1)_{B-L}$}

If we ignore the geometrical - i.e. Lorentz transformation in $1+ 3$ related - components in the group 
$SO(1,7)$
we are just considering the $SO(4)\times SO(6)$ group at the intermediate level. Note, that
a priori in breaking $SO(10)$ one can have each of these subgroups $SO(4)$ and 
$SO(6)$ breaking further themselves at different scales of energy, thus providing parameters to be fitted
at these scales. This situation means that we have as the unavoidable predictions mainly the relations 
between the ``contributions'' to the (inverse) weak hypercharge gauge coupling squared $\alpha_1$ coming from 
respectively the $SO(6)$ - a contribution unifying with the $SU(3)$ coupling $\alpha_3=\alpha_S$ - and the 
$SO(4)$ unifying with the weak isospin coupling $\alpha_2=\alpha_W$.

\section{Normalization of coupling constants}

Let us here perform a little trivial calculation of the parametrization of the Standard 
Model gauge couplings - in terms of those of $SO(4)$ and $SO(6)$ - as they would appear under the assumptions of:

a) the gauge group $SO(4)\times SO(6)$, the subgroup of $SO(10)$ (lying behind the group of $SO(1,13)$)
giving rise to respectively  
$SU(2)\times U(1)_{SO(4)}$ and $SU(3)\times U(1)_{SO(6)}$,

b) and then having the two $U(1)$'s broken into only one in such a way that the surviving 
$U(1)$ identifies with the Standard Model $U(1)$ as embedded into conventional $SO(10)$ as it
could be containing the mentioned group $SO(4)\times SO(6)$.

Before we shall fix the combination of U(1)-charges to be identified with the Standard Model
weak hyper-charge  we should choose a normalization of the $U(1)$-group couplings for the $U(1)$'s
embedded into respectively $SO(4)$ and $SO(6)$ as 
respectively $U(1)_{SO(4)}\times SU(2)_{SO(4)} \subseteq SO(4)$ and  $U(1)_{SO(6)}\times SU(3)_{SO(6)}
\subseteq SU(4)= SO(6)$
and we shall do that so that $SO(4)$ and $SO(6)$ symmetry ensure the same coupling for
$U(1)_{SO(4)}$ and $SO(4)$ and $SU(2)$ inside it  as well as for $U(1)_{SO(6)}$  and $SO(6)$ and $SU(3)$.

To do that we shall first rearrange the generators $S^{ab}$ of the group $SO(10)$ so that they will manifest 
the subgroups $SO(4)$ and $SO(6)$ structure\cite{norma92so,norma93so,norma01so}. Making a choice that indices 
$k \in 1,2,3,4$ belong to the subgroup $SO(4)$ and $h \in 5,6,7,8,9,10$ to the subgroup $SO(6)$, we find
the $SU(2)\times SU(2)$ substructure of $SO(4)$ and make a choice that the generators
\begin{eqnarray}
\tau^{11}: &=& \frac{1}{2}(S^{14} - S^{23}), \quad \tau^{12}: = \frac{1}{2}(S^{13} + S^{24}), \quad
\tau^{13}: = \frac{1}{2}(S^{12} - S^{34}), \nonumber\\
\tau^{2}: &=& \frac{1}{2}(S^{12} + S^{34}) =: Q_{U(1), SO(4)} 
\label{su2u1}
\end{eqnarray}
close the algebra of the group $SU(2)$ and $U(1)$, respectively
\begin{eqnarray}
[\tau^{Ai},  \tau^{Bj}] = i f^{Aijk} \tau^{Ak} \delta^{AB},  
\label{su2u1com}
\end{eqnarray}
with $f^{1ijk}= \varepsilon^{ijk}$ and $f^{2ijk}=0$, while $\tau^{31}: = \frac{1}{2}(S^{14} + S^{23})$,
$\tau^{32}: = \frac{1}{2}(S^{13} - S^{25})$ together with $\tau^{2}$ form the algebra of the group $SU(2)$ as well.

We express the generators of the subgroups $SU(3)$ and $U(1)$ embedded into $SO(6)$ in terms of $S^{ab}$ as follows
\begin{eqnarray}
\tau^{41}: &=& \frac{1}{2}(S^{58} - S^{67}), \quad \tau^{42}: = \frac{1}{2}(S^{57} + S^{68}), \quad
\tau^{43}: = \frac{1}{2}(S^{12} - S^{34}), \nonumber\\
\tau^{44}: &=& \frac{1}{2}(S^{5 10} - S^{6 9}), \quad \tau^{45}: = \frac{1}{2}(S^{5 9} + S^{6 10}),\quad
\tau^{46}: = \frac{1}{2}(S^{7 10} - S^{8 9}), \nonumber\\
\tau^{47}: &=& \frac{1}{2}(S^{7 9} + S^{8 10}),\quad \tau^{48}: = \frac{1}{2\sqrt{3}}(S^{5 6} + S^{7 8} - 2 S^{9 10})
\nonumber\\
\tau^{5}: &=& (-)\frac{1}{3}(S^{5 6} + S^{7 8} + S^{9 10}) / (\sqrt{2/3}), \quad  \tau'^5 = \sqrt{2/3} \cdot \tau^5
= :Q_{U(1), SO(6)}
\label{su3u1}
\end{eqnarray}
with commutation relations defined in Eq.(\ref{su2u1com}) and with $f^{4ijk}$ which are the standard structure constants
of the group $SU(3)$, while $f^{5ijk}=0$.
We have introduced besides the operator $\tau^5$ also the operator $\tau'^5$, since the second one  plays the role of the
operator determining the number of baryons minus the number of leptons 
\begin{eqnarray}
(B-L):  = 2\cdot \tau'^5 
\label{B-L}
\end{eqnarray}
 and in addition, it is 
also the operator which, together with the operator $\tau^2$, defines the operator 
$Y^1$, which corresponds to 1/2 of the hypercharge $ y $ appearing in the Standard Model
\begin{eqnarray}
y/2: = Y^1=
(\tau^2 + \tau'^5).
\label{hypercharge}
\end{eqnarray}
We also may define accordingly the operator $Y^2$ as an additional hypercharge, 
since $Y^1$ and $Y^2$ are the two orthogonal charges,
while $Y^1$ and $(B-L)$ are not 
\begin{eqnarray}
Y^2=
(-\tau^2 + \tau'^5)
\label{hypercharge2}.
\end{eqnarray}

The covariant derivative for spinor representations in the model, which starts with $SO(10)$ (in the 
proposed model it really starts with $SO(1,13)$), is as follows
\begin{eqnarray}
D_{\mu}= \partial_{\mu} -  g\frac{1}{2}S^{ab} \omega_{ab\mu},
\label{covders}
\end{eqnarray}
where $g$ is the coupling constant of the group $SO(10).$

Taking into account only the $SU(2)\times U(1)_{SO(4)}\times SU(3)\times U(1)_{SO(6)}$ substructure of the 
group $SO(10)$ (or the $SO(1,3) \times SU(2) \times U(1)_{SO(4)}\times SU(3)\times U(1)_{SO(6)}$ substructure 
 of the group $SO(1,13)$) we can write for spinor representations  the corresponding covariant derivative 
 (in the case of $SO(1,13)$ this follows, if gravity in the four dimensional subspace is neglected)
\begin{eqnarray}
D_{\mu} = \partial_{\mu}  - \frac{g}{\sqrt{2}}\{\tau^{2}  A_{\mu}^{2} +
 \tau^5  A_{\mu}^{5} + \vec{\tau^1}\cdot \vec{A}_{\mu}^{W}
+ \vec{\tau^4}\cdot \vec{A}_{\mu}^{S}\} \label{covder}
\end{eqnarray}
and the corresponding Lagrange density 
\begin{eqnarray}
{\cal L} = \bar{\psi}\gamma^{\mu} P_{\mu}  \psi
\end{eqnarray}
with $P_{\mu}= i D_{\mu}$. The (only one) coupling constant in Eq.(\ref{covder}) is  clearly related to the 
coupling constant in Eq.(\ref{covders}), since fields $A^{Ai}_{\mu}$ are orthonormal superpositions of 
$\omega_{ab\mu}$. ( We present here two particular  examples, which are 
$A^{11}_{\mu} =  (\omega_{14 \mu} - \omega_{23 \mu})/\sqrt{2}$, and $A^{48}_{\mu} = 
(\omega_{56 \mu} + \omega_{23 \mu} -2 \omega_{9 10 \mu})/\sqrt{6}$. The rest of the fields $A^{Ai}$, expressed
in terms of  $\omega_{ab \mu}$, can be found in ref. \cite{norma01so} with the normalization factor $1/2$ rather  
than $1/\sqrt{2}$ as here).

We also can write the covariant derivative presented in Eq.(\ref{covder}) in a way, which points out the Standard 
Model structure of the group $SO(10)$
\begin{eqnarray}
D_{\mu} &=& \partial_{\mu} -  g_{U(1), SO(4)} \;\tau^{2}  A_{\mu}^{2} -
g_{U(1), SO(6)} \;\tau'^5  A_{\mu}^{5} \nonumber \\
 &{\ }& - g_W \; \vec{\tau^1}\cdot \vec{A}_{\mu}^{W}
- g_S\; \vec{\tau^4}\cdot \vec{A}_{\mu}^{S},
\label{covderSM}
\end{eqnarray}
with 
\begin{eqnarray}
g_W = g_S = g/\sqrt{2}= g_{U(1), SO(4)},\;\; g_{U(1),SO(6)}= \sqrt{3/2}\; g_{U(1),SO(4)},
\label{newgu1}
\end{eqnarray}
in agreement with Eq.\ref{su3u1}.

The notation points out that the coupling constants of the two $U(1)$ subgroups belong to the two 
different starting subgroups, namely to $SO(4)$ and $SO(6)$. 

From Eq.(\ref{newgu1}) it follows that
\begin{eqnarray}
 (g_{U(1), SO(6)})^2 : (g_{U(1), SO(4)})^2 = 3:2.
\label{grel}
\end{eqnarray}
Introducing the notation
\begin{eqnarray}
 \hat{A}^{Ai}: =  g_{A} A^{Ai}_{\mu}, \quad A=\{ 1,2,4,5\}, i=\{1, n_{a}\}, 
\label{hatA}
\end{eqnarray}
with $n_1 =3, n_2 =1, n_4 = 8, n_5 =1$,
we may write the Lagrange density for the gauge fields
\begin{eqnarray}
 {\cal L} = - \frac{1}{4(g_{A})^2} Tr ( \hat{F}^{ Ai \mu \nu} \hat{F}^{Ai}_{\mu \nu}),
\label{FFlagrange}
\end{eqnarray}
with summation performed over $A$ and $i$ and with 
\begin{eqnarray}
\hat{F}^{Ai}_{ \mu \nu}:= g_{A} \tau^{Ai} \; ( \;  \partial_{\mu} A^{Ai}_{\nu} - \partial_{\nu} A^{Ai}_{\mu} - 
 \;f^{Aijk} A^{Aj}_{\mu} A^{Ak}_{\nu}\;  ) .
\label{hatF}
\end{eqnarray}

We present in Table~\ref{table1so} quantum numbers of one irreducible representation of the subgroup $SO(1,7)$ of the group
$(SO(1,13))$. It coincides with one generation of fermions (left handed weak charged doublets and right handed weak
chargeless singlets) of the Standard Model. We find them  using the technique 
presented in refs.\cite{norma01so,normaholgertec2002}, since with the help of this technique 
one easily finds vectors and the corresponding quantum numbers of one generation. 

\begin{table}
\begin{center}
\begin{tabular}{||c|c|c|cccccc|cccccc||}
\hline
\hline
 \multicolumn{3}{||c|} {}&
 \multicolumn{6}{c|}{SU(2) doublets} & 
 \multicolumn{6}{c||}{SU(2) singlets}\\ 
 \multicolumn{3}{||c|} {} & $\tau^{13}$& $
\tau^{2} $& $\tau'^5$ &$Y_1$&$Y_2$ &
$\Gamma^{(4)}$ & $\tau^{13}$& 
$\tau^{2}$ & $\tau'^{5}$ & $Y_1 $& $Y_2$&
$\Gamma^{(4)}$ \\ \hline
SU(3)&$\tau^{4\;3}\;\;$  =  & 
$u_i$& 
1/2 & 0 & 1/6 & 1/6 & 1/6 & - 1 & 0 & 1/2 & 1/6 & 2/3 & -1/3 & 1
\\ 
tri-&  $(  \frac{1}{2},$ $ -\frac{1}{2},\;\;$
 $ 0\;$  )& &\multicolumn{6} {c|} {} &\multicolumn{6}{c||}{}\\
plets&$\tau^{4\;8}\;\;$  =  & 
$d_i$ & -1/2 & 0 & 1/6 & 1/6 & 1/6 & -1 & 0 & -1/2 & 1/6 & -1/3
& 2/3 & 1 \\
& $( \frac{1}{2 \sqrt{3}},$
$\frac{1}{2 \sqrt{3}},$  $-\frac{1}{ \sqrt{3}}$  )&&\multicolumn{6} {c|} {}&\multicolumn{6} {c||} {}\\  \hline 
SU(3)&$\tau^{4\; 3} =  0$ & $\nu_i$ & 1/2 & 0 & -1/2 & 
-1/2 & -1/2 & -1 & 0 & 1/2 & -1/2 & 0 & -1 & 1\\
singlets&$\tau^{4\; 8} = 0$ & 
$e_i$ & -1/2 & 0 & -1/2 & 
-1/2 & -1/2 & -1 & 0 & -1/2 & -1/2 & -1 & 0 & 1\\ \hline
\hline
\end{tabular}
\end{center}
\caption{%
Expectation values of the Cartan subalgebra of the group $SU(3)\times U(1)\subset SU(6)$ and of the group 
$SU(2)\times U(1) \times SO(1,3) \subset SO(1,7)$, 
following ref.\cite{norma01so}.
We chose the  representations to be eigenvectors of the operators $\tau^{13}$ (Eq.\ref{su2u1}) and
$\tau^2$ (Eq.\ref{su2u1}), $\tau^{43}$ and $\tau^{48}$ (Eq.\ref{su3u1})  and $\tau'^5$ (Eq.\ref{su3u1}) 
as well as the operator of handedness
$\Gamma^{(4)}$ ($\Gamma^{(4)} = \mp 1$, representing  left and right handed spinors, respectively). 
We also present the sum and the difference of the two $U(1)$ operators $\tau^5$ and $\tau^2$, which are $Y^1$
and $Y^2$, respectively.} 
\label{table1so}
\end{table}

According to the quantum numbers, presented in Table~\ref{table1so} , we easily see that the operators $\tau^{1i}$ of the group 
$SU(2),$  the operators $\tau^{4i}$ of the group $SU(3)$ and the operators $\tau^2$ are normalized so that
\begin{eqnarray}
Tr\{\tau^{Ai}\tau^{Bj}\} = 2\delta^{AB}\delta^{ij}, 
\label{normal}
\end{eqnarray}
since we find if following Table~\ref{table1so} for, let say,  $Tr\{(\tau^{13})^2 \} = 3 \cdot (1/2)^2 + 3 
\cdot (-1/2)^2 + 3\cdot (0)^2
+3 \cdot(0)^2 + 1\cdot (1/2)^2 + 1\cdot (-1/2)^2 + 1\cdot (0)^2 + 1 \cdot(0)^2= 2 $, for, let say, 
$Tr\{(\tau^{48})^2 \} = 4 \cdot (1/(2\sqrt{3}))^2 + 4 \cdot (1/(2\sqrt{3}))^2 + 4 \cdot (-1/\sqrt{3})^2 + 
4 \cdot (0)^2
=2$ and for $Tr (\tau^2)^2 = 3 \cdot (0)^2 + 3 \cdot (0)^2  + 1\cdot (0)^2 + 
1\cdot (0)^2 +  3 \cdot (1/2)^2 + 3 \cdot (-1/2)^2  + 1\cdot (1/2)^2 + 
1\cdot (-1/2)^2 =2$. While the operator $\tau^5$ is also normalized according to Eq.(\ref{normal}) to $2$, the operator 
$\tau'^5$ is normalized instead to $4/3$. 
We see this, since
according to Table~\ref{table1so} we find that  $Tr\{ (\tau'^5)^2\} =   3 \cdot (1/6)^2 + 3 \cdot (1/6)^2  + 1\cdot (-1/2)^2 + 
1\cdot (-1/2)^2 +  3 \cdot (1/6)^2 + 3 \cdot (1/6)^2  + 1\cdot (-1/2)^2 + 
1\cdot (-1/2)^2 = 2\cdot 2/3$. Similarly we find that both $Tr\{ (Y^1)^2 \} = 2 \cdot 5/3$ $(Y^1 = \tau^2 + \tau^{'5} )$ 
and $Tr\{(Y^2)^2 = 2 \cdot 5/3, \; ( Y^2= -\tau^2+\tau^{'5}) \}. $ We namely find that $Tr\{ (Y^1)^2 \}  = 3 \cdot (1/6)^2 + 
3 \cdot (1/6)^2  + 1\cdot (-1/2)^2 + 1\cdot (-1/2)^2 +  3 \cdot (2/3)^2 + 3 \cdot (-1/3)^2  + 1\cdot (0)^2 + 
1\cdot (-1)^2 = 2 \cdot 5/3 = Tr\{ (Y^2)^2\}$.

\section{Breaking of  (B-L) gauge symmetry}
\label{breaking}

We recognize, as we said, in the quantum number $Y^1 = y/2$ the hypercharge of the Standard Model and in 
$2 \tau'^5 = (B-L) $ the baryon minus lepton quantum number. We could instead of the 
$(B-L)$ quantum number as well use the quantum number $Y^2$
as the additional quantum number to $Y^1$ (Eq.(\ref{hypercharge2})). The first one is the sum of the two $U(1)$
generators  belonging to the subgroups $SO(6)$ and $SO(4)$, respectively, of the group SO(10), namely 
($\tau^2 + \tau'^5$) (Eq.(\ref{hypercharge})), while $(B-L) = 2\tau'^5$ and $Y^2 =(\tau^2 + \tau'^5)  $ 
(Eq.(\ref{hypercharge2})).
 
Let us now come back to the covariant derivative (Eq.(\ref{covder})) manifesting the Standard Model 
structure of the group $SO(10)$ and let us   assume that there is a Higgs field 
%
%
%(**HOLGER, I AM OFFERING THE ALTERNATIVE $Y^2$ INSTEAD OF $(B-L)$, WHICH WORKS AS WELL, 
%LEADING TO THE SAME REQUIREMENT FOR THE
%TWO FIELDS $\hat{A}^5_{\mu} = \hat{A}^2_{\mu}$. I KNOW, $(B-L)$ IS MORE POPULAR, WHILE $Y^1$ AND $Y^2$ ARE
%LINEARLY INDEPENDENT. I LEAVE YOU TO DECODE  EITHER $(B-L)$ OR $Y^2$.**)
%
 $\phi_{Y^2}$ %(or  $\phi_{(B-L)}$)
with the vacuum expectation value much larger than the weak scale, coupling alone  to the $Y^2$
%( or to the $(B-L)$ $=2 \cdot \tau'^5$)
hypercharge
quantum number but not to the $Y^1=y/2$ hypercharge [which means that the vacuum expectation  value of 
$Y^2,$ $<\phi|Y^2|\phi> \neq 0$ %(or $(B-L),$ $<\phi|(B-L)|\phi> \neq 0$),
while  $<\phi|y|\phi>=0$]  since it would make say $Z^0_{\mu}$ too heavy. 
 ( Let us remaind the reader that the quantum numbers of the Standard Model Higgs just agree, up to  the spin and
  accordingly the handedness, with the
 quantum numbers of a spinorial 
field, which is ( a left handed) $SU(2)$ doublet 
and a $SU(3)$ singlet ( Table~\ref{table1so})).

Since 
\begin{eqnarray}
\hat{A}^{Y^2}_{\mu} = ( \hat{A^5}_{\mu} - \hat{A^2}_{\mu})/\sqrt{2} ,
\label{slave0}
\end{eqnarray}
with 
$A^{2}_{\mu}$ and $ A^{5}_{\mu}$ defined in Eq.\ref{hatA}, at low energy it must be - due to the sea-gull term
for the Higgs field $\phi_{Y^2}$ ($\phi_{B-L}$)- that
\begin{eqnarray}
\hat{A^5}_{\mu} = \hat{A^2}_{\mu}.
\label{slave1}
\end{eqnarray}
As long as the relation of Eq.(\ref{slave1}) is valid, the only charge among the two abelian charges ($Y^1$ and $Y^2$),
which matters, is $Y^1= y/2$. Effectively, $\tau'^5 = (B-L)/2$ may also be put to zero, so that effectively
\begin{eqnarray}
(\frac{y}{2})_{{\rm effective}} = \frac{1}{2}Q_{U(1),SO(4)},
\label{slave}
\end{eqnarray}
but with an effective Lagrange density ${\cal L}_{(\frac{y}{2})_{effective}}$
\begin{eqnarray}
({\cal L}_{\frac{y}{2}})_{{\rm effective}} = -\frac{1}{4 (g_{U(1),SO(4)})^2} \cdot (1 + \frac{2}{3})
\hat{F}^{U(1),SO(4)}_{\mu \nu}.
\label{slave2}
\end{eqnarray}
Denoting the $Y^2$ (the $(B-L)$) gauge breaking scale weak hypercharge coupling by $g'= g_{\frac{y}{2}}$,
the covariant derivative has the form
\begin{eqnarray}
D_{\mu}= \partial_{\mu} - (y/2) \hat{A'^y}_{\mu} -
  \vec{\tau^W}\cdot \vec{\hat{A}_{\mu}^{W}}
  -\vec{\tau^S}\cdot \vec{\hat{A}_{\mu}^{S}},
\label{covderstan}
\end{eqnarray}
with  $\tau^{Wi} = \tau^{1i} $, $\tau^{Si} = \tau^{4i} $ and with  the correspondingly defined 
$\hat{A'}^{y}_{\mu} = \hat{A'}^{Y^1}_{\mu},  \hat{A}^{Wi}_{\mu} = \hat{A}^{1i}_{\mu}, 
\hat{A}^{Si}_{\mu} = \hat{A}^{4i}_{\mu}$.

Due to the slavery following from equation (\ref{slave1}) we have for the $y$-part of the Lagrange density
\begin{eqnarray}
 {\cal L}_{(y/2)_{effective}} &=& -\frac{5}{3}\;\frac{1}{4}\; \frac{1}{(g_{U(1),SO(4)})^2} Tr(\hat{F}^{U(1),SO(4) \mu \nu} 
 \hat{F}^{U(1),SO(4)}_{\mu \nu})
 \nonumber\\
 &=& - \frac{1}{4}\frac{1}{(g')^2} Tr(\hat{F'}^{ \mu \nu} \hat{F'}_{\mu \nu}),
\label{FFy/2}
\end{eqnarray}
with
\begin{eqnarray}
\frac{1}{(g')^2} = \frac{1}{(g_{U(1),SO(4)})^2} + \frac{1}{(g_{U(1),SO(6)})^2} = \frac{5}{3} 
\frac{1}{(g_{U(1),SO(4)})^2}
\label{slaveg}
\end{eqnarray}
and accordingly $\hat{A'}^{Y^1}_{\mu} = g' {A'}^{Y^1}_{\mu}$.

This is just the usual $SU(5) \subset SO(10)$ prediction for the $U(1)$ coupling constant, assuming 
that the break occurred at the unification scale, the same for $SO(4)$ and $SO(6)$. After the $Y^2$  
and accordingly the $(B-L)$ is broken,
the Standard Model coupling constants 
%$(\alpha_{A})^{-1} = 4 \pi / (g_A)^2 $, with $A= (U(1),SO(4))($or $2),
%(U(1),SO(6)) ($ or $3)$
%or prime $(')$,
further run, leading to the three lines crossing at one point if $SU(5) ( \subset SO(10))$ would work
\begin{eqnarray}
\frac{1}{\alpha'_{SU(5)}} \dot{=} \frac{3}{5} \frac{1}{\alpha'}.
\label{slavealpha}
\end{eqnarray}
Instead we have
\begin{eqnarray}
\frac{1}{\alpha'_{SU(5)}} = \frac{3}{5} (\frac{1}{\alpha_{U(1),SO(4)}} + \frac{1}{\alpha_{U(1),SO(6)}})=
 \frac{3}{5} \frac{1}{\alpha_2} + \frac{2}{5}  \frac{1}{\alpha_3}.
\label{slavealphaSM}
\end{eqnarray}

\begin{figure}
\centering
\includegraphics[width=12cm]{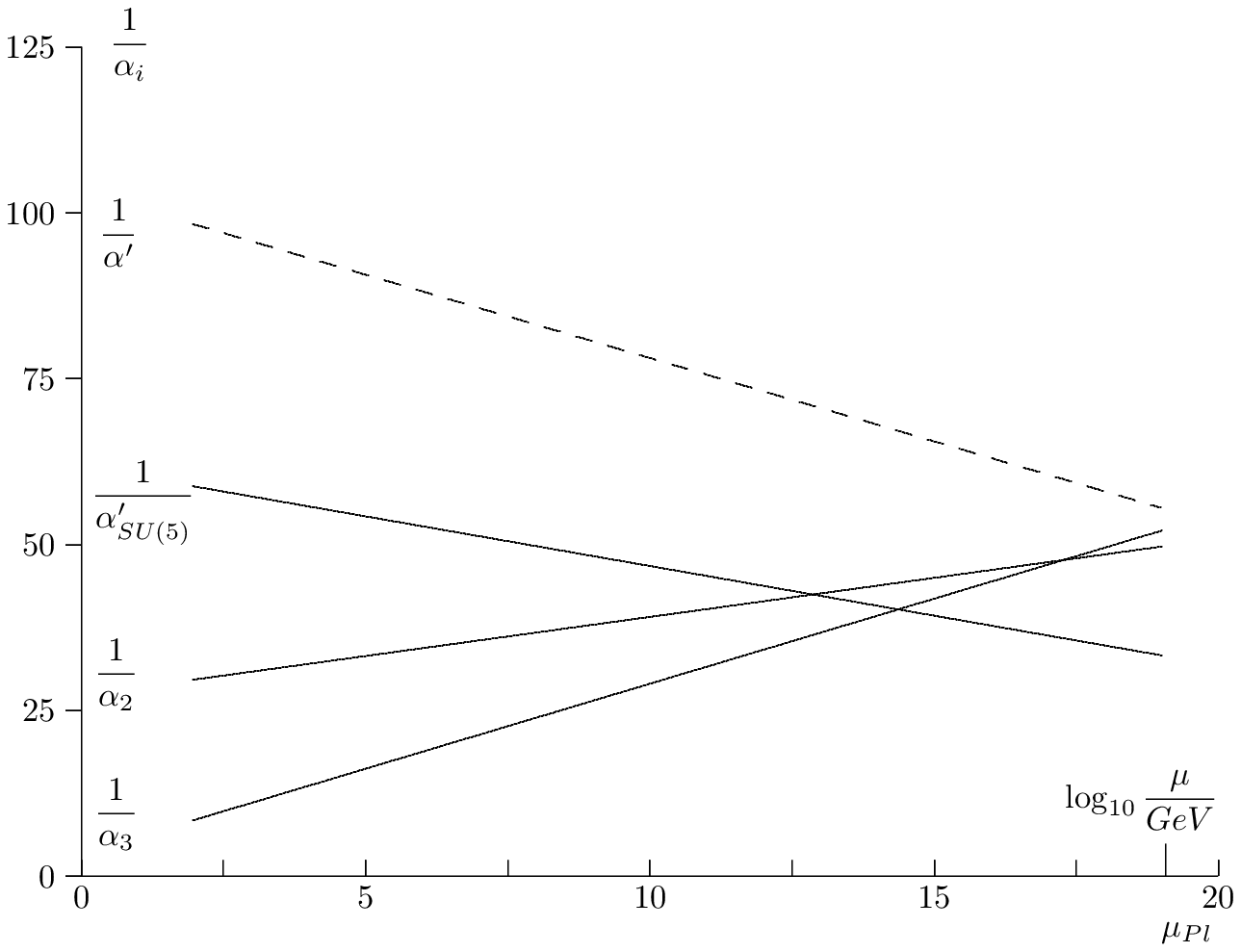}
\caption{%
The running coupling constants are presented, extrapolated from the experimental values
by the assumption that they run independently up to the unification scale. They would meet in one 
point, if the $SU(5)$ unification would work.}
\label{nmb-so-f1}
\end{figure}

We present a plot of experimental values of $(\alpha')^{-1}, (\alpha_2)^{-1}$ and $(\alpha_3)^{-1}$, as well as
of $(\alpha'_{SU(5)})^{-1}= \frac{3}{5}(\alpha')^{-1}$ on Fig.~\ref{nmb-so-f1}. 

The main point of this paper is, however, to see what happens, if breaking went differently.
We may assume that above some scale, say $10^{17} {\rm GeV}$, it is only one group and accordingly
only one coupling constant, the one 
of the group $SO(1,13)$ or rather $SO(10)$, since in this paper we do not pay attention to the 
gravitational (that is the spin ) part 
$SO(1,3)$ of the internal space. At $10^{17} {\rm GeV}$ then $SO(1,13)$ breaks to $SO(1,7)$ and $SO(6)$. 
The $SO(4)$ part of the group 
$SO(1,7)$ runs now with a slower rate than $SO(6)$ (which runs approximately as $SU(4)$). Then $SU(4)$ 
at around say $10^{16}$ further breaks to $SU(3)\times U(1)$ and accordingly slows the rate, 
while $SO(4)$ breaks to $SU(2)\times SU(2)$ without changing the rate (since 
$SU(2)\times SU(2)$ are the invariant subgroups of the group SO(4) - with the same number of generators).
One of the $SU(2)$ further breaks to $U(1)_{SO(4)}$, this $U(1)_{SO(4)}$ determining with $U(1)_{SO(6)}$ the 
hypercharge $y = Y^1$ (Eq.\ref{hypercharge}) of the Standard Model, while the second hypercharge 
$Y^2$ (Eq.\ref{hypercharge2}) breaks, say, at around $3\cdot 10^{13} {\rm GeV}$. 
%which is above when $SO(4)$ and $SO(6)$  break  at orders of magnitude 
%different scales. 
The break of the hypercharge $Y^2$ changes the slope of the $(\alpha'_{SU(5)})^{-1}$ and of $(\alpha_3)^{-1}$
but not of $(\alpha_2)^{-1}$. 

Since the $SO(4)$ stayed unbroken down to much lower energy scale than $SO(6)$,
which broke close to the $SO(10)$ unification scale,  (for this kind of assumption speaks the 
fact that one $SO(1,7)$ multiplet contains left handed $SU(2)$ doublets and right handed $SU(2)$ 
singlets as it can be found in ref\cite{norma01so,pikanormaholger2002}), 
the $\alpha _{U(1),SO(4)}$ would at the lower scales
be stronger relative to $\alpha _{U(1),SO(6)}$ than the ideal $SO(10)$ unification requires
($\alpha _{U(1),SO(6)} : \alpha _{U(1),SO(4)} = 2:3$) and the unification  can be made so that
all three coupling constants meet at the grand unification scale at arround $10^{17} GeV$. 
%We assume that $Y^2$ (or $(B-L)$)  broke at the  
%energy scale around $10^{13}$ GeV ($=log_{10}\frac{\mu}{GeV}$).

\begin{figure}
\centering
\includegraphics[width=12cm]{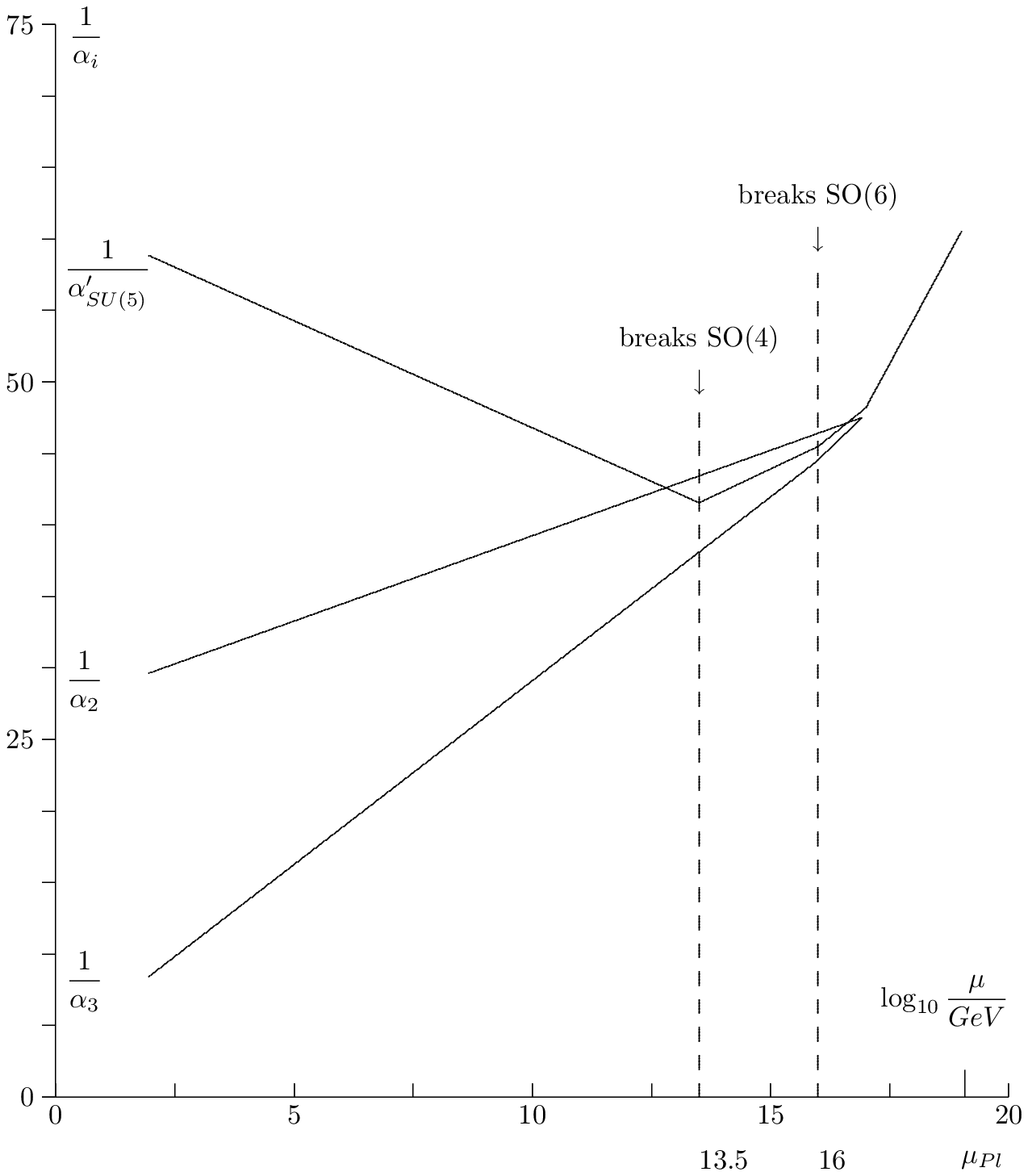}
\caption{%
The running coupling constants are presented, extrapolated from the experimental values
by the assumption that the gauge group $SO(4)$ breaks at much lower  scale ( at around $10^{13}$ GeV)
than the gauge group $SO(6)$ (which breaks at around $10^{17}$ GeV). The $SU(2)$ gauge group coupling constant
does not change when running together with $(\alpha_{U(1),SO(4)})^{-1}$.  The three coupling constants can
meet at the same point and then run together as $SO(1,13)$ (rather SO(10)).}
\label{nmb-so-f2}
\end{figure}

This kind of proposed unification can be seen on Fig.~\ref{nmb-so-f2}.

\section{Numerology for almost viable nice picture}
\label{numerology}

From the fitting to the gauge coupling constants we got that the ratio of the scale of
the $SO(10)$ breaking, which really
means the scale where $SO(1,13)$ breaks to $SO(1,7) \times SO(6)$ ( so that the $SO(10)$
gets restored above the scale in question) to the scale 
 of $SO(1,7)$ breaking,  is  $\frac{ 10^{17}GeV}{3\cdot 10^{13}GeV} \approx 3000$.
Over this scale ratio we have to have the $4$ extra dimensions implied by the $SO(1,7)$ rather than $SO(1,3)$.
This means that for instance the Newton constant $G$ is getting weakened by a factor $3000^4$
where the number 4 is the number of extra dimensions. That means in terms of the Planck scale, which is
$1/\sqrt{G}$, that it goes up compared to the ``fundamental scale '' by a factor $3000^{4/2}$ $\approx$ $10^{7}$.
In other words with the by the phenomenologically given effective Planck scale of $2\cdot 10^{19} GeV$
we should have a fundamental scale of $2*10^{12}$ GeV. This matches badly with our 
$SO(10)$ unifying scale
which became $10^{17}$ GeV. In fact we should of course expect the fundamental scale to be above the 
breaking scale of $SO(1,13)$ which is in the model representing the breaking scale of the ( hidden ) $SO(10)$.
So really it looks like the inequality expected is not fulfilled.

Similarly to the weakening of the gravitational coupling we also have of course a weakening of the 
gauge couplings. Here it must though be admitted that we do not find it attractive with 
a weakening factor of the order of $3000^4$ for the fine structure constants. It would namely mean that 
at the fundamental scale and measured in fundamental units - remember they have dimension in other
dimensions than 4 - the fine structure constants would be about $1/50 * 3000^4$ $\approx   10^{12}$
which seems a very strong coupling that would probably confine or become strange and unbelievable otherwise.
Rather we should take say that the couplings have $\alpha$'s of order unity at the fundamental level,
and then we would rather expect an weakening factor of $50$. That would with $4$ extra dimensions 
mean the fourth root of $50 = 2.7$ scale ratio of $SO(1,7)$ breaking to that of $SO(1,13)$.

This naive scaling calculation does not take into account the running coupling constants in higher dimensions
properly, it treats them naively - as in four dimensional space. This fact could of course be reason for the
above result.

\section{Proton decay}

The scale of $SO(10)$ or $SO(1,13)$ breaking becoming to $SO(1,7) \times SO(6)$ occurs in the proposed approach
at around $10 ^{17}$ GeV. Further breaking of $SO(6)$ to $SU(3)\times U(1)$ occurs at around 
$10^{16}$ GeV, while $SO(4)$ breaks at around $3 \cdot 10^{13}$ GeV. Since neither a $SO(4)$ multiplet nor
a $SO(6)$ multiplet includes both quarks and antiquarks, while the $SO(1,13)$ multiplet 
does (to see this the reader can have a look in Table I of the ref.\cite{pikanormaholger2002}, 
presented in this Proceedings), 
a proton decay may occur at the $SO(1,13)$ ($SO(10)$) unification scale. The 
decay rates corresponding to a massive  gauge particle exchange with a mass $M$ of 
this last scale ($10^{17}$ GeV).
Since the proton lifetime is approximately proportional to $(1/\alpha^2_3)_{{\rm at} 10^{17}} \times M^4 /
m_p^5 $, while the best $SU(5)$ GUT fitting with the unification scale $10^{15}$ GeV  and $1/\alpha  \approx 43$ 
gives the life time of a proton 
equal to $10^{33}$ years, the unification scale  at $10^{17}$ and the  inverse coupling constant  $\approx 47$
makes the life time of the proton for more than $8$ order of magnitude longer, and accordingly causes no problem
in the approach unifying spins and charges, at least not in this simplified version.

\section{Conclusion}

In this paper we have demonstrated that the approach of one of us, unifying spins and charges, suggests the
breaking of $SO(1,13)$ which leads to the unification scale for the three coupling constants at around 
$10^{17} GeV$ and accordingly to acceptable (not yet measurable) life time of the proton (since neither of the subgroups
$SO(4)$ and $SO(6)$ includes both coloured and anticoloured representations and could  correspondingly cause
no proton decay at lower energy  scale when $SO(4)$ alone or  $SO(6)$ alone is the unifying group). 
But it is also true that the model unifying spins and charges  proposed by one of us allows for enough many 
parameters so that the prediction is indeed an inequality prediction. This 
inequality prediction happens to be fulfilled experimentally, i.e.
by the measured fine structure constants, and thus the model passed 
a little test that with fifty percent probability could have gone wrong. 
The model like other unifying models, however, tends to get so low 
unification scale (for at least $SU(2)\times U(1)$ into $SO(4)$) that proton decay  could become a problem: 
it could become a 
difficulty to the non-observation of proton decay. But this is not happening in this model. 
Further there appears a related problem: Because extra dimensions are 
in principle to appear in the model,  an assumption of the Newton constant being 
of order unity at the fundamental scale in fundamental units leads to a 
fundamental scale lower in energy than the usual Planck scale. That brings 
this fundamental scale bellow of the unification of the whole group. This is not what would support the model,
unless this effect is
an artifact of the too simplified treatment of the running coupling constants in more than
four dimensional spaces.

%When one has extra dimensions for instance the fine structure constants
%that in 4 dimensions - happen to be - is dimensionless gets
%dimensionized.
%In fact the dimension of the inverse fine structure constant $4\pi/g^2$
%is in d dimensions $mass^{d-4}$. When you therefore describe its size
%in terms of the renormalization scale $\mu$ its size come in these units
%to vary as a power over those scales where one has the higher dimension.
%When the dimension is above four the running of the size of the inverse
%fine structure constant is so that the coupling gets weaker towards
%smaller $\mu$ by a factor 16 say for each factor two in the extra
%dimension size.  

\section{Acknowledgement } 

This work was supported by Ministry of Education, 
Science and Sport of Slovenia and Ministry of Science of Denmark.

%
%%%%%%%%%%%%%%%%%%%%%%%%%%%%%%%%%%%%%%%%%%%%%%%%%%%%%%%%%%%%%%%%%%%%%

\end{document}